\documentclass[preprint,12pt]{elsarticle}
\pdfoutput=1 
\usepackage[]{graphicx}

\usepackage{amssymb}
\usepackage{amsmath,amsfonts}
\journal{Mechanical Systems and Signal Processing}

\begin{document}

\begin{frontmatter}



\title{Transient Propagation and Scattering of Quasi-Rayleigh Waves in Plates: Quantitative comparison between Pulsed TV-Holography Measurements and  FC(Gram) elastodynamic simulations}


\author[1]{Faisal Amlani}
\address[1]{Department of Aerospace and Mechanical Engineering, University of Southern California, Los Angeles, CA 90007 (USA)}

\author[2]{Oscar P. Bruno}
\address[2]{Applied and Computational Mathematics, California Institute of Technology 217-50, Pasadena, California 91125 (USA)}

\author[3]{Jos\'e Carlos L\'opez-V\'azquez\corref{cor1}}
\ead{jclopez@uvigo.es}
\author[3]{Cristina Trillo}
\author[3]{\'Angel F. Doval}
\author[3]{Jos\'e L. Fern\'andez}
\author[3]{Pablo Rodr\'iguez-G\'omez} 
\address[3]{Departamento de F\'isica Aplicada, Universidade de Vigo, Escola de Enxe\~ner\'ia Industrial, Rua Maxwell, s/n, Campus Universitario, E36310 Vigo, Spain}

\cortext[cor1]{Corresponding author}

\begin{abstract}
We study the scattering of transient, high-frequency, narrow-band quasi-Rayleigh elastic waves by through-thickness holes in aluminum plates, in the framework of ultrasonic nondestructive testing (NDT) based on full-field optical detection. Sequences of the instantaneous two-dimensional (2-D) out-of-plane displacement  scattering maps are measured with a self-developed Pulsed TV Holography (PTVH) system. The corresponding simulated sequences are obtained by means of a FC(Gram) elastodynamic solver introduced recently, which implements a full three-dimensional (3D) vector formulation of the direct linear-elasticity scattering problem. A detailed quantitative comparison between these experimental and numerical sequences, which is presented here for the first time, shows very good agreement both in the amplitude and the phase of the acoustic field in the forward, lateral and backscattering areas. It is thus suggested that the combination of the PTVH system and the FC(Gram) elastodynamic solver provides an effective ultrasonic inspection tool for plate-like structures, with a significant potential for ultrasonic NDT applications.
\newline
\end{abstract}


\end{frontmatter}

\newpage

\section{Introduction}
\label{section1}
We study the scattering of transient, high-frequency, narrow-band quasi-Rayleigh elastic waves by through-thickness holes in aluminum plates, in the framework of ultrasonic nondestructive testing (NDT) based on full-field optical detection. Our work combines an innovative experimental pulsed TV holography (PTVH) system~\cite{fdezijo07} and a state-of-the-art three-dimensional (3D) FC(Gram) numerical elastodynamic solver~\cite{amlani2016fc} to obtain corresponding sequences of the instantaneous out-of-plane displacements at the fully two-dimensional (2D) top surface of the plates. For the first time, a detailed quantitative comparison between these experimental and numerical sequences is presented, showing very good agreement both for the amplitude and the phase of the acoustic field in the forward, lateral and backscattering areas. It is thus suggested that the combination of the PTVH system and the FC(Gram) elastodynamic solver provides an effective ultrasonic inspection tool for plate-like structures.

As it is well-known, ultrasonic techniques are a mature and powerful tool for NDT and evaluation in industry~\cite{HNDT_Bricks} with well-known benefits (deep-penetration capability with high degree of interaction with flaws or inhomogeneities, wide temporal bandwidth, high information content,...) and a wide range of different applications, particularly, in the case of ultrasonic guided waves NDT~\cite{Rose_hot,ews302}. Its basic principle can be described in three basic steps: first, an adequate ultrasonic field is insonified in the part to be inspected; second, interaction with defects in the material produce scattering of the insonified field; third, some physical quantity associated with the acoustic field is measured providing information about the associated scattering phenomena. 

This general scheme is implemented in classical pointwise techniques employing pulse-echo or pitch-catch configurations~\cite{HNDT_Bricks} and using excitation and detection at one or two points with contact transducers (typically PZT), which renders outputs signals with high resolution in time and frequency, but normally with low spatial information content. From this well-established classical schemes, ultrasonic NDT technology has progressed in the last decades in two main complementary directions: on the one hand, developing non-contact inspection approaches~\cite{noncontact} (f.i. air-coupled devices~\cite{aircoupled}, electromagnetic acoustic transducers~\cite{EMATS,inplane} or pointwise optical generation and probing of ultrasound~\cite{Scruby,int35}) and, on the other hand, obtaining more spatial information content about the ultrasonic field in contact techniques (f.i. pointwise techniques with phased array transducers to control the directivity of excited or detected waves~\cite{phase} or combined with some kind of scanning in detection~\cite{Knuuttila_IJO} or excitation~\cite{tom35} or techniques with transducers spatially distributed or scanned over the part to be inspected combined with tomographic approaches~\cite{ews57b}).

In particular, optical probing of ultrasound with full-field techniques (f.i. interferometry~\cite{Nakano_IJO}, holographic interferometry~\cite{Blackshire_IJO,ews59}, speckle photography~\cite{kum01}, shearography~\cite{ews68} or TV-holography~\cite{trilloao03}), are intrinsically non-contact and specially well-suited for obtaining spatial information by registering images---i.e. avoiding any type of scanning over the part to be inspected--- with high lateral spatial resolution and irradiance distributions correlated with the acoustic field at a given time interval, and allow the inspection or remote or inaccessible areas maintaining the previously mentioned classical benefits of NDT with ultrasound. Within this framework, those of us at the University of Vigo have demonstrated that ultrasonic non-destructive inspection can be performed in plates using a self-developed pulsed TV-holography (PTVH) system~\cite{fdezijo07,trilloao03,trillomst03,cernadasmssp06} that records the two-dimensional (2D) spatial distribution of the acoustic field corresponding to the  instantaneous out-of-plane displacements over the surface of the plate. As it is characteristic of full-field optical techniques, our PTVH output has high spatial resolution with a large number of spatial samples (about $10^{6}$), even though the number of temporal samples is low (typically lower than 128). Also, in our case, the field of view for the detected 2D scattering pattern, which can be easily adapted simply by changing the imaging optics (zooming), includes tens of ultrasonic wavelengths both in the near-field and far-field zones, which means that the corresponding scattering phenomena are in the high frequency range~\cite{scs213}. The short acquisition time of each snapshot provides true transient analysis capability and the total measurement time to generate a complete sequence that reproduce the time evolution of 2D spatial distribution of the acoustic field,  including repetitive data acquisition and processing, is typically one or two orders of magnitude shorter than with other ultrasound pointwise probing technologies. The minimum measurable out-of-plane acoustic displacement in the current prototype is of the order of one nanometer, a modest figure compared f.i. to the typical performance of laser velocimetry but adequate enough to detect the wedge-generated guided waves employed in this study. 

As in any other ultrasound-based NDT approach, to extract useful information from the 
2D patterns measured with our PTVH system an additional fourth assessment step is required, namely, processing and analysis of the output signal---leading to subsequent detection, location and characterization of any existing flaws~\cite{ews287}. Generally speaking, this assessment process implies, implicitly or explicitly, some kind of modelling of wave propagation and scattering~\cite{review_Bond}, which is usually developed in the framework of the rigorous three-dimensional (3D) vector linear elasticity theory, in conjunction with a great variety of analytical or numerical schemes (e.g. normal mode expansion method~\cite{ews91}, finite difference method~\cite{ews291}, finite element method (FEM)~\cite{ews292}, boundary element method~\cite{Cho_Rose_ews71}, discrete point source method~\cite{ews79}, local interaction simulation approach~\cite{ews290}, various types of combinations or hybrid methods~\cite{Liu_ews72,ews254}, etc.). In order to avoid the high computational cost typically associated to standard 3D simulation  approaches in the high frequency regime, approximate theories (i.e. plate theories~\cite{ews197,Wang_Chang_2005} or other even more highly simplified appoximations~\cite{mast-gordon}), which are valid only for a limited number of situations, but that occur often in practice, have been utilized frequently~\cite{ews199,OEng2010,OEng2013}.

The computational solvers employed in this contribution are based on a
novel ``spectral'' approach: the recently introduced
Fourier-Continuation (FC) method presented
in~\cite{amlani2016fc,bruno2010high,albin2011spectral,albin2012fourier}.
Like regular spectral solvers~\cite{boyd2001chebyshev}, the FC methods
represent solutions by means of Fourier series, and they therefore
make it possible to dramatically reduce the problematic {\em
  dispersion errors}---which accumulate in the finite-difference or
finite-element approximation of the solution over domains spanning
many wavelengths. But, unlike the classical Fourier-based method, the
FC approach is applicable to general domains. And, unlike the
Chebyshev-based methods, the FC approach utilizes regular Cartesian
grids, and it therefore does not suffer from the severe Courant-Friedrichs-Levy  (CFL) constraints that stem from the point-clustering inherent in
Chebyshev-based explicit solvers.

As detailed in the aforementioned references, the ability of the FC
method to provide high-order, low-dispersion, spectral accuracy, for
general domains, stems from two main strategies, namely,
1)~decomposition of the computational domain into a number of
overlapping patches (following~\cite{henshaw2008parallel}); and
2)~Fourier expansion along each coordinate axis, in each one of the
patches, by creation of periodic functions (the continuation
functions) defined in adequately chosen domains beyond the physical
patch.  In view of its low dispersion, the elastic FC solver has
provided improvements by factors of hundreds, or even thousands and
beyond, over the computing times required by other methods, for a
given accuracy, for the problem of propagation of high-frequency
elastic waves~\cite{amlani2016fc}.

As mentioned above, this contribution compares experimental values
measured by PTVH with results obtained from the aforementioned
FC-based elastic-wave solver, for the problem of scattering of
transient quasi-Rayleigh waves by through-thickness holes. The paper
is organized as follows. Section~\ref{section2} establishes the
nomenclature and the theoretical formulation of the direct scattering
problem within the framework of the linear elasticity. Section~\ref{section3} then describes details concerning the
experimental methods (for excitation of quasi-Rayleigh transient
wavetrains, measurement of the displacement field by PTVH and the
two-step spatio-temporal Fourier transform method for deriving the
final complex displacement field) as well as the numerical
implementation of the FC elastic-wave solver approach. In both cases
only the essential points are included for completeness: the PTVH
technique was described in detail in previous contributions (see
references~\citenum{fdezijo07}, ~\citenum{trilloao03},
~\citenum{trillomst03} and ~\citenum{trillonimes06}), while a detailed
description of the FC(Gram) elastic solver was presented
in~\cite{amlani2016fc}. The experimental and simulated maps are
presented in section~\ref{section4}, including comparisons over various
specific areas in the 2-D image of the acoustic field, and the
agreement between theory and experiment is discussed taking into
account the pixel-wise matching of the spatial field distribution, the
profiles of amplitude and phase, and the corresponding values of the
global (root-mean-square) relative errors. To the best of our
knowledge this is the first time that such a quantitative comparison
is reported for transient waves both in near- and far-field zones,
apart from our corresponding recent studies in which this work is
based~\cite{amlani2016fc,Tesis_PRG}.  A full 3D sequence generated by
the FC(Gram) solver, displaying the temporal and spatial variation of
all three components of the displacement vector both in the interior
and the lateral boundaries of the plate, is also included in this
section---providing useful insights on rarely observed mechanisms
concerning elastic scattering. Section~\ref{section5}, finally,
present our closing remarks.

\section{Theoretical framework}
\label{section2}

We assume that for modelling wave propagation and scattering in our case it is adequate to describe plates as linear isotropic solids in the framework of linear elasticity theory within small   displacement regime and linear stress-strain relationship~\cite{rose00,Graff}. In this context, first in subsection~\ref{subsection2_A} we introduce the notation and establish the basic equations that define the initial boundary value problem (IBVP) to be solved for the scattering of waves in plates with stress-free boundaries in transient regime. After that, in subsection~\ref{subsection2_B} we consider the particular properties that can be used in the case of the scattering of quasi-Rayleigh wavetrains.

\subsection{Formulation of the direct scattering problem}
\label{subsection2_A}
\begin{figure}
\centering
\begin{tabular}{
c@{\hspace{0.2 cm}}
c
}
    \raisebox{10 mm}{
    \includegraphics 
     [width=60 mm,keepaspectratio=true]   
     {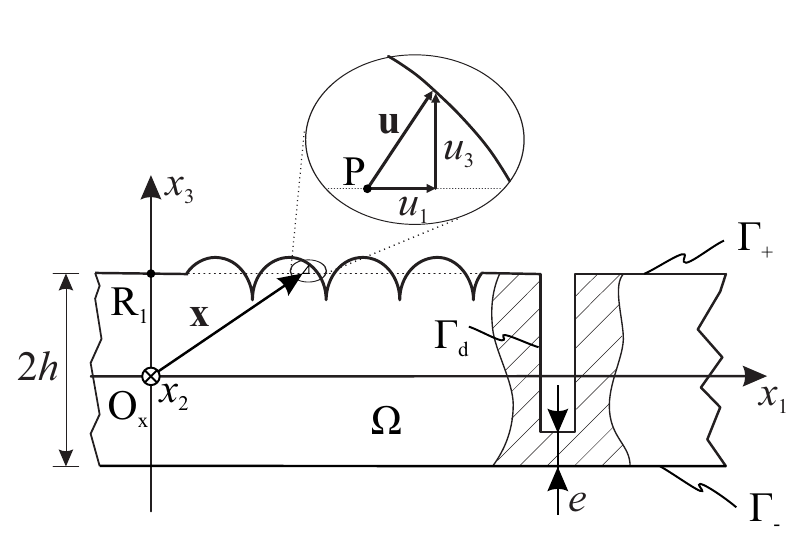}
     }
    & \includegraphics 
    [width=60 mm,keepaspectratio=true]   
    {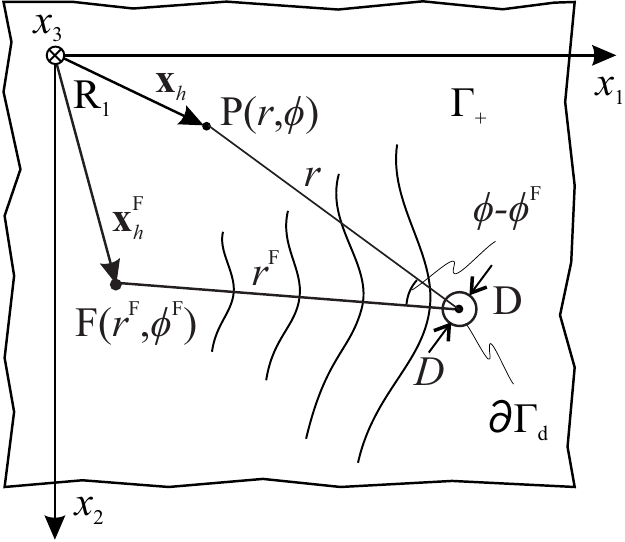}
\\

   (a)
   & (b)
\\
\end{tabular}
	\caption{Scheme of a plate of thickness $2h$ with a cylindrical hole of diameter $D$ and residual depth $e$: (a) section of the plate showing the detail of the instantaneous surface displacement associated to a harmonic guided wave with in-plane component $u_{1}$ and out-of-plane component $u_{3}$ that propagates along the $x_{1}$ axis, (b) top side  of the plate, showing the position of point $\mathrm{F}$, the intersection of the virtual line-source parallel to $x_{3}$ axis associated to the quasi-cylindrical wavetrain of the incident field.}
	\label{esquema_plate}
\end{figure}

\begin{figure}
\centering
\begin{tabular}{
c@{\hspace{0.2cm}}
c
}  
		\includegraphics 
		[width=60 mm,keepaspectratio=true]   
		 {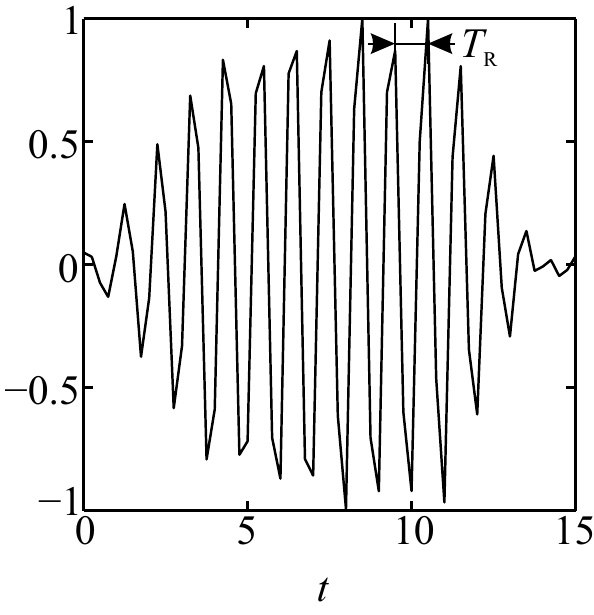}
		& 		\includegraphics
		[width=60 mm,keepaspectratio=true]   
		 {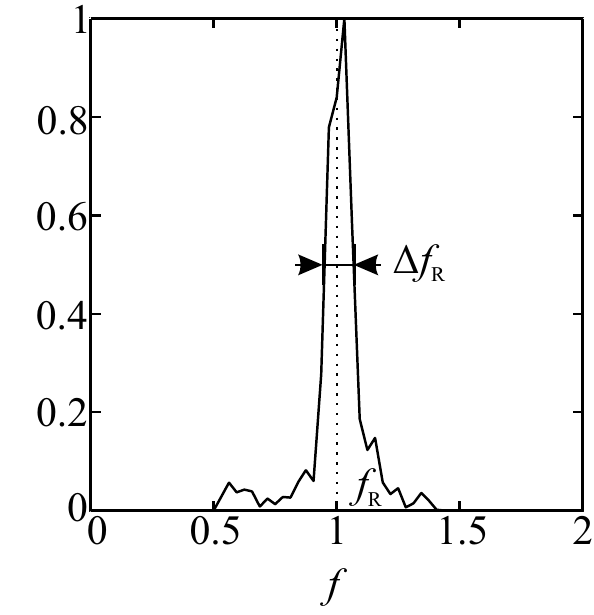}
		\\
		(a)
		& (b)
\\
\end{tabular}
	\caption{(a) Typical temporal record of the out-of-plane component of the incident field $u_{3}^{\mathrm{i}}\left(\mathbf{x}_{h},t\right)$ at a given point over the top surface of the plate $\mathbf{x}_{h}$ for a narrow-band quasi-Rayleigh wavetrain with central excitation frequency $f_{\mathrm{R}}=1.00$ MHz and bandwidth $\Delta f_{\mathrm{R}}<<f_{\mathrm{R}}$ ($T_{\mathrm{R}}=1/f_{\mathrm{R}}$), (b) amplitude of the temporal spectrum of the profile $u_{3}^{\mathrm{i}}\left(\mathbf{x}_{h},t\right)$ in (a). In both cases we show the normalized profiles as a function of $t$ in $\mathrm{\mu}$s and $f$ in MHz, respectively.}
	\label{pulso_QR}
\end{figure}

Using the Cartesian reference frame of figure~\ref{esquema_plate}, with origin at point $\mathrm{O_{x}}$ and orthonormal base vectors $\mathbf{a}_{1}$, $\mathbf{a}_{2}$ and $\mathbf{a}_{3}$, the position of a material point of the plate $\mathrm{P}$ in the reference configuration is characterized by its Cartesian coordinates $\left(x_{1},x_{2},x_{3}\right)$ which are the components of its position vector \linebreak $\mathbf{x}=x_{1}\mathbf{a}_{1}+x_{2}\mathbf{a}_{2}+x_{3}\mathbf{a}_{3}$ (for the shake brevity we simply write $\mathbf{x}=\left( x_{1},x_{2},x_{3}\right)$). When a wave perturbation propagates  in the plate, the associated movement of $\mathrm{P}$ with respect to its reference position $\mathbf{x}$ at a given time instant $t$ is specified by the displacement field 
%EQ.1************************************************************
%*******************************************************************
\begin{equation} \label{eq_u}
\mathbf{u}\left(\mathbf{x},t\right) =
           \sum_{i=1}^{3}u_{i}\left(\mathbf{x},t\right)\mathbf{a}_{i}         
\end{equation}
with components that can always be written as
%*******************************************************************
%*******************************************************************
%EQ.2************************************************************
%*******************************************************************
\begin{equation}\label{eq_ui}
u_{i}\left(\mathbf{x},t\right)=
u_{i\mathrm{m}}\left(\mathbf{x},t\right)\cos \left[\varphi_{i\mathrm{m}}\left(\mathbf{x},t\right)-2\pi ft \right],  \quad i=1,2,3         
\end{equation}
%*******************************************************************
%*******************************************************************
where $u_{i\mathrm{m}}\left(\mathbf{x},t\right)$ and $\varphi_{i\mathrm{m}}\left(\mathbf{x},t\right)$ are the  acoustic amplitude and the acoustic phase, respectively ($\varphi_{i\mathrm{M}}\left(\mathbf{x},t\right)=\varphi_{i\mathrm{m}}\left(\mathbf{x},t\right)-2\pi ft$ is referred as the total acoustic phase). Despite the fact expression~(\ref{eq_ui}) is formally valid in general, it is particularly useful in two important particular cases: on the one hand, for perturbations with harmonic time-dependence at a given temporal frequency $f$ with displacement field $\mathbf{u}_{f}\left(\mathbf{x},t\right)$ and harmonic components $u_{if}\left(\mathbf{x},t\right)$, because in this case the associated acoustic amplitudes $u_{if\mathrm{m}}\left(\mathbf{x}\right)$ and  acoustic phases $\varphi_{if\mathrm{m}}\left(\mathbf{x}\right)$ are truly time-independent fields; on the other hand, for transient narrow-band perturbations with a central temporal frequency $f_\mathrm{R}$, as those employed in our experiments, because the associated acoustic amplitudes and acoustic phases are nearly constant within any time interval of the order of the central period of the wavetrain $T_\mathrm{R}=1/f_\mathrm{R}$ in such a way that they can be understood in close analogy with the harmonic case, except perhaps in areas that are in the leading and trailing edges of the wavetrain (see figure~\ref{pulso_QR}).

% figura 1: esquema_plate******************************************
%*****************************************************************
%***************************************************************
%***************************************************************

Considering that the wave perturbation can be described using the small strain regime, we have an associated strain tensor $\boldsymbol{\epsilon}\left(\mathbf{x},t\right)$ with components
%EQ. 3***********************************************************
%*******************************************************************
\begin{equation} \label{eq_epsilon_ij}
  \epsilon_{ij}\left(\mathbf{x},t\right)= \frac{1}{2}\left[
                                  \frac{\partial{}u_{i}\left(\mathbf{x},t\right)}{\partial{}x_{j}}
                                 +\frac{\partial{}u_{j}\left(\mathbf{x},t\right)}{\partial{}x_{i}}
                            \right],  \quad  i,j=1,2,3
\end{equation}
%*******************************************************************
%*******************************************************************

% figura 2: pulso y espectro QR*********************************
%*****************************************************************
%*****************************************************************
%*****************************************************************

Assuming that the plate is a linear homogeneous isotropic solid with volume mass density $\rho$ and a linear elastic response characterized by the Lam\'e constants $\lambda_{\mathrm{e}}$ and $\mu_{\mathrm{e}}$, the linear stress-strain relationship can be described writing the components of the stress tensor $\boldsymbol{\tau}\left(\mathbf{x},t\right)$ as
%EQ. 4**********************************************************
%*******************************************************************
\begin{equation} \label{eq_ley_Hook_ij}
	\tau_{ij}\left(\mathbf{x},t\right)=\lambda_{\mathrm{e}}\delta_{ij}	          \sum_{k=1}^3 \epsilon_{kk}\left(\mathbf{x},t\right)+2\mu_{\mathrm{e}}\epsilon_{ij}\left(\mathbf{x},t\right)
	          ,  \quad i,j=1,2,3
\end{equation} 
%*******************************************************************
%*******************************************************************
where $\delta_{ij}$ the Kronecker delta. In this context, the dynamic behavior in the interior volume of the plate $\Omega$ is described by the Lam\'e-Navier equations  
%EQ. 5**********************************************************
%*******************************************************************
\begin{multline}
\label{eq_Lame-Navier_ij}
\frac{\left(\lambda_{\mathrm{e}}+\mu_{\mathrm{e}}\right)}{\rho}
    \sum_{j=1}^{3}\frac{\partial^{2}u_{i}\left(\mathbf{x},t\right)}
    					{\partial x_{i}\partial x_{j}}
	+ \frac{\mu_{\mathrm{e}}}{\rho}
	\sum_{j=1}^{3}\frac{\partial^{2} u_{i}\left(\mathbf{x},t\right)}
						{\partial x_{j}^{2}}
						\\	
	=\frac{\partial^{2}u_{i}\left(\mathbf{x},t\right)}{\partial t^{2}}, 
	\quad i=1,2,3 \quad \mathbf{x}\in\Omega
\end{multline}
%*******************************************************************
%*******************************************************************
which can be derived employing equations~(\ref{eq_epsilon_ij}) and (\ref{eq_ley_Hook_ij}) in the balance of momentum equation~\cite{Graff} and assuming that the interior of the plate is a region free of body forces. The components of the displacement $u_{i}\left(\mathbf{x},t\right)$  that are solutions of equation~(\ref{eq_Lame-Navier_ij}) can be discomposed in longitudinal (L) and transversal (T) wave components that propagate, respectively, with phase velocities 
%EQ.6***********************************************************
%*****************************************************************
\begin{equation}\label{eq_cLcT}
c_{\mathrm{L}}=\sqrt{\left(\lambda_{\mathrm{e}}+2\mu_{\mathrm{e}}\right)/{\rho}}  
\quad \mathrm{and} \quad
 c_{\mathrm{T}}=\sqrt{\mu_{\mathrm{e}}/{\rho}}.
\end{equation}
%*******************************************************************
%*******************************************************************
Using~(\ref{eq_cLcT}) we get 
$\left(\lambda_{\mathrm{e}}+\mu_{\mathrm{e}}\right)/\rho=\left(c_{\mathrm{L}}^{2}-c_{\mathrm{T}}^{2}\right)$ and $\mu_{\mathrm{e}}/\rho= c_{\mathrm{T}}^{2}$ in such a way that we can express the constants of the Lam\'e-Navier equations exclusively in terms of the phase velocities.

Any solution of the Lam\'e-Navier equation in plates with stress-free \linebreak boundaries (as it is our case) has to verify over its top side $\Gamma_{+}$, its bottom side $\Gamma_{-}$ and the hole border $\Gamma_{\mathrm{d}}$ the stress-free boundary condition
%EQ. 7*********************************************************
%*******************************************************************
\begin{equation} \label{eq_BC_tau_ij}
    \sum_{j=1}^{3}\tau_{ij}\left(\mathbf{x},t\right)n_{j}\left(\mathbf{x}\right)=0, \quad i=1,2,3 \quad \mathbf{x}\in\partial\Omega
\end{equation} 
%*******************************************************************
%*******************************************************************
being $\partial\Omega=\Gamma_{+}\cup\Gamma_{-}\cup\Gamma_{\mathrm{d}}$ and $n_{j}\left(\mathbf{x}\right)$ the components of the normal unit vector $\mathbf{n}\left(\mathbf{x}\right)$ at point $\mathbf{x}\in\partial\Omega$ pointing outwards the plate interior $\Omega$. Equations~(\ref{eq_Lame-Navier_ij}-\ref{eq_BC_tau_ij}), with additional initial conditions for displacement and velocity in the plate interior volume $\Omega$, are the essential elements that define the initial boundary value problem (IBVP) to be solved for the scattering of waves in plates with stress-free boundaries in the transient case.

\subsection{Scattering of quasi-Rayleigh waves by through-thickness holes}
\label{subsection2_B}
In the following we circumscribe the analysis to the particular case of propagation and scattering of narrow-band quasi-Rayleigh wavetrains by through-thickness holes, with residual depth $e=0$ (figure~\ref{esquema_plate}) considering how to obtain the incident field inside the plate  from the value of the out-of-plane component of the incident field at the top surface of the plate. More specifically, we use the complex representation of the displacement field (see Appendix A) to present the strategy 
to reconstruct the complex Cartesian components of the displacement of the incident field $\hat{u}_{i}^{\mathrm{i}}\left(\mathbf{x},t\right)$ at any point $\mathbf{x}=\left(x_{1},x_{2},x_{3}\right)$ inside the plate from the value of the complex out-of-plane component of the displacement $\hat{u}_{3}^{\mathrm{i}}\left(\mathbf{x}_{h},t\right)$ at the corresponding point $\mathbf{x}_{h}=\left(x_{1},x_{2},h\right)$ a the top surface of the plate,  
which is a quantity that can be measured by our PTVH system (see figure~\ref{pulso_QR}.a and subsection \ref{subsection3_A}).

% figura 3: espectro_Lamb****************************************
%*****************************************************************
\begin{figure}
	\centering
		\includegraphics 
		[width=120 mm,keepaspectratio=true]   
		{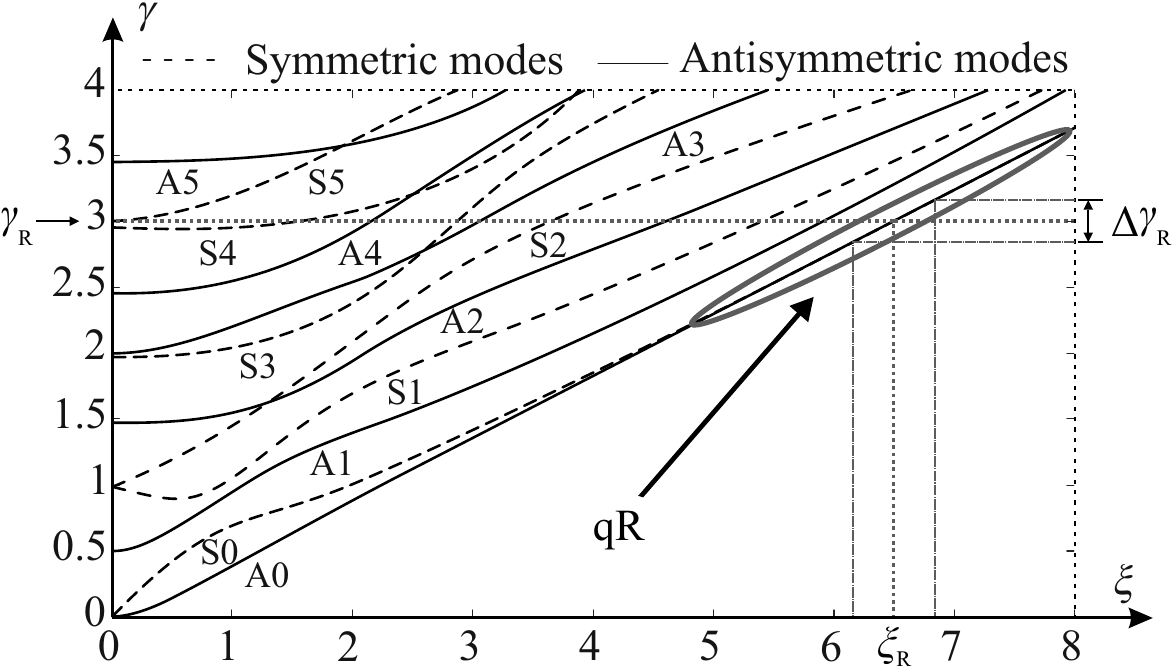}
	\caption{Frequency spectrum of Lamb propagating modes for an aluminum plate with stress-free boundaries and Poisson's ratio $\nu=0.33$. $\gamma=4hf/c_{\mathrm{L}}$ and $\xi=2hk/\pi$ are the normalized frequency and wavenumber respectively being $c_{\mathrm{L}}$ the longitudinal wave velocity. The location of normalized central excitation frequency $\gamma_{\mathrm{R}}=4hf_{\mathrm{R}}/c_{\mathrm{L}}$, normalized wavenumber $\xi_{\mathrm{R}}=4h/\lambda_{\mathrm{R}}$ and normalized value of the temporal bandwidth of the pulse $\Delta\gamma_{\mathrm{R}}=4h\Delta f_{\mathrm{R}}/c_{\mathrm{L}}$ in our case (see section~\ref{section4}) are schematically represented. The branches that cross the horizontal line at $\gamma_{\mathrm{R}}$ correspond to the propagating Lamb modes that could be excited at the corresponding normalized temporal frequency. qR identifies the region of the spectrum that corresponds to quasi-Rayleigh waves.}
	\label{espectro_Lamb}
\end{figure}
%*****************************************************************
%*****************************************************************

Quasi-Rayleigh waves result from particular combinations of Lamb modes in the so-called quasi-Rayleigh region of the Lamb spectrum where the associated branches of the S0 and A0 modes nearly overlap (see qR region with $\gamma\geq1.5$ in figure~\ref{espectro_Lamb}). A quasi-Rayleigh harmonic guided wave is just the superposition of a symmetric S0 Lamb mode and an anti-symmetric A0 Lamb mode with the same temporal frequency within the qR range and with the same value of their amplitudes at the top surface of the plate. In this conditions, the harmonic A0 and S0 modes have practically the same propagating wavenumber and a common phase velocity $c_{\mathrm{R}}$ (characterized by the common slope of the branches in the qR region) which is the Rayleigh phase velocity of the plate material. The Rayleigh  equation
%EQ.8***********************************************************
%*****************************************************************
\begin{equation}\label{eq_cR}
    \chi_{\mathrm{RT}}^{6}-
   8\chi_{\mathrm{RT}}^{4}+
   8\left(3-2\chi_{\mathrm{TL}}^{2}\right) 
    \chi_{\mathrm{RT}}^{2}-
  16\left(1-\chi_{\mathrm{TL}}^{2}\right)=0,
\end{equation}
%************************************************************
%*****************************************************************
where $\chi_{\mathrm{RT}}=c_{\mathrm{R}}/c_{\mathrm{T}}$ and $\chi_{\mathrm{TL}}=c_{\mathrm{T}}/c_{\mathrm{L}}$, 
shows that $c_{\mathrm{R}}$ can be obtained as a function of the longitudinal and transversal phase velocities $c_{\mathrm{L}}$ and  $c_{\mathrm{T}}$, i.e. that only two of the three phase velocities $c_{\mathrm{R}}$, $c_{\mathrm{L}}$ and  $c_{\mathrm{T}}$ are independent. If only harmonic quasi-Rayleigh waves of this type are efficiently excited by the ultrasonic source, the complex components of the complex displacement of the quasi-Rayleigh incident wavetrain (figure~\ref{pulso_QR}.a) can be written as the linear superposition
%EQ. 9**************************************************************
%*******************************************************************
\begin{equation}\label{eq_uiQR}
  \hat{u}_{i}^{\mathrm{i}}\left(\mathbf{x},t\right) = {\int}^{\infty}_{0} w\left(f\right) \hat{u}_{if}^{\mathrm{i}}\left(\mathbf{x},t\right)  df \quad i=1,2,3
\end{equation}
%*******************************************************************
where each quasi-Rayleigh harmonic complex component of frequency $f$ can be discomposed as
%EQ 10*************************************************************
%*******************************************************************
\begin{equation}\label{eq_uifQR}
\hat{u}_{if}^{\mathrm{i}}\left(\mathbf{x},t\right) = \hat{u}_{if}^{\mathrm{iS0}}\left(\mathbf{x},t\right)+\hat{u}_{if}^{\mathrm{iA0}}\left(\mathbf{x},t\right) \quad i=1,2,3
\end{equation}
%*******************************************************************
%*******************************************************************
being $\hat{u}_{if}^{\mathrm{iS0}}\left(\mathbf{x},t\right)$ and $\hat{u}_{if}^{\mathrm{iA0}} \left(\mathbf{x},t\right)$, respectively, the corresponding harmonic complex components of the S0 and A0 modes with temporal frequency $f$ and $w\left(f\right)$ a scalar function that specifies the weight of each harmonic quasi-Rayleigh displacement of frequency $f$ within the qR range spectrum of the quasi-Rayleigh pulse. We will assume that $w\left(f\right)$ is peaked around a temporal frequency $f_{\mathrm{R}}$ and takes non zero values over a bandwidth interval $\Delta f_{\mathrm{R}}<<f_{\mathrm{R}}$ (figure~\ref{pulso_QR}.b), in such a way that each harmonic component of the pulse has a wavenumber that is very close to the central wavenumber of the train $k_{\mathrm{R}}=2\pi f_{\mathrm{R}}/c_{\mathrm{R}}$ within the quasi-Rayleigh region qR of the Lamb spectrum (in our experiments we employ an excitation with a normalized central temporal frequency and central wavenumber of the order of 3 and 6 respectively -see below subsection~\ref{section4} and figure~\ref{espectro_Lamb}). Provided that in the scattering problem the hole is located in the far-field region of the source it is justified, on the one hand,
to assume that the phase and the amplitude of the incident field can be described as a quasi-cylindrical inhomogeneous wave that diverges from a line-source, parallel to the $x_{3}$ axis (which intersect the top surface of the plate at point $\mathrm{F}$ with position vector $\mathbf{x}_{h}^{\mathrm{F}}=\left({x}_{1}^{\mathrm{F}},{x}_{2}^{\mathrm{F}},h\right)$, see figure~\ref{esquema_plate}.b) and, on the other hand, to neglect the contributions of non-propagating modes in such a way that expressions~(\ref{eq_uiQR}-\ref{eq_uifQR}) give an adequate description of the components of the displacement for the a transient narrow-band quasi-Rayleigh incident field.
In addition, as the normalized transversal distribution of the complex components of the displacement field of each harmonic quasi-Rayleigh contribution is a function only of its wavenumber and the thickness of the plate we can also assume, in a first approximation, that all the harmonic quasi-Rayleigh components of the pulse share the same normalized transversal distribution for their displacements (see Appendix B). In these conditions, the value of the complex components of the complex displacement of the quasi-Rayleigh wavetrain inside the volume of the plate can be written as 
%EQ.11**************************************************************
%*******************************************************************
\begin{subequations}\label{eq_uqR_uqRh}
\begin{align}
  \hat{u}_{i}^{\mathrm{i}}\left(\mathbf{x},t\right)
   = &
   \left[
   \underline{\hat{u}}_{\parallel f_{\mathrm{R}}\mathrm{m}}^{\mathrm{S0}}
   \left(x_{3}\right)
   + 
   \underline{\hat{u}}_{\parallel f_{\mathrm{R}}\mathrm{m}}^{\mathrm{A0}}
   \left(x_{3}\right)
   \right] 
   \nonumber \\
   & \times \left[
   \frac
   {\mathbf{x}_{h}-\mathbf{x}^{\mathrm{^F}}_{h}}
   {\vert \mathbf{x}_{h}-\mathbf{x}^{\mathrm{^F}}_{h}\vert}
   \cdot 
   \mathbf{a}_{i}
   \right]    
   \hat{u}_{3}^{\mathrm{i}}\left(\mathbf{x}_{h},t\right) \quad i=1,2
\\
   \hat{u}_{3}^{\mathrm{i}}\left(\mathbf{x},t\right)
   = & \left[ \underline{\hat{u}}_{\perp f_{\mathrm{R}}\mathrm{m}}^{\mathrm{S0}}\left(x_{3}\right)+ \underline{\hat{u}}_{\perp f_{\mathrm{R}}\mathrm{m}}^{\mathrm{A0}}\left(x_{3}\right)\right] \hat{u}_{3}^{\mathrm{i}}\left(\mathbf{x}_{h},t\right)
\end{align}
\end{subequations}
%*******************************************************************
%*******************************************************************
in terms of the complex out-of-plane component $\hat{u}_{3}^{\mathrm{i}}\left(\mathbf{x}_{h},t\right)$, being $\underline{\hat{u}}_{\parallel f_{\mathrm{R}}\mathrm{m}}^{\mathrm{S0}}\left(x_{3}\right)$ and $\underline{\hat{u}}_{\perp f_{\mathrm{R}}\mathrm{m}}^{\mathrm{S0}}\left(x_{3}\right)$ the normalized transversal distributions of the complex amplitude of the in-plane and out-of-plane components of the S0 mode with frequency $f_{\mathrm{R}}$ and wavenumber $k_{\mathrm{R}}$ and $\underline{\hat{u}}_{\parallel f_{\mathrm{R}}\mathrm{m}}^{\mathrm{A0}}\left(x_{3}\right)$ and $\underline{\hat{u}}_{\perp f_{\mathrm{R}}\mathrm{m}}^{\mathrm{A0}}\left(x_{3}\right)$ the corresponding transversal distributions for the A0 mode. Extracting the real part of the result of expression (\ref{eq_uqR_uqRh}) we finally obtain the Cartesian components of the displacement of the incident field $u_{i}^{\mathrm{i}}\left(\mathbf{x},t\right)$ at any point $\mathbf{x}=\left(x_{1},x_{2},x_{3}\right)$ inside the plate.

\section{Materials and methods}
\label{section3}

In the following, we first describe the experimental set-up employed to generate and to detect quasi-Rayleigh wavetrains in plates (subsection~\ref{subsection3_A}), then we outline the foundations of the new FC(Gram) numerical method (subsection~\ref{subsection3_B}) and finally we present the discrete formulation of the direct scattering problem (subsection~\ref{subsection3_C}) and the procedure employed for obtaining an analytic approximation to the incident field (subsection~\ref{subsection3_D}).

\subsection{Experimental methods}
\label{subsection3_A}
\begin{figure}
	\centering
		\includegraphics 
		[width=70 mm,keepaspectratio=true]   
		{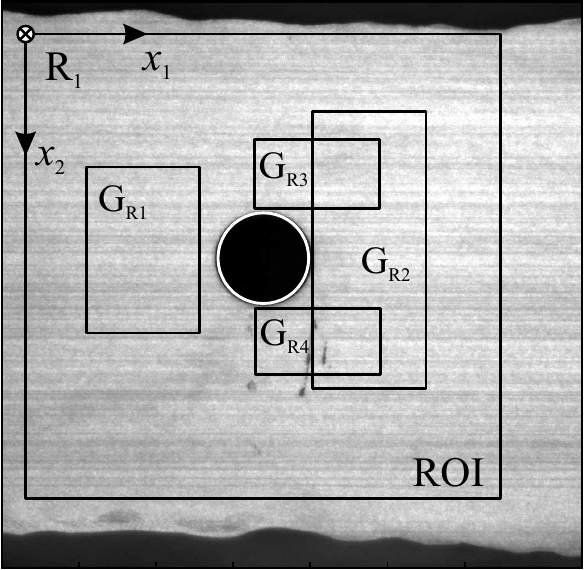}
	\caption{Reference image for a through-thickness hole ($e=0$) of diameter $D=12\;\mathrm{mm}$ showing the region of interest (ROI) employed in the images and profiles of numerical and experimental ultrasonic fields over the plane $\left(x_{1},x_{2},h\right)$ and the subzones of the ROI $\mathrm{G}_{\mathrm{R}1}, \mathrm{G}_{\mathrm{R}2}, \mathrm{G}_{\mathrm{R}3}$ and $\mathrm{G}_{\mathrm{R}4}$ used for quantitative assessment of the relative global error between theory and experiment (see section~\ref{section4}).}
	\label{ROI}
\end{figure}

Several through-thickness holes, with nominal diameters $D$ in the range $4\;\mathrm{mm}$ to $12\;\mathrm{mm}$, have been adequately prepared in aluminum plates with nominal thickness $2h=10\;\mathrm{mm}$. In all of the cases the plates have been supported so that the constraints at their surface are minimized and plasticine has been used as acoustic absorber at the edges of the plates to minimize reflections of the incident and scattered waves that could disturb the measured acoustic fields inside the region of interest (ROI) (figure~\ref{ROI}). The average thickness of the plates and the average diameter of the holes have been determined experimentally with a relative error lower than $1\%$ by averaging several direct measurements obtained with a caliper with resolution of 0.05 mm and a metalographic microscope with a translation stage with resolution of 0.01 mm. The longitudinal wave velocity in the plates has been measured by means of the classical pulse-echo method resulting $c_{\mathrm{L}}=6360\left(1\pm0.5\%\right)\;\mathrm{m}/\mathrm{s}$.

%% figura 4: ROI*************************************************
%%*****************************************************************
%*****************************************************************
%*****************************************************************

% figura 5: setup*************************************************
%*****************************************************************
\begin{figure}
	\centering
		\includegraphics
		[width=100 mm,keepaspectratio=true]   
		{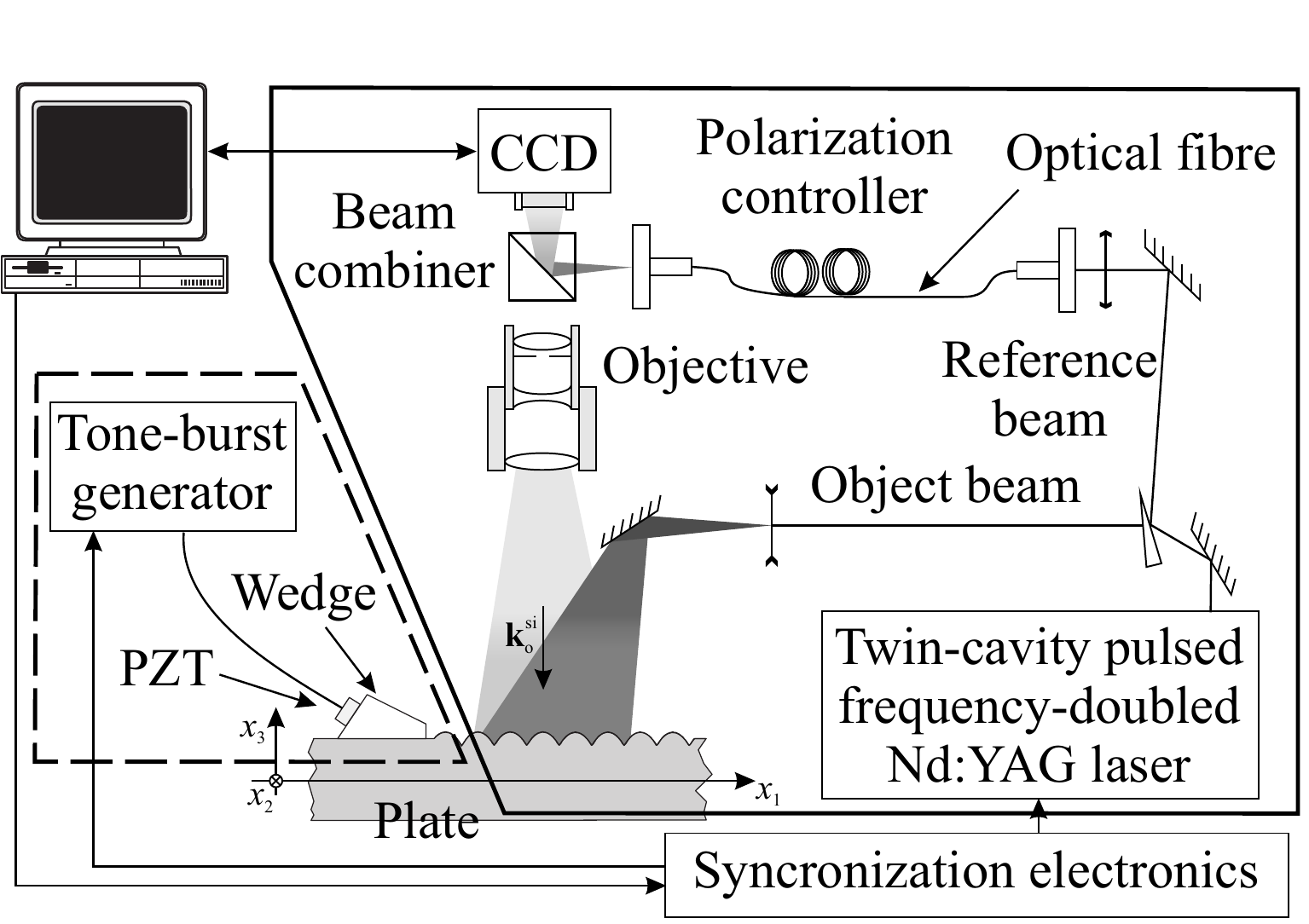}
 	\caption{Experimental setup}
	\label{setup}
\end{figure}
%*****************************************************************
%*****************************************************************

The experimental system used to generate and to detect the elastic waves in this plates is schematically represented in figure~\ref{setup}. Quasi-Rayleigh transient wavetrains have been generated by means of the classical wedge method using a short tone-burst with central frequency $f_{\mathrm{R}}=1.00\;\mathrm{MHz}$ and an integer number of cycles (between 5 and 10). The instantaneous out-of-plane acoustic field at the top surface of the plate $u_{3}^{\mathrm{e}}\left(\mathbf{x}_{h},t\right)$  associated to the wavetrain during propagation and scattering has been measured with our self-developed double-pulsed TV holography system~\cite{trilloao03}. As is common in TV holography techniques~\cite{Doval2000} we employ an image-hologram configuration, sensitive to the out-of-plane component of the displacement of the surface points, with the image sensor of a video camera as recording medium. There is not optical reconstruction of the recorded holograms but instead their intensity distribution is electronically processed to render the optical phase-difference map, which depends on the displacements of the specimen surface. A twin-cavity pulsed, injection seeded and frequency doubled Nd:YAG laser (Spectron SL404T), the core of the TV holography system, emits two laser pulses, with a duration of $20\;\mathrm{ns}$ and a adjustable temporal inter-pulse delay $\tau$, that are employed to record two correlograms in separate frames of a CCD camera (PCO Sensicam Double-Shutter). Each correlogram corresponds to the interference of the reference beam and the object beam scattered back by the plate surface. A processing procedure based on the spatial Fourier transform method has been applied to the correlograms~\cite{trillomst03}, which renders the so-called optical phase-change map
%EQ. 12**************************************************************
%*******************************************************************
\begin{equation} \label{eq_OPC}
	\Delta\Phi^{\mathrm{e}}\left(\mathbf{x}_{h},t\right)=\frac{8\pi}{\lambda_{\mathrm{o}}}u_{3}^{\mathrm{e}}\left(\mathbf{x}_{h},t\right) \quad \mathbf{x}_{h}\in\Gamma_{+}
\end{equation}
%*******************************************************************
%*******************************************************************
proportional to the instantaneous out-of-plane acoustic displacement field  $u_{3}^{\mathrm{e}}\left(\mathbf{x}_{h},t\right)$ being $\lambda_{\mathrm{o}}=532\;\mathrm{nm}$ the wavelength of the laser and $t$ the instant of emission of the first laser pulse. The usual scaling factor ($4\pi/   \lambda_{\mathrm{o}}$) between optical phase-change and displacement is doubled in equation~(\ref{eq_OPC}) by selecting $\tau=1.5\;\mathrm{\mu s}$ ($3/2$ of the period $T_{\mathrm{R}}$ associated to central excitation frequency $f_{\mathrm{R}}=1.00\;\mathrm{MHz}$, that is the minimum number of odd half-periods for which the camera can record the two correlograms in different frames).

With this scheme we have performed measurements in successive instants of the wavetrain propagation ($t_{0}$, $t_{1}$, $\ldots$, $t_{n}$, $\ldots$, $t_{N_{\mathrm{o}}-1}$  with $t_{n+1}-t_{n}=\Delta t$) to obtain a complete sequence of $N_{\mathrm{o}}$ optical phase-change maps $\Delta\Phi
^{\mathrm{e}}\left(\mathbf{x}_{h},t_{n}\right)$ (we have selected $N_{\mathrm{o}}=128$ with $\Delta t=250\;\mathrm{ns}$, i.e., a quarter of the period associated to the central excitation frequency of the wavetrain). Even though this sequence of maps represents by itself a useful means to assess the interaction of the quasi-Rayleigh wavetrain and the defect for every time instant, it has been processed as a whole employing the spatio-temporal 3D Fourier transform method~\cite{trillonimes06} to obtain a sequence of $N_{\mathrm{o}}$ experimental complex displacement fields $\hat{u}_{3}^{\mathrm{e}}\left(\mathbf{x}_{h},t_{n}\right)$, from which the acoustic amplitude $u_{3\mathrm{m}}^{\mathrm{e}}\left(\mathbf{x}_{h},t_{n}\right):=\left|\hat{u}_{3}^{\mathrm{e}}\left(\mathbf{x}_{h},t_{n}\right)\right|$ and the total acoustic phase $\varphi_{3\mathrm{M}}^{\mathrm{e}}\left(\mathbf{x}_{h},t_{n}\right):=\mathrm{arg}\left[\hat{u}_{3}^{\mathrm{e}}\left(\mathbf{x}_{h},t_{n}\right)\right]$ can be retrieved independently and a  reconstructed out-of-plane displacement $u_{3}^{\mathrm{e}}\left(\mathbf{x}_{h},t_{n}\right)=\mathrm{Re}\left[\hat{u}_{3}^{\mathrm{e}}\left(\mathbf{x}_{h},t_{n}\right)\right]$ is obtained with an improved signal-to-noise ratio. The sequence of $N_{\mathrm{o}}$ complex displacement fields $\hat{u}_{3}^{\mathrm{e}}\left(\mathbf{x}_{h},t_{n}\right)$ is the raw experimental data set for the comparison with numerical simulations and it is used for obtaining an analytic approximation to the incident complex field (see subsection~\ref{subsection3_C}), which is employed to define a non-homogeneous Dirichlet boundary condition in the discrete formulation of the direct scattering problem that is introduced in the following. 

\subsection{Numerical Method}
\label{subsection3_B} 

In view of their simplicity and versatility, classical differentiation
methods such as those based on use of finite differences (FD) or finite
elements are often used as part of partial differential equation (PDE) discretization schemes. When
applied to differential equations governing wave-like phenomena,
however, approaches of this type do give rise to high numerical
dispersion: phase errors accumulate over subsequent periods of a wave
field, and thus FD and FEM solvers require increasing numbers of
discretization points per wavelength to resolve the solution within a
given accuracy tolerance as the size of the problem grows. As is well
known, methods based on use of Fourier series do not suffer from this
problem: a fixed number of discretization points per wavelength
suffices for such methods to produce wave-like solutions with a fixed
accuracy in arbitrarily large domains (i.e. domains that can contain an arbitrarily large number of wave periods).

Unfortunately, however, classical Fourier methods are only applicable
to periodic functions---since Fourier expansion of a non-periodic
functions gives rise to the well-known ``ringing effect'' (the Gibb's
phenomenon~\cite{gocken}): the jump that results in the periodic
extension of an inherently non-periodic function leads to an
``overshoot'' in the Fourier representation at points of discontinuity
of the periodic continuation of the given function. This inaccuracy
does not subside as more terms are added and, in fact, it results in
extremely slow pointwise convergence throughout the interior of the
computational domain---in addition to the unmitigated error at domain
boundaries.  

\begin{figure}[!h]
\centering
{\center\includegraphics[width=1.0\textwidth]{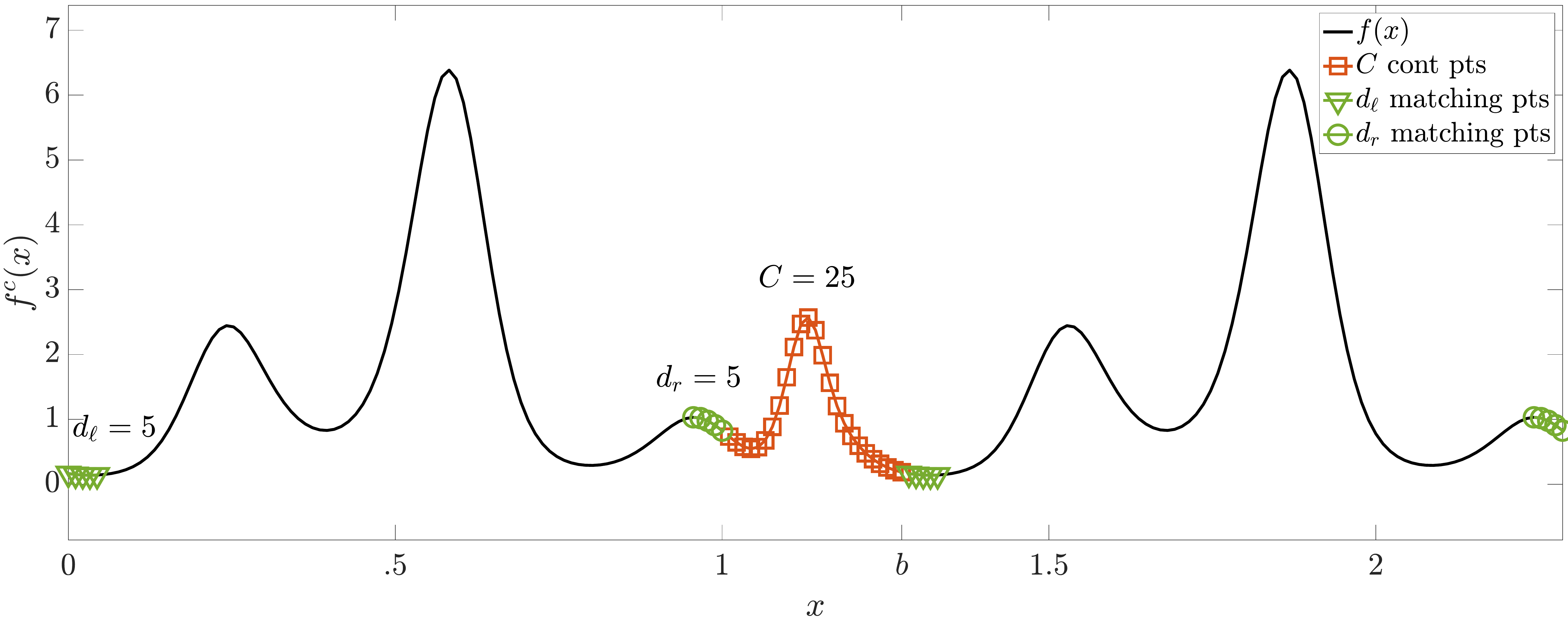}}
\caption{Fourier continuation of the non-periodic function given by $f(x) = 
\exp\left[\sin(5.4\pi x - 2.7\pi)-\cos(2\pi x)\right] .$ Triangles/circles and squares represent $d_\ell=d_r=5$ matching points and $C=25$ continuation points, respectively.}
\label{fig:fcexample}
\end{figure}

A goal to extend the applicability of Fourier methods (together with
its inherent excellent qualities, most notably dispersionless-ness and
high-order accuracy) to general non-periodic configurations has lead
to the development of the FC methods that form the basis of the
numerical solvers utilized in this work.  The Fourier continuation
(FC) method produces a rapidly-convergent interpolating Fourier series
representation of a given function on a region larger than the given
physical domain. This is accomplished by relying on a ``periodic
extension'' of the given function, that closely approximates the given
function values in the original domain, but which is periodic (albeit
in a slightly enlarged domain) and thus suitable for Fourier
approximation. The procedure is illustrated in
figure~\ref{fig:fcexample}. In detail, given a function $f$ which is
defined, without loss of generality, on the unit interval as
\[f(x) : [0,1] \subset \mathbb{R} \to \mathbb{R},\] the FC method
produces a periodic function $f^c$ defined on an extended
interval,
\[f^c(x) : [0,b]\subset \mathbb{R} \to \mathbb{R}, \quad b>1\] which
closely approximates $f(x)$ on the original interval $[0,1].$

% figura 6:
%figura 1 Oscar: FC******************************************
%*****************************************************************
%***************************************************************
%***************************************************************

The proposed fully-discrete Fourier continuation algorithms proceed as
follows: letting $N$ be the number of discretization points over the
unit interval (yielding a uniform grid $x_i = i h, i=0,\dots,N-1$,
$h = 1/(N-1)$) together with point values $f(x_i)$ of the function of
interest, the Fourier continuation method produces a $b$-periodic
trigonometric polynomial $f^c$ of the form
%EQ. 13************************************************************
%*******************************************************************
\begin{equation}\label{eq:fcseries} f^c(x) = \displaystyle\sum_{k=-M}^{M} a_k \exp\left(\mathrm{j} \frac{2\pi k x}{b}\right),\end{equation}
%EQ. **************************************************************
%*******************************************************************
that matches the given discrete values of $f$:
$f^c(x_i) = f(x_i), i=0,...,N-1$, and which, for smooth functions $f$,
rapidly approaches $f$ as the $N$ grows. Derivatives of the function can
then be easily and accurately be computed through term-by-term
differentiation, e.g.
%EQ. 14************************************************************
%*******************************************************************
\begin{equation}
\begin{aligned}
f_x(x) &=&f^c_x(x) &=& \displaystyle\sum_{k=-M}^{M} \left( \mathrm{j} \frac{2\pi  k}{b}\right) a_k \exp\left(\mathrm{j} \frac{2\pi k x}{b}\right),\\
f_{xx}(x) &=&f^c_{xx}(x) &=& \displaystyle-\sum_{k=-M}^{M} \left(\frac{2\pi k^2}{b^2}\right)a_k \exp\left(\mathrm{j} \frac{2\pi k x}{b}\right),\quad\mbox{etc.}
\end{aligned}
\end{equation}
%**************************************************************
%*******************************************************************
In the simplest treatment~\cite{brunohan,boyd,brunospringer}, the
coefficients $a_k$ of (\ref{eq:fcseries}) are found by invoking the Singular Value Decomposition (SVD) to solve the least squares system
%EQ. 15***********************************************************
%*******************************************************************
\begin{equation}
\displaystyle\min_{a_k} \sum_{i=0}^{N-1} \left|f^c(x_i) - f(x_i)\right|.
\end{equation}
%**************************************************************
%*******************************************************************

Although straightforward, repeated applications of the SVD for \linebreak
time-dependent function values on each line of a three-dimensional
mesh in a PDE is, in general, unacceptably costly. To circumvent this
difficulty, an FFT accelerated
method~\cite{bruno2010high,lyonbrunoII,albin2011spectral} can be
used. Relying on numbers of $d_\ell$ and $d_r$ of function values
nearest to the left and right endpoints of the interval
(e.g. $d_\ell=d_r=5$), the accelerated FC approach proceeds by
projecting the known near-endpoint sets of function values onto certain
basis of Gram discretely orthogonal polynomials. Each one of these
orthogonal polynomials possesses a discrete continuation, and, thus,
the near boundary values, which are given by certain linear
combinations of the orthogonal-polynomial basis, can easily be
continued as well.  Once discrete periodic continuation functions have been
determined, the corresponding Fourier coefficients can be obtained via
an application of the FFT. The overall FC approximation procedure is
demonstrated in Figure~\ref{fig:fcexample}.

% figura 7
% figura 2 Oscar: dipersion**************************************
%*****************************************************************
\begin{figure}[!h]
\centering
{\center
\includegraphics[width=0.70\textwidth]{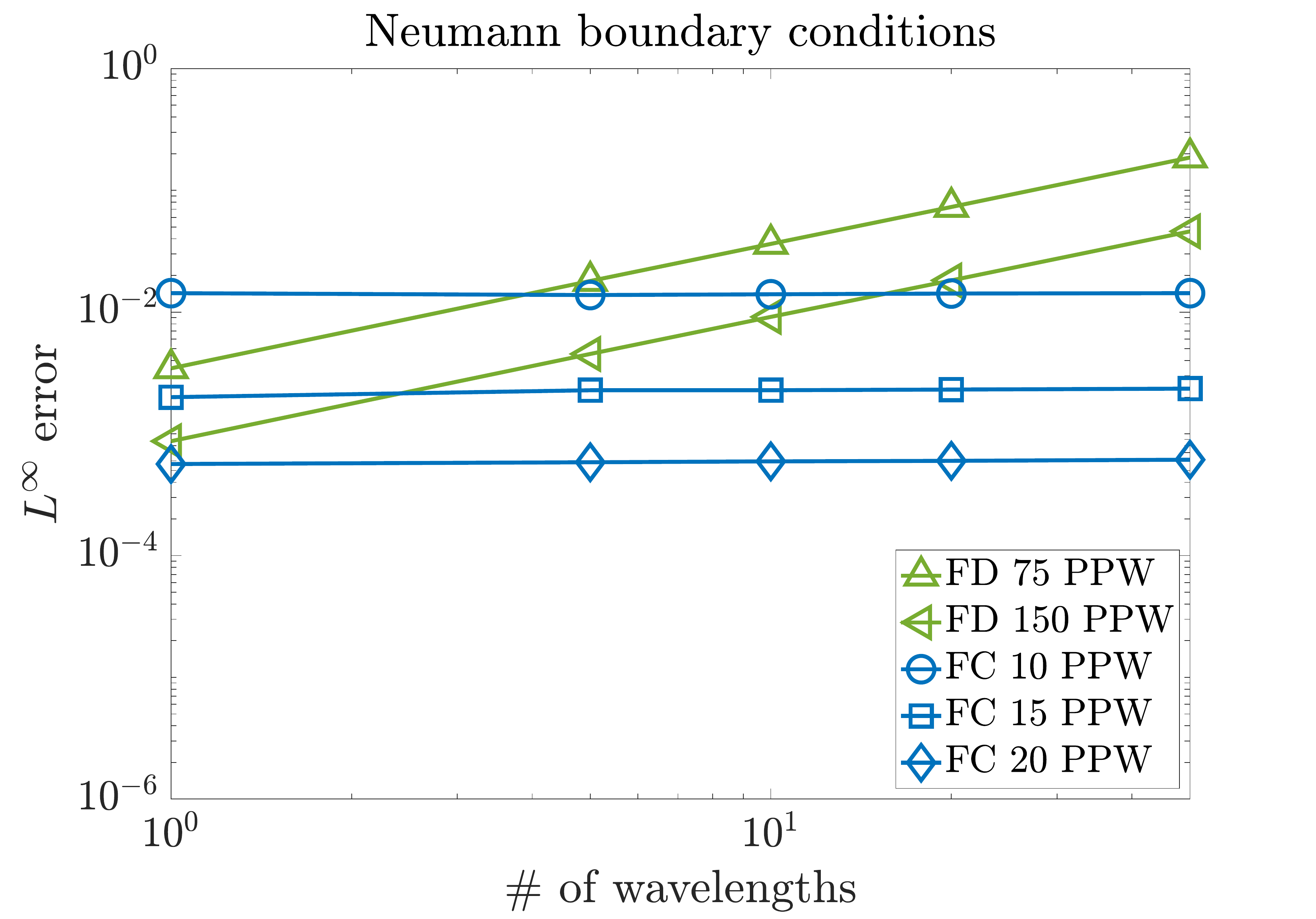}
} \caption{Plate-propagation test. Maximum numerical errors over all space and over one full temporal cycle (defined as the time required for any one crest to travel the length of the plate) are presented for a plane wave solution with increasing number of wavelengths for the 3D elastic wave equation with Neumann boundary conditions. }
\label{fig:fcdispersion}
\end{figure}
%***************************************************************
%***************************************************************

% figura 8
% figura 3 Oscar: close-up CD************************************
%*****************************************************************
\begin{figure}[!h]
\centering
{\center
\includegraphics[width=0.70\textwidth]{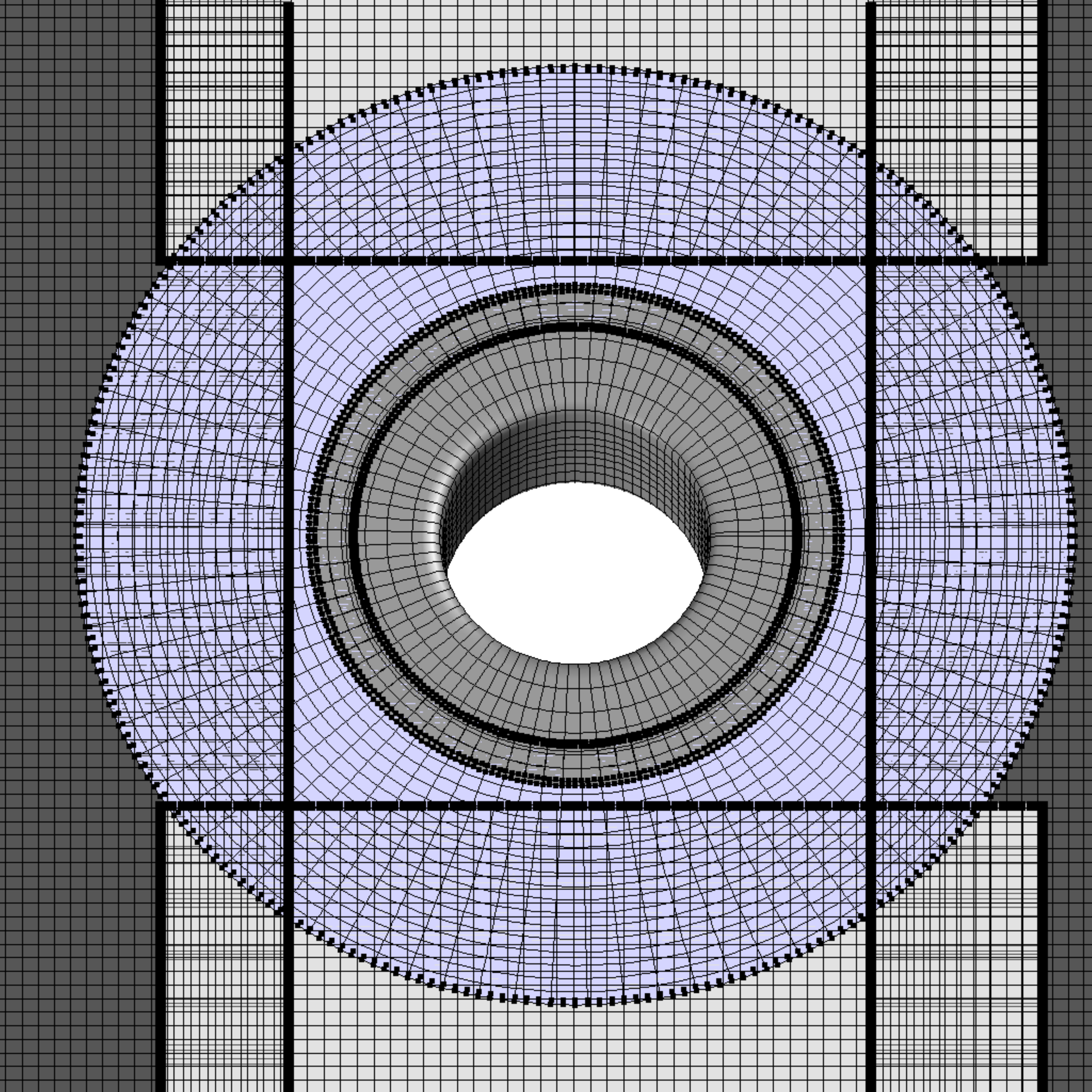}
}
\caption{Close up of computational model, composed of six overlapping patches, for a through-hole on an elastic plate.}
\label{fig:comp_model}
\end{figure}

Using the FC method to evaluate function values and their derivatives,
a time-domain PDE solver can be completed simply by incorporating an
adequate ordinary differential equation (ODE) solver. Following~\cite{amlani2016fc} we utilize the
fourth-order Adams-Bashforth (AB4) algorithm to effect the
timestepping.  The errors produced by the FC procedure for a simple
rectangular plate are illustrated in
figure~\ref{fig:fcdispersion}. For comparison errors resulting from
use of a finite-difference method of second order are also included. The figure shows that the FC errors remain constant as the number of wavelengths increase, while the FD errors grow---in a classical manifestation of the dispersion error. Note that, additionally, even for an acoustically small plate (one wavelength along the length of the plate) a total of 150 FD discretization points per wavelength are necessary to achieve an error comparable to that resulting from the FC approach on the basis of only 20 FC points. Other comparisons in the figure provide more details in these regards. For a given accuracy tolerance, the reduced FC meshes lead to substantially reduced computing times in comparison with those required by the FD method.  Figure~\ref{fig:comp_model} illustrates the  overlapping-patch geometrical description used in conjunction with the FC approach. Full details concerning accuracy, geometry treatment and time-stepping methods used can be found in~\cite{amlani2016fc}.

\subsection{Computational specification of the experimental configuration}
\label{subsection3_C} 

\begin{figure}
	\centering
  	\includegraphics
  	[width=70 mm,keepaspectratio=true]   
  	{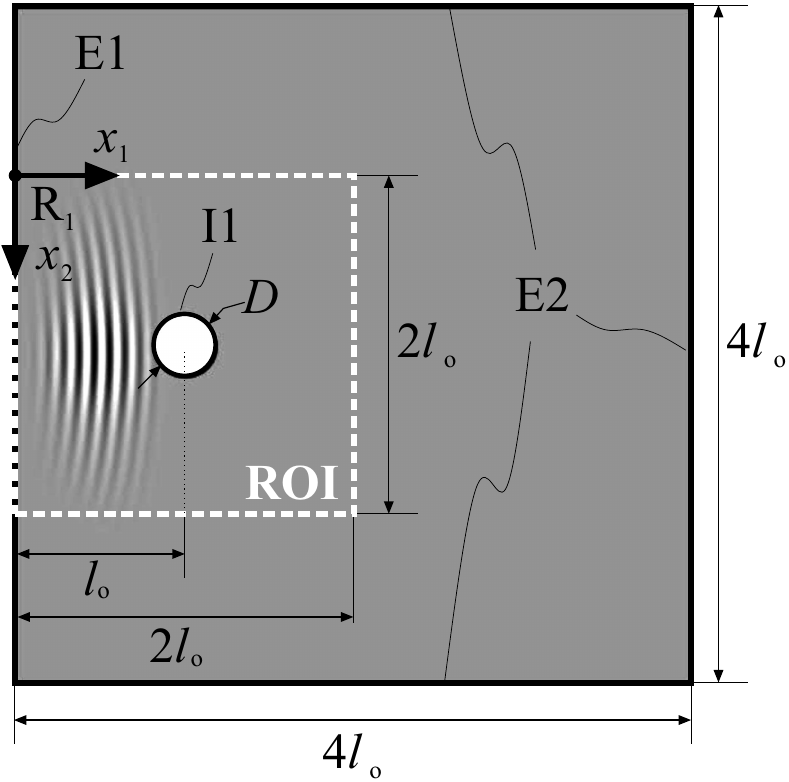}
	\caption {Top side of the prismatic computational domain, defined by two parallel planes at $x_{3}=\pm h$ (the top and bottom sides of the plate) and the plane lateral boundaries E1 and E2, with a through-thickness void, defined by the lateral cylindrical boundary I1 (the hole border). The relative position of the ROI and the reference system employed are shown.}
	\label{com_dom}
\end{figure}

A discrete version of the IVBP problem specified by equations~(\ref{eq_Lame-Navier_ij}-\ref{eq_BC_tau_ij}) has been developed in a prismatic computational domain with interior volume $\mathrm{G}$ (figure~\ref{com_dom}) using a self-developed high-precision MATLAB pre-computation for the FC(Gram) basis and a self-developed C/C++ implementation of the full 3D FC(Gram)-based elasticity solver described in subsection~\ref{subsection3_B}. Lam\'e-Navier
equation~(\ref{eq_Lame-Navier_ij}-\ref{eq_cLcT}) has been employed in
$\mathrm{G}$ and stress-free boundary condition~(\ref{eq_BC_tau_ij})
has been applied over the top and bottom sides of the prismatic domain
(which correspond to a portion of the top $\Gamma_{+}$ and bottom
$\Gamma_{-}$ surfaces of the plate) and over the cylindrical boundary
I1 (which corresponds to the hole border $\Gamma_{\mathrm{d}}$). On
top of that, the Dirichlet boundary conditions
%EQ.16************************************************************
%*****************************************************************
\begin{subequations}\label{eq_BC_E1}
	\begin{align} 
	u_{i}\left(\mathbf{x}^{\mathrm{E1}},t\right) & = u_{i}^{\mathrm{ai}}\left(\mathbf{x}^{\mathrm{E1}},t\right) 
	& \,i=1,2,3 \quad & \mathbf{x}^{\mathrm{E1}}\in    
	 \mbox{E1} 
	\\
	u_{i}\left(\mathbf{x}^{\mathrm{E2}},t\right) & = 0 
	& \, i=1,2,3 \quad & \mathbf{x}^{\mathrm{E2}}\in   
	 \mbox{E2},
 \end{align} 
\end{subequations}
%***********************************************************
%*****************************************************************
have been imposed over the lateral exterior boundary of $\mathrm{G}$,  non homogeneous ones for introducing the incident field at E1 and homogeneous ones in the remaining part E2. The thickness of the computational domain has been set equal to the nominal thickness of the plates ($2h=10\;\mathrm{mm}$). The height and the width of the top and bottom sides of the computational domain have been adequately selected in order to avoid interference in the ROI region between direct scattering phenomena and waves reflected by the exterior boundary. Null initial conditions 
%EQ.17***********************************************************
%*****************************************************************
\begin{subequations}\label{eq_ini}
	\begin{align} 
	u_{i}\left(\mathbf{x},0\right) 
	& =  0 
	& \quad i=1,2,3 \quad\mathbf{x}\in\mathrm{G}
	\\
	\frac{\partial u_{i}}{\partial t}\left(\mathbf{x},0\right)
	& = 0 
	& \quad i=1,2,3 \quad \mathbf{x}\in\mathrm{G}
 \end{align} 
\end{subequations}
%**********************************************************
%*****************************************************************
have been assumed for both displacement and velocity in $\mathrm{G}$.

% figura 9: com_dom%**********************************************
%*****************************************************************
%*****************************************************************
%*****************************************************************

\subsection{Analytic approximation to the incident field}
\label{subsection3_D} 

For specifying the analytic approximation $u_{i}^{\mathrm{ai}}\left(\mathbf{x}^{\mathrm{E1}},t\right)$ to the Cartesian components of the incident field used in the non-homogeneous Dirichlet boundary condition at E1, where $\mathbf{x}^{\mathrm{E1}}=\left(0,x_{2},x_{3}\right)$, we use a two-stage procedure: first, we obtain a complex analytic approximation $\hat{u}_{3}^{\mathrm{ai}}\left(\mathbf{x}_{h}^{\mathrm{E1}},t\right)$ to the experimental measured out-of-plane complex component of the incident field at the top surface of the plate at E1, where $\mathbf{x}_{h}^{\mathrm{E1}}=\left(0,x_{2},h\right)$. Then, from this complex field we reconstruct the Cartesian components of the real displacement of the incident field at any point over the whole section of the plate in E1 using the properties of quasi-Rayleigh wavetrains stated in subsection~\ref{subsection2_B}. 
 
 % tabla 1: Tab_1*****************************************************
%***************************************************************
\begin{table*}
\caption{\label{Tab_1}Models employed to obtain the analytic approximation $\hat{u}_{3}^{\mathrm{ai}}\left(\mathbf{x}_{h}^{\mathrm{E1}},t\right)$ to the out-of-plane component of the incident field at the top surface of the plate. The best-fit values of the parameters of the three models are obtained using a least squares procedure over three selected experimental profiles along the $x_{1}$, $x_{2}$ and $t$ directions, respectively. The best-fit values $\left[a_{1}\right]$ and $\left[c_{1}\right]$ of the parameters $a_{1}$ and $c_{1}$ are used to obtain the central Rayleigh wavenumber $k_{\mathrm{R}}$ and the central temporal frequency $f_{\mathrm{R}}$, respectively.}
%\begin{indented}
%\item[]
\begin{tabular}{
@{}
c@{\hspace{0.05cm}}
c@{\hspace{0.05cm}}
c@{}}
%\br
\hline
 complex field
 & amplitude
 & phase
\\
\hline   
   $\hat{U}_{1}\left(x_{1}\right)
=U_{1\mathrm{m}}\left(x_{1}\right)
\exp\left[\mathrm{j} \varphi_{1\mathrm{m}}\left(x_{1}\right)\right]$ 
  & not used  
  &$a_{1}x_{1}+a_{2}$
  \\
\hline   
   $\hat{U}_{t}\left(t\right)
=U_{t\mathrm{m}}\left(t\right)
\exp\left[\mathrm{j} \varphi_{t\mathrm{m}}\left(t\right)\right]$ 
  &\(\displaystyle 
\sum_{i=1}^{2} b_{1i}\exp\left[-\left( \frac{t-b_{2i}}{b_{3i}}\right)^2 \right]\) 
  &$c_{1}t+c_{2}$
  \\
%\mr
\hline  
  $\hat{U}_{2}\left(x_{2}\right)
=U_{2\mathrm{m}}\left(x_{2}\right)
\exp\left[\mathrm{j} \varphi_{2\mathrm{m}}\left(x_{2}\right)\right]$ 
  &\(\displaystyle 
\frac{p_{1}}{x_{2}^{2}+q_{1}x_{2}+q_{2}}\) 
%\(\displaystyle\)
  &$e_{1}x_{2}^{2}+e_{2}x_{2}+e_{3}$
  \\
  \hline  
%\br
\end{tabular}
%\end{indented}
\end{table*}
%***************************************************************
%***************************************************************

\begin{figure*}
\centering
\begin{tabular}{
c@{\hspace{0.2cm}}
c
}
     \includegraphics 
     [height=54 mm,keepaspectratio=true] 
     {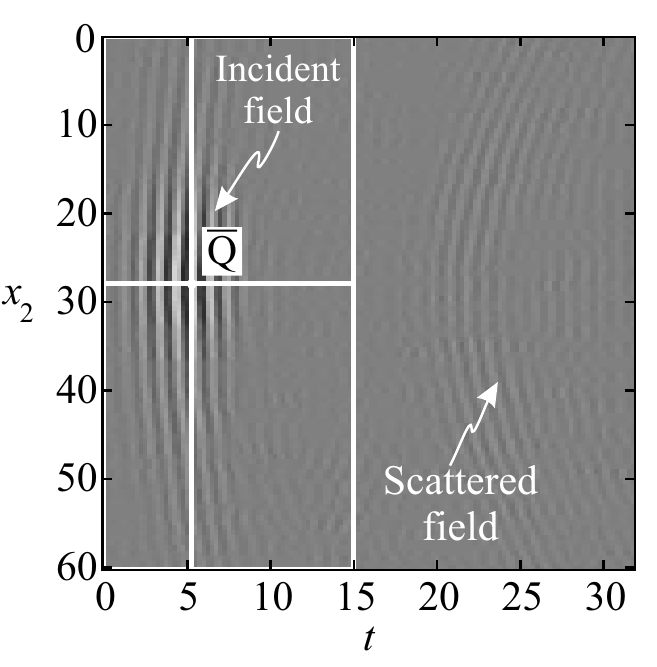}
  &  \includegraphics 
	 [height=54 mm,keepaspectratio=true]   
     {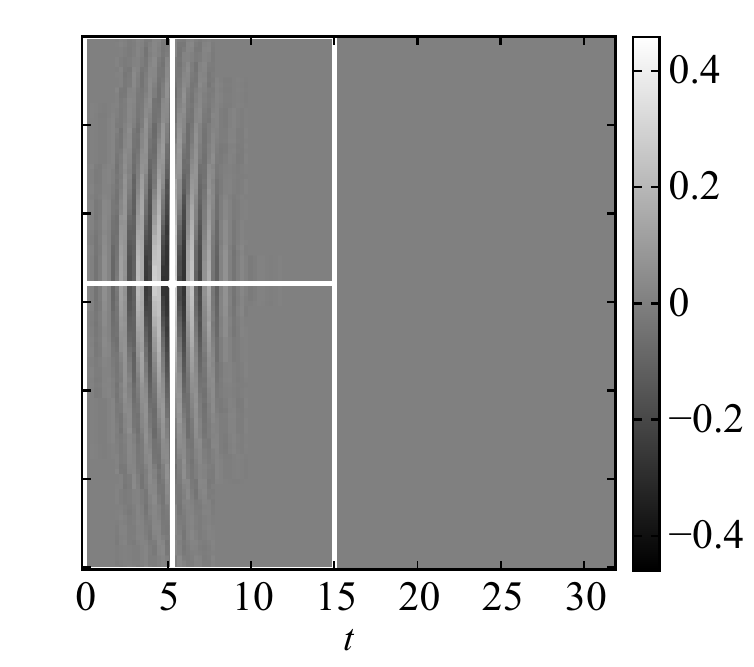}
\\
     (a)
   & (b)
\\
\end{tabular}	
	\caption{Temporal record of the out-of-plane displacement at boundary E1 ($\mathbf{x}_{h}^{\mathrm{E1}}=\left(0,x_{2},h\right)$): (a) experiment $u_{3}^{\mathrm{e}}\left(\mathbf{x}_{h}^{\mathrm{E1}},t\right)$, (b) reconstructed incident field employing the analytic approximation $u_{3}^{\mathrm{ai}}\left(\mathbf{x}_{h}^{\mathrm{E1}},t\right)$. Point $\overline{\mathrm{Q}}$ with coordinates $\bar{x}_{2}\,=\,0.028\,\mathrm{mm}$ and $\bar{t}\,=\,5.25\,\mathrm{\mu s}$ is marked in white. Only the rectangular subzone on the left in (a) has been employed for obtaining the analytic approximation to the non-homogeneous Dirichlet boundary condition in (b). Displacement in units of $\lambda_{\mathrm{o}}/8\pi$. $x_{2}$ in millimeters and $t$ in microseconds. The axis numbering corresponds to that of figure~\ref{ROI}. Midgray level represents zero.} 
\label{analitic_aprox}
\end{figure*}

To obtain $\hat{u}_{3}^{\mathrm{ai}}\left(\mathbf{x}_{h}^{\mathrm{E1}},t\right)$ in the first stage, on the one hand, we use the part of the temporal record of the out-of-plane displacement at the top surface of the plate in E1 $\hat{u}_{3}^{\mathrm{e}}\left(\mathbf{x}_{h}^{\mathrm{E1}},t\right)$ which corresponds to the incident field (see figure~\ref{analitic_aprox}.a). On the other hand, from the  experimental sequence 
(see f.i. figure~\ref{res_secuencia}),
we select a particular frame $\hat{u}_{3}^{\mathrm{e}}\left(\mathbf{x}_{h},\overline{t}\right)$ at  reference time instant $\overline{t}$ in which the wavetrain is completely inside the ROI but without interaction with the hole. From the spatial  distribution of the amplitude and phase in this particular frame we can assume that
the phase and the amplitude of the incident field can be described as a quasi-cylindrical inhomogeneous wave that diverges from a line-source, parallel to the $x_{3}$ axis, which intersect the top surface of the plate at point $\mathrm{F}$ (see figure~\ref{esquema_plate}.b). In this framework, we derive the complex analytic approximation $\hat{u}_{3}^{\mathrm{ai}}\left(\mathbf{x}_{h}^{\mathrm{E1}},t\right)$ by following an objective matching procedure in 6 steps~\cite{Tesis_PRG}: 1) the position $\mathbf{x}_{h}^{\mathrm{F}}=\left({x}_{1}^{\mathrm{F}},{x}_{2}^{\mathrm{F}},h\right)$ of the virtual source $\mathrm{F}$ at the top surface of the plate is obtained as the center of curvature of the wavefronts of the incident field in the selected frame at $\overline{t}$; 2) using $\overline{x}_{2}={x}_{2}^{\mathrm{F}}$ over this  frame we select the complex profile $\hat{u}_{3}^{\mathrm{e}}\left(x_{1},\overline{x}_{2},h,\overline{t}\right)$ and from the best-fit value $\left[a_{1}\right]$ of the slope of a linear fit to its acoustic phase (the phase of the field $\hat{U}_{1}\left(x_{1}\right)$ in table~\ref{Tab_1}) we estimate the central wavenumber of the train $k_{\mathrm{R}}=\left[a_{1}\right]$; 3) over the temporal record  $\hat{u}_{3}^{\mathrm{e}}\left(\mathbf{x}_{h}^{\mathrm{E1}},t\right)$  
we select a reference point $\overline{\mathrm{Q}}$ with coordinates
$\left(\overline{x}_{2},\overline{t}\right)$ located roughly at the center of the part of the temporal record corresponding to the incident field (figure~\ref{analitic_aprox}.a), obtaining the profiles
$\hat{u}_{3}^{\mathrm{e}}\left(\overline{x}_{1},\overline{x}_{2},h,t\right)$ and 
$\hat{u}_{3}^{\mathrm{e}}\left(\overline{x}_{1},x_{2},h,\overline{t}\right)$ which characterize at $\overline{\mathrm{Q}}$ the distribution of the incident field in the $t$ and $x_{2}$ directions, respectively; 4) it is assumed that the the out-of-plane component of the incident field at the top surface of the plate at E1 can be factorized as
% equation 18********************************************
%***************************************************************
\begin{equation}\label{eq_factor}
\hat{u}_{3}^{\mathrm{ai}}\left(\mathbf{x}_{h}^{\mathrm{E1}},t\right):=\hat{u}_{3}^{\mathrm{ai}}\left(0,x_{2},h,t\right)
=
u_{30}\exp\left(\mathrm{j} \varphi_{30}\right)
\hat{U}_{t}\left(t\right)\hat{U}_{2}\left(x_{2}\right),
\end{equation}
%************************************************************
%***************************************************************
where the models for $\hat{U}_{t}\left(t\right)$ and $\hat{U}_{2}\left(x_{2}\right)$ are congruent with the usual description of the phase of a cylindrical wave in the Fresnel approximation (table~\ref{Tab_1}); 5) the best-fit values of the parameters of these models are found by matching $\hat{U}_{t}\left(t\right)$ to $\hat{u}_{3}^{\mathrm{e}}\left(\overline{x}_{1},\overline{x}_{2},h,t\right)$ and $\hat{U}_{2}\left(x_{2}\right)$ to
$\hat{u}_{3}^{\mathrm{e}}\left(\overline{x}_{1},x_{2},h,\overline{t}\right)$  using a least squares procedure. In particular, the best-fit value $\left[c_{1}\right]$ (see table~\ref{Tab_1}) divided by $2\pi$ gives an estimate of the central temporal frequency $f_{\mathrm{R}}$ of the wavetrain, that must be close to the nominal value of the selected excitation frequency. From $f_{\mathrm{R}}$ and $k_{\mathrm{R}}$ we obtain the Rayleigh velocity of the plate as $c_{\mathrm{R}}=2\pi f_{\mathrm{R}}/k_{\mathrm{R}}=\left[c_{1}\right]/\left[a_{1}\right]$ that combined with $c_{\mathrm{L}}$ allows us to determine any other relevant parameter that could be used to characterize the wave propagation in the plate (e.g. the phase velocity of transversal waves $c_{\mathrm{T}}$, the Poisson's ratio of the plate $\nu$  and the associated normalized spectrum of Lamb modes, etc.); 6) $u_{30}$ and $\varphi_{30}$ are global renormalization factors in amplitude and phase that provide, if necessary, additional degrees of freedom to obtain a better global matching of the analytical approximation considering all the values of the complex amplitude of the incident field within the temporal record $\hat{u}_{3}^{\mathrm{e}}\left(\mathbf{x}_{h}^{\mathrm{E1}},t\right)$. 

% figura 10
% figura 7 Carlos: analitical aproximation incident field****************
%*****************************************************************

In the second stage we use expression~(\ref{eq_uqR_uqRh}) to reconstruct the components of the complex displacement of the incident field at any point over the whole section of the plate in E1 as
%EQ.19**************************************************************
%*******************************************************************
\begin{subequations}\label{eq_uai_c}
\begin{align}
   \hat{u}_{i}^{\mathrm{ai}}\left(\mathbf{x}^{\mathrm{E1}},t\right)
   & =
   \left[
   \underline{\hat{u}}_{\parallel f_{\mathrm{R}}\mathrm{m}}^{\mathrm{S0}}
   \left(x_{3}\right)
   + 
   \underline{\hat{u}}_{\parallel f_{\mathrm{R}}\mathrm{m}}^{\mathrm{A0}}
   \left(x_{3}\right)
   \right]
  \nonumber \\ 
   & \times
   \left[
   \frac
   {\mathbf{x}_{h}^{\mathrm{E1}}-\mathbf{x}_{h}^{\mathrm{^F}}}
   {\vert \mathbf{x}_{h}^{\mathrm{E1}}-\mathbf{x}_{h}^{\mathrm{^F}}\vert}
   \cdot 
   \mathbf{a}_{i}
   \right]    
   \hat{u}_{3}^{\mathrm{ai}}\left(\mathbf{x}_{h}^{\mathrm{E1}},t\right) \quad i=1,2
\\
   \hat{u}_{3}^{\mathrm{ai}}\left(\mathbf{x}^{\mathrm{E1}},t\right)
  & = \left[ \underline{\hat{u}}_{\perp f_{\mathrm{R}}\mathrm{m}}^{\mathrm{S0}}\left(x_{3}\right)+ \underline{\hat{u}}_{\perp f_{\mathrm{R}}\mathrm{m}}^{\mathrm{A0}}\left(x_{3}\right)\right] \hat{u}_{3}^{\mathrm{ai}}\left(\mathbf{x}_{h}^{\mathrm{E1}},t\right) \quad \quad \quad \quad
\end{align}
\end{subequations}
%*******************************************************************
%*******************
from which we derive the analytic approximation to the real displacement incident field as 
%EQ.20***********************************************************
%*****************************************************************
\begin{equation}\label{eq_uai_def}
u_{i}^{\mathrm{ai}}\left(\mathbf{x}^{\mathrm{E1}},t\right)=
\mathrm{Re}\left[\hat{u}_{i}^{\mathrm{ai}}\left(\mathbf{x}^{\mathrm{E1}},t\right)\right]
\times E_{t}(t)E_{2}(x_{2}), \quad i=1,2,3 \quad \mathbf{x}^{\mathrm{E1}}\in   
	 \mbox{E1} 
\end{equation}
%*******************************************************************
%*****************************
Auxiliary real envelopes $E_{t}(t)$ and $E_{2}(x_{2})$ (with pulse profiles with a smooth transition in amplitude from one to zero along $t$ and $x_{2}$, respectively) have been employed in expression (\ref{eq_uai_def}) to guarantee proper matching of the inhomogeneous boundary condition at E1 with the lateral homogeneous boundary condition at E2 and the zero initial values in the computational domain imposed by expression (\ref{eq_ini}). Practical implementation and processing of numerical data based on this approach has been developed employing self-developed routines in Matlab$^{\ooalign{\hfil\raise.07ex\hbox{$\scriptstyle\rm\text{R}$}\hfil\crcr\mathhexbox20D}}$.

%******************************
%*****************************
\section{Results and discussion}
\label{section4}
We have employed the previously described methodology to analyze the experimental and numerical sequences corresponding to the scattering of quasi-Rayleigh wavetrains by four through-thickness cylindrical holes with nominal diameters $12\;\mathrm{mm}$, $8\;\mathrm{mm}$, $6\;\mathrm{mm}$ and $4\;\mathrm{mm}$. After obtaining the experimental sequence for each hole $\hat{u}_{3}^{\mathrm{e}}\left(\mathbf{x}_{h},t\right)$ using the procedure described in subsection \ref{subsection3_A}, we obtained the  analytical approximation to the incident field and the propagation parameters that characterize the wave propagation in the plates using the procedure described in subsection \ref{subsection3_D}. We found that the analytical approximation matches the experimental incident field in the corresponding temporal record in all cases, even though pixel to pixel agreement (see figure~\ref{analitic_aprox}) presents errors at some points in concordance with previously measured noise levels in our case, that are typically around $0.1$ of the mean signal level, both in amplitude and phase~\cite{Tesis_PRG}. In addition, the values of the propagation parameters for each hole and their associated sample averages (table \ref{Tab_2}) have coherent and reasonable values: on the one hand, the measured central temporal frequency $f_{\mathrm{R}}$ matches the nominal excitation frequency and the values of the Rayleigh phase velocity, the transversal phase velocity and the Poisson's ratio are within the typical values reported for 
aluminum; on the other hand, the corresponding normalized values of the frequency and wavenumber of the wavetrains are localized in the qR region of the spectrum (see f.i. the point associated to the averaged normalized frequency $\gamma_{\mathrm{R}} \sim 3.1$ an the associated normalized wavenumber $\xi_{\mathrm{R}} \sim 6.5$ in figure~\ref{espectro_Lamb}). Then, using the analytic approximation and the associated  propagation parameters, we solved the IBVP for each hole following the procedure outlined in section \ref{subsection3_B} to obtain the corresponding numerical sequences $\hat{u}_{3}^{\mathrm{n}}\left(\mathbf{x}_{h},t\right)$. The typical computational time was of the order of 90 min using a computer with 512 cores for 40000 time-steps of $0.5$ ns and a discretization of the computational domain that leads to $5$ million of unknowns.
% tabla 2: Tab_2*****************************************************
%***************************************************************
\begin{table*}
\caption{\label{Tab_2}Values of the propagation parameters obtained from the coefficients of the analytic approximation  $\hat{u}_{3\mathrm{m}}^{\mathrm{ai}}\left(\mathbf{x}_{h}^{\mathrm{E1}},t\right)$ to the out-of-plane component of the incident field at the top surface of the plate. The dispersion around the sample mean is specified after the $\pm$ sign by the rounded-up value of the standard deviation of the mean.}
%\begin{indented}
%\item[]
\begin{tabular}{
@{}
c@{\hspace{0.05cm}}
c@{\hspace{0.05cm}}
c@{\hspace{0.3cm}}
c@{\hspace{0.3cm}}
c@{\hspace{0.3cm}}
c@{\hspace{0.3cm}}
c@{\hspace{0.3cm}}
c@{}}
%\br
\hline
parameter
 & expression
 & unit
 & \multicolumn{4}{c} {$D/\mathrm{mm}$}
 & sample mean
\\
 
 & 
 & 
 & $12$
 & $8$
 & $6$
 & $4$
 & 
\\
%\mr
\hline  
   $k_\mathrm{R}$ 
  & $\left[a_{1}\right]$ 
  & rad/m
  & $2033$
  & $2061$
  & $2064$
  & $2067$
  & $2057\pm8$
  \\
\hline   
   $\omega_\mathrm{R}$ 
  & $\left[c_{1}\right]$ 
  & (rad/s) $10^6$
  & $6.113$
  & $6.132$
  & $6.165$
  & $6.161$
  & $6.143\pm0.002$
\\
\hline 
\hline
$\lambda_\mathrm{R}$ 
  & $2\pi/k_{\mathrm{R}}$
  & mm
  & $3.09$
  & $3.05$
  & $3.04$
  & $3.04$
  & $3.05\pm0.02$
 \\
\hline
$f_\mathrm{R}$ 
  & $\omega_{\mathrm{R}}/2\pi$
  & s $10^6$
  & $0.973$
  & $0.976$
  & $0.981$
  & $0.981$
  & $0.977\pm0.002$
 \\
\hline
\hline
$c_\mathrm{R}$ 
  & $f_{\mathrm{R}}\lambda_{\mathrm{R}}$
  & m/s
  & $3005$
  & $2975$
  & $2986$
  & $2980$
  & $2987\pm7$
 \\
\hline
$c_\mathrm{T}$ 
  & equation (\ref{eq_cR})
  & m/s
  & $3227$
  & $3191$
  & $3204$
  & $3197$
  & $3205\pm8$
 \\
\hline
$\nu$ 
  & $\frac{\left(2{\chi_{\mathrm{TL}}^2}-1\right)}
  {\left(2{\chi_{\mathrm{TL}}^2}-2\right)}$ 
  & -
  & $0.327$
  & $0.332$
  & $0.330$
  & $0.331$
  & $0.330\pm0.002$
 \\
\hline
\hline
$\xi_\mathrm{R}$ 
  & $2hk_\mathrm{R}/\pi$
  & -
  & $6.47$
  & $6.56$
  & $6.57$
  & $6.58$
  & $6.55\pm0.03$
\\
\hline 
$\gamma_\mathrm{R}$ 
  & $4hf_\mathrm{R}/c_{\mathrm{L}}$
  & -
  & $3.059$
  & $3.069$
  & $3.085$
  & $3.083$
  & $3.074\pm0.007$
 \\
 \hline 
%\br
\end{tabular}
%\end{indented}
\end{table*}
%***************************************************************
%***************************************************************

% figura 11
% figura 8 Carlos sequence****************************************
%***************************************************************
\begin{figure*}
\centering
    \includegraphics 
     [width=120 mm,keepaspectratio=true]  
     {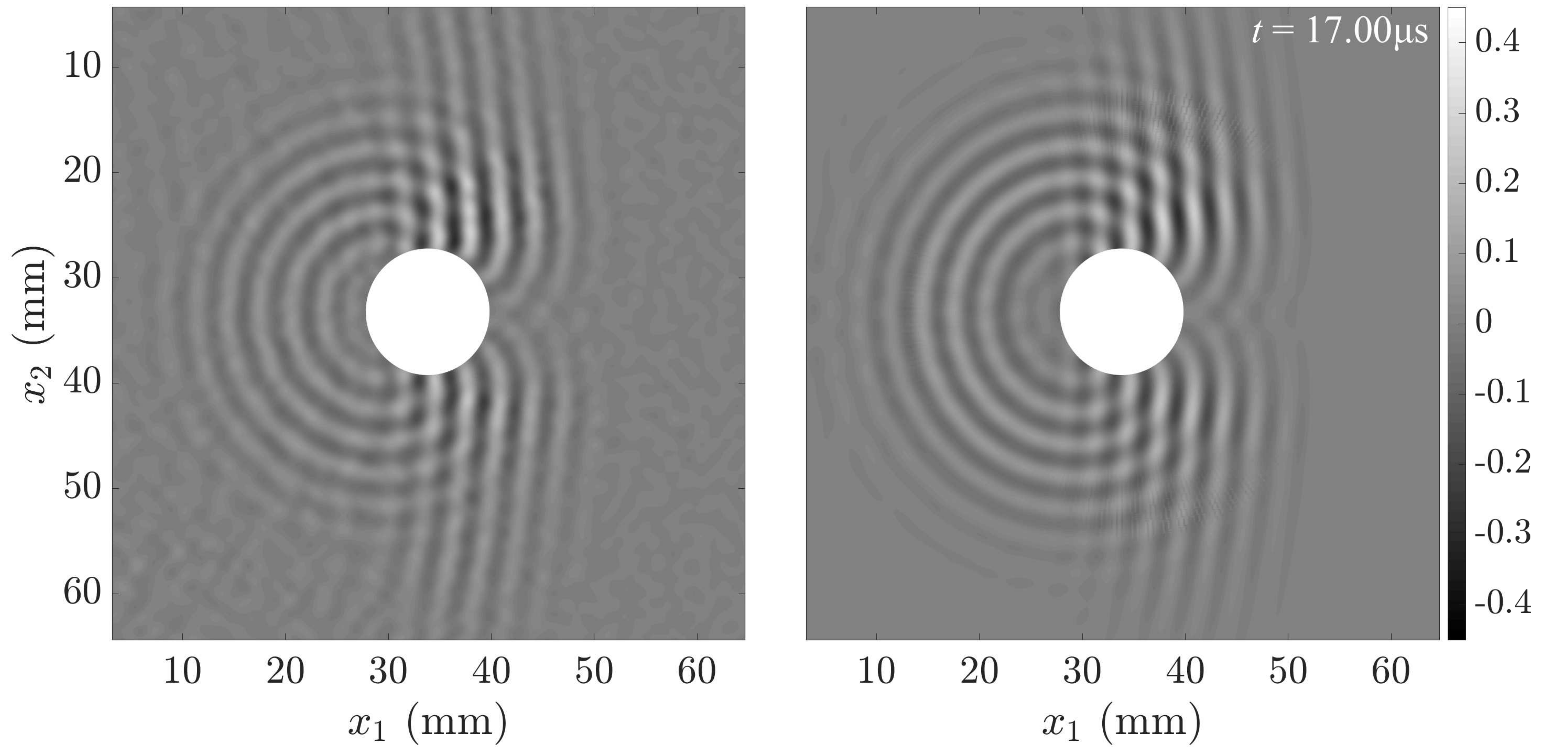}
\caption{Frame of the video sequence of the out-of-plane displacement field for the scattering of a transient narrow-band quasi-Rayleigh wave by a cylindrical hole with $D\,=\,12\,\mathrm{mm}$: reference field (experimental) on the left and problem field (numerical) on the right. Amplitude in units of $\lambda_{\mathrm{o}}/8\pi$. Lengths in millimeters. The axis numbering corresponds to that of figure~\ref{ROI}. 
Full sequence in Video 1, MPEG, 1 MB, [video\_1\_figure\_11.mp4].} 
\label{res_secuencia}
\end{figure*}
%*****************************************************************
%*****************************************************************

\begin{figure*}
\centering
\begin{tabular}{
c@{\hspace{0.2cm}}
c@{\hspace{0.2cm}}
c
}
   \raisebox{18mm}{(a)}
  &  \includegraphics 
     [width=42 mm,keepaspectratio=true] 
     {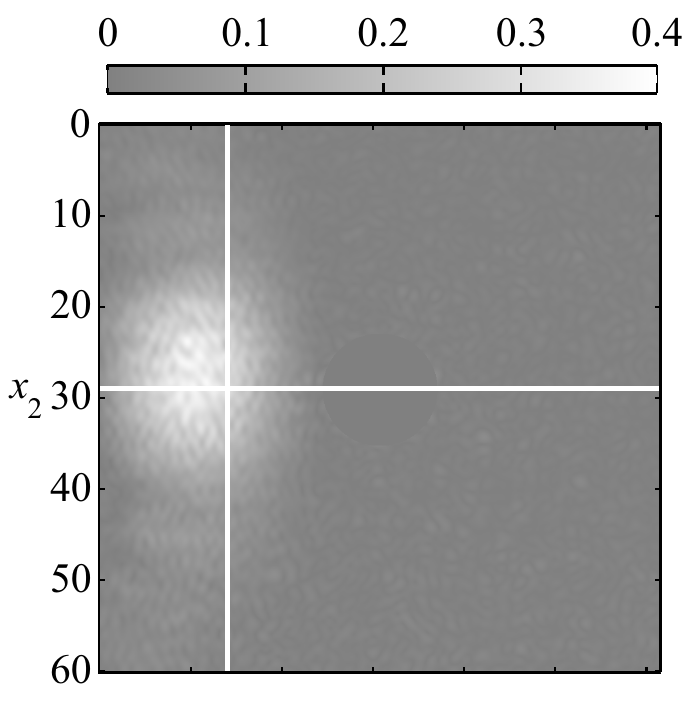}
  &  \includegraphics 
	 [width=41 mm,keepaspectratio=true]   	 
  	 {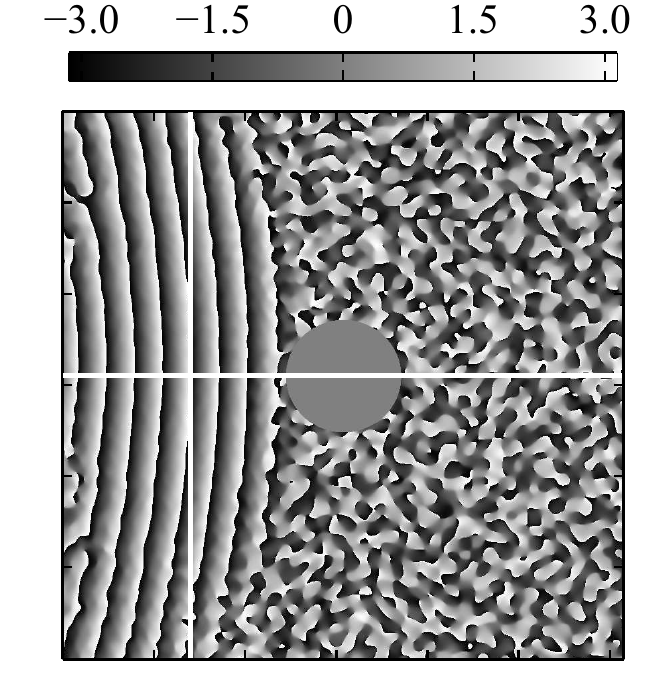}
  \\
    \raisebox{23mm}{(b)}
  & \includegraphics 
	[width=42 mm,keepaspectratio=true]   	
  	{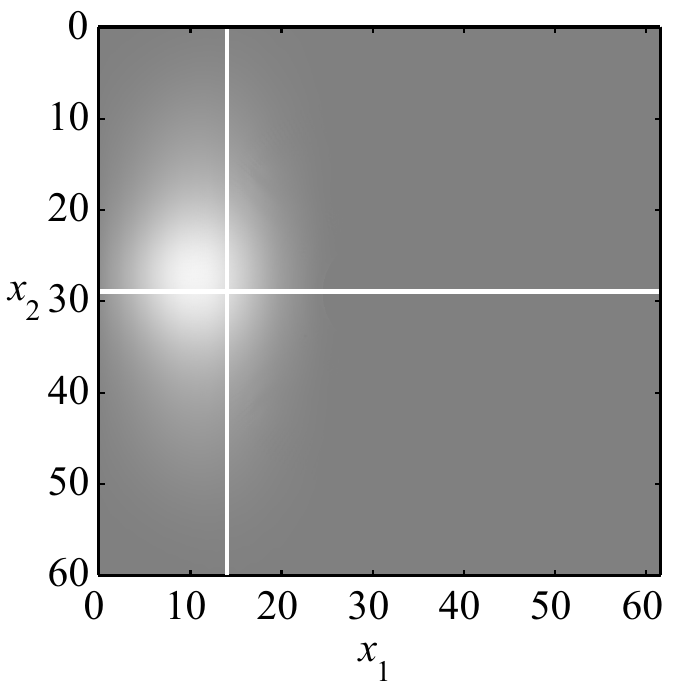}
  & \includegraphics 
	[width=41 mm,keepaspectratio=true]    	
  	{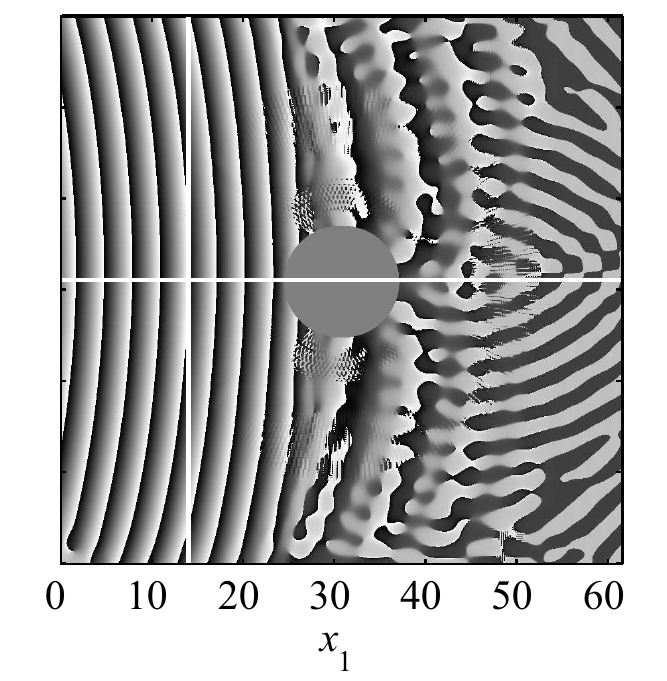}
   \\
    \raisebox{15mm}{(c)}
  &  \includegraphics 
	 [width=41 mm,keepaspectratio=true]    	
  	{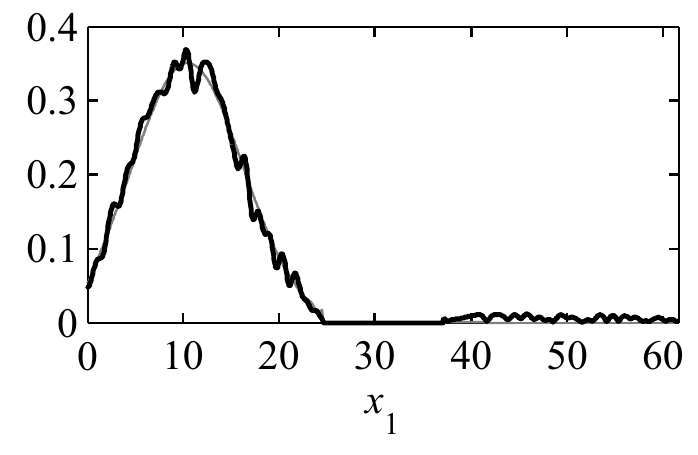}
  &  \includegraphics 
     [width=41 mm,keepaspectratio=true]      
    {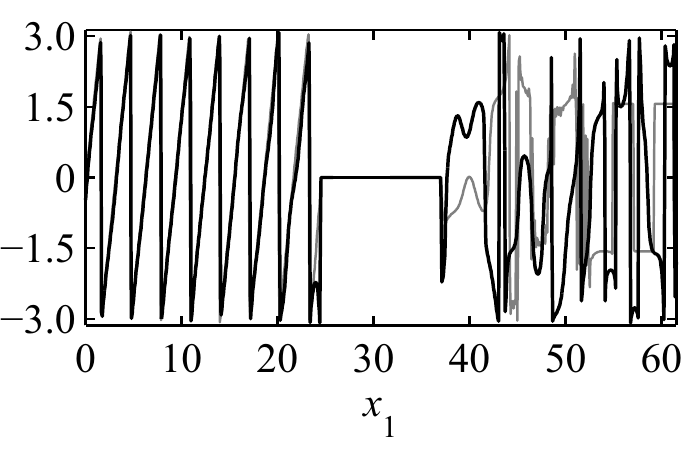}
   \\
   \raisebox{15mm}{(d)}
  &  \includegraphics 
    [width=41 mm,keepaspectratio=true]    	
  	{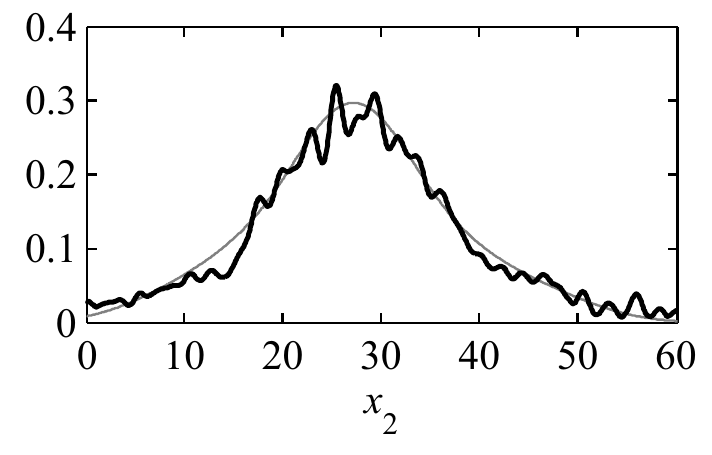}
  &  \includegraphics 
	[width=41 mm,keepaspectratio=true]    
    {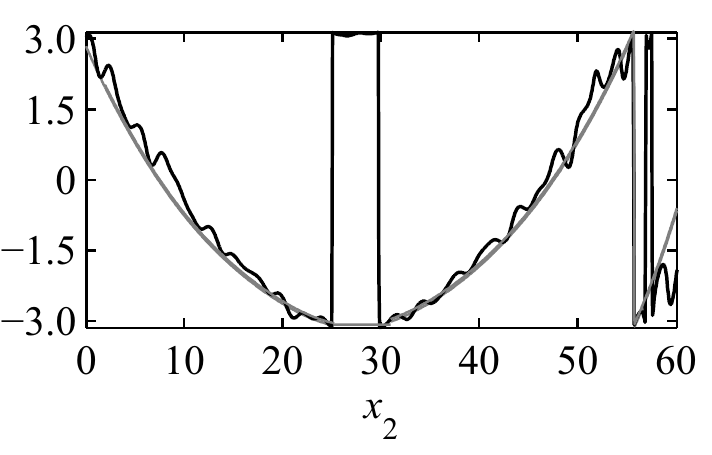}
   \\
   
  &  Amplitude(C1)
  &  Phase(C2)
  \\
\end{tabular}
	\caption{Scattering of a transient narrow-band quasi-Rayleigh wave by a cylindrical hole with $D\;=\;12\;\mathrm{mm}$ at $t_{1}\;=\;8.50\;\mathrm{\mu s}$: a) experimental field, (b) numerical field, (c) and (d) profiles along the lines in the $x_{1}$ and $x_{2}$ directions marked in white in (a) (thick) and (b) (thin). Amplitude in units of $\lambda_{\mathrm{o}}/8\pi$. Phase in radians. Lengths in millimeters. The axis numbering corresponds to that of figure~\ref{ROI}.} % The complete sequence of 2D maps (from $t\;=\;0$ to $t\;=\;18.75\;\mathrm{\mu s}$) is available as the MPEG file \textit{sequence.mpg}.	
\label{res_tran_t1}
\end{figure*}

\begin{figure*}
\centering
\begin{tabular}{
c@{\hspace{0.2cm}}
c@{\hspace{0.2cm}}
c
}
   \raisebox{18mm}{(a)}
  & \includegraphics 
    [width=42 mm,keepaspectratio=true]   
    {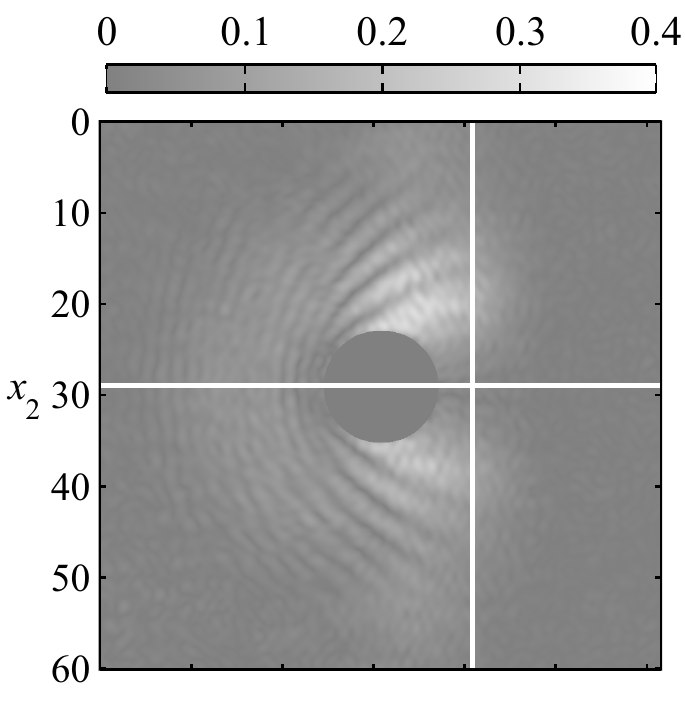}
  & \includegraphics 
    [width=41 mm,keepaspectratio=true]   
    {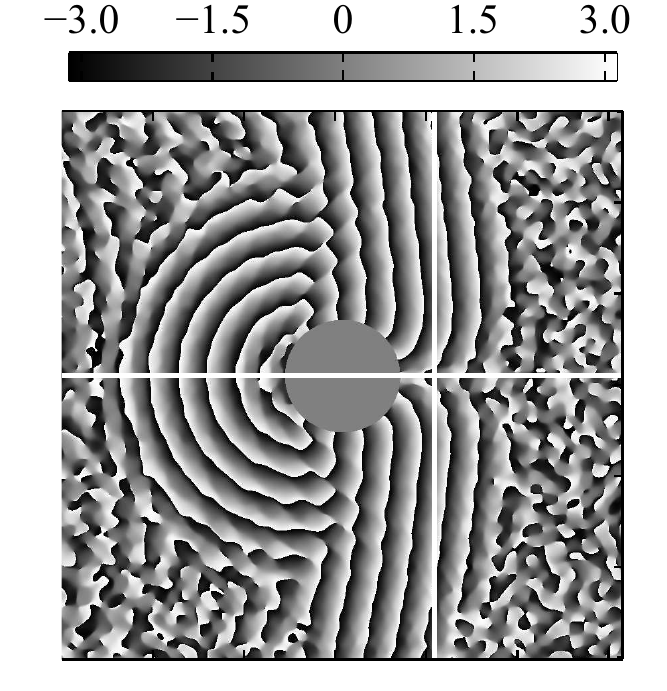}
  \\
    \raisebox{23mm}{(b)}
  & \includegraphics 
    [width=42 mm,keepaspectratio=true]   
    {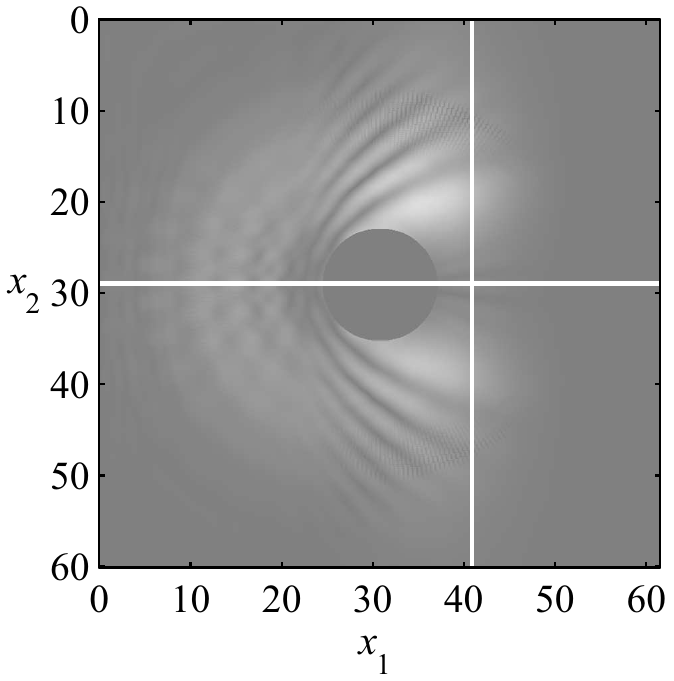}
  & \includegraphics 
    [width=41 mm,keepaspectratio=true]   
    {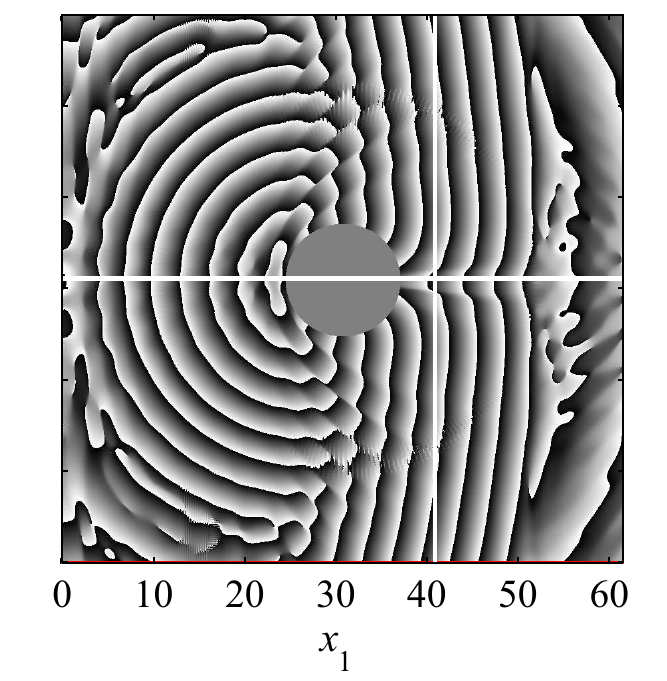}
   \\
    \raisebox{15mm}{(c)}
  & \includegraphics
    [width=41 mm,keepaspectratio=true]   
    {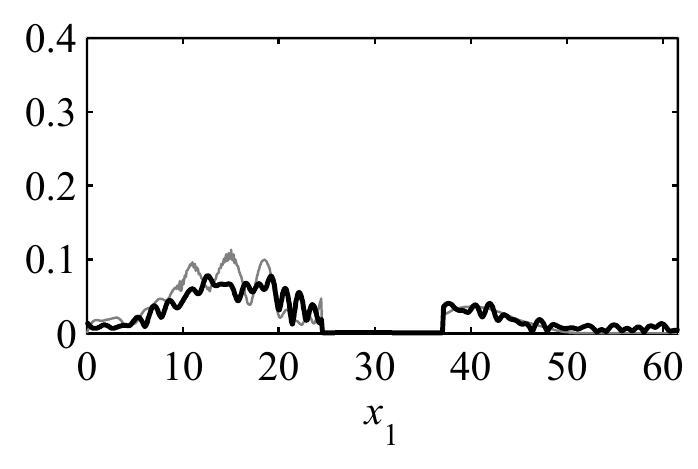}
  & \includegraphics 
    [width=41 mm,keepaspectratio=true]  
    {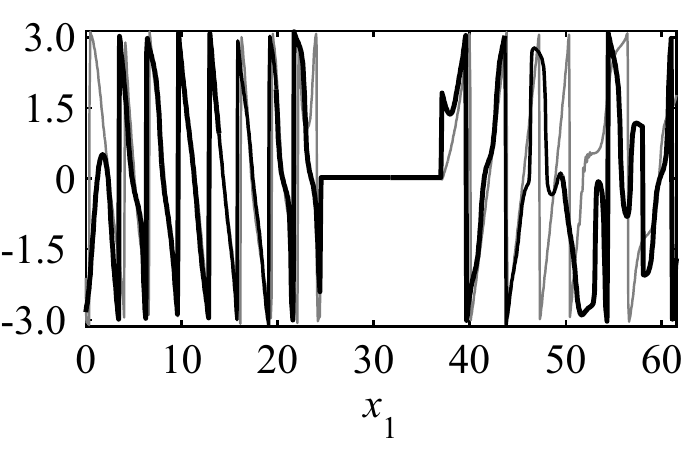}
   \\
   \raisebox{15mm}{(d)}
  & \includegraphics 
    [width=41 mm,keepaspectratio=true]    
    {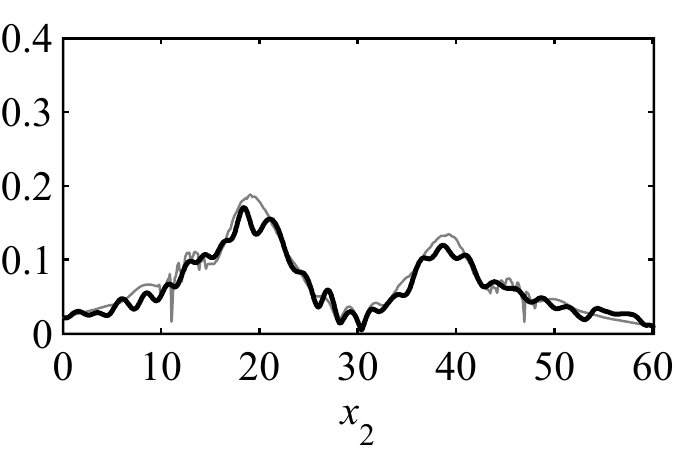}
  & \includegraphics 
    [width=41 mm,keepaspectratio=true]  
    {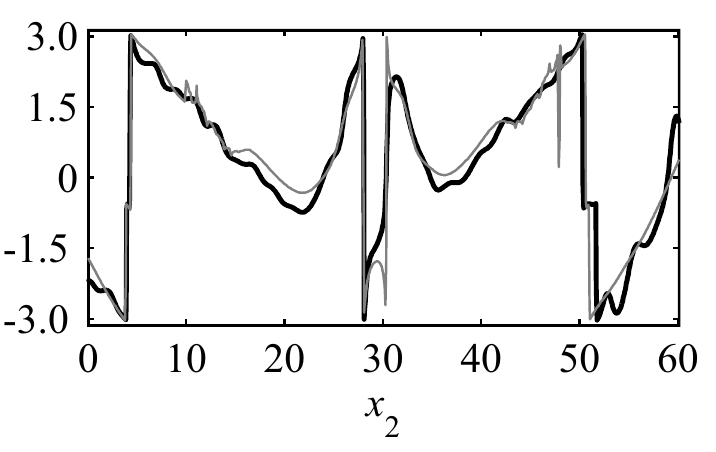}
   \\
  &  Amplitude(C1)
  &  Phase(C2)
  \\
\end{tabular}
	\caption{Scattering of a transient narrow-band quasi-Rayleigh wave by a cylindrical hole with $D\;=\;12\;\mathrm{mm}$ at $t_{2}\;=\; 17.00\;\mathrm{\mu s}$: a) experimental field, (b) numerical field, (c) and (d) profiles along the lines in the $x_{1}$ and $x_{2}$ directions marked in white in (a) (thick) and (b) (thin). Amplitude in units of $\lambda_{\mathrm{o}}/8\pi$. Phase in radians. Lengths in millimeters. The axis numbering corresponds to that of figure~\ref{ROI}.} % The complete sequence of 2D maps (from $t\;=\;0$ to $t\;=\;18.75\;\mathrm{\mu s}$) is available as the MPEG file \textit{sequence.mpg}.	
\label{res_tran_t2}
\end{figure*}

To illustrate the obtained results in the comparison between experimental and numerical data we use representative sequences corresponding to experimental and numerical displacement fields for the cylindrical hole with nominal diameter $D\;=\;12\;\mathrm{mm}$ (figure~\ref{res_secuencia}). A visual comparison of the sequences evidences a close agreement in the spatial distribution of the experimental and numerical values of the real displacement. This agreement is confirmed with a more careful visual comparison frame by frame as it is illustrated in figures~\ref{res_tran_t1} and \ref{res_tran_t2} where, using a common grey level scale for the images and a common vertical scale for the profiles, the amplitude and phase of the experimental and numerical displacement are presented for two representative instants in the sequences: $t_{1}\;=\;8.50\;\mathrm{\mu s}$ corresponding to the incident field before the interaction with the hole (figure~\ref{res_tran_t1}), and $t_{2}\;=\;17.00\;\mathrm{\mu s}$ corresponding to the scattered field after the interaction with the hole (figure~\ref{res_tran_t2}). From these figures, apart from local differences associated to the fluctuations in the experimental field due to presence of speckle, we can conclude that there is a very good matching of the experimental and numerical values. This good matching can be found all along the sequence both for the incident field, which confirms that the employed approach for introducing the incident field using an analytic approximation works properly, and the scattered field, including the forward, lateral and backscattering areas in the image.

\begin{figure*}
\centering
\begin{tabular}{
c@{\hspace{0.2cm}}
c@{\hspace{0.2cm}}
c
}
   \raisebox{16mm}{(a)}
  & \includegraphics 
    [width=55 mm,keepaspectratio=true]  
    {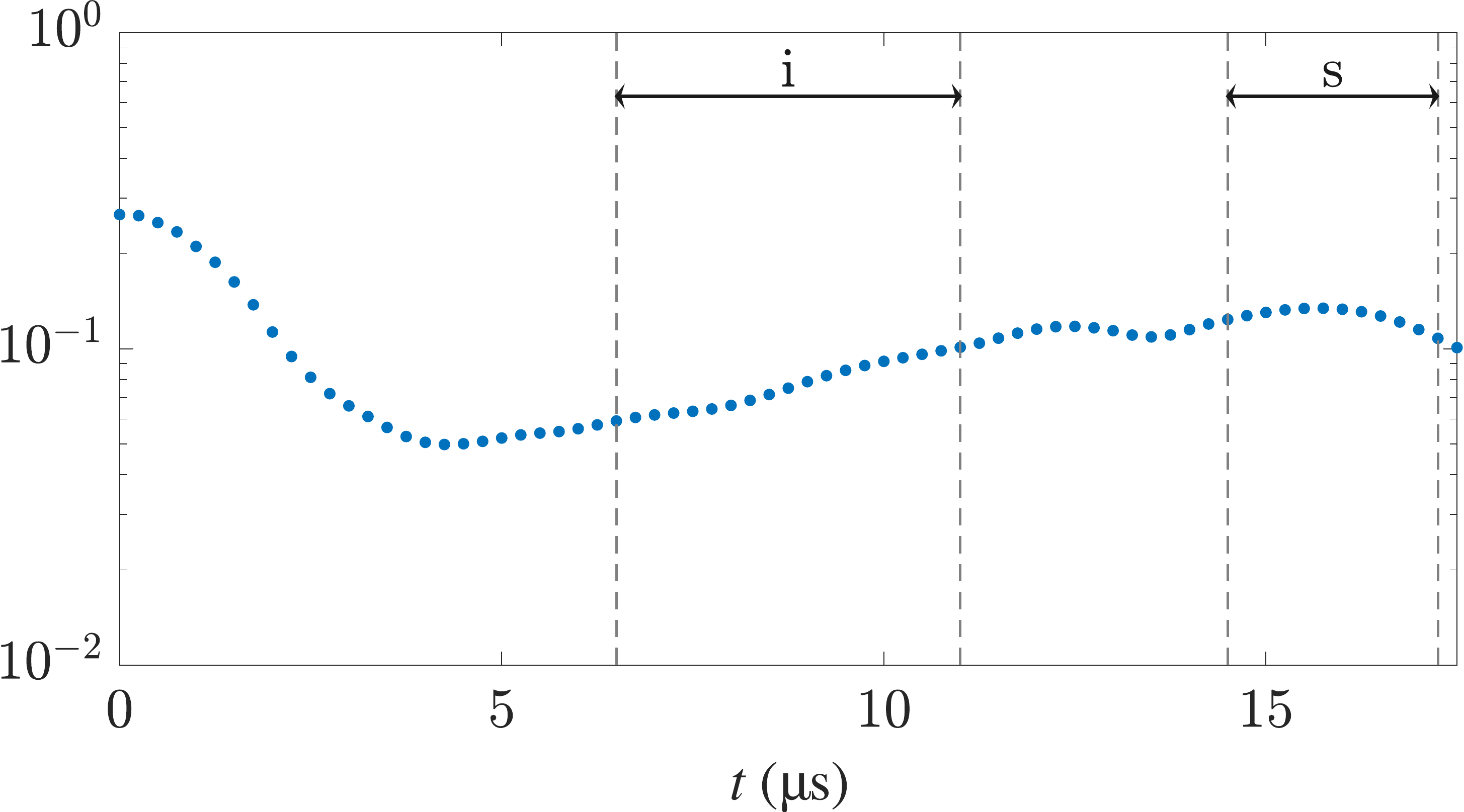}
  & \includegraphics 
    [width=55 mm,keepaspectratio=true]  
    {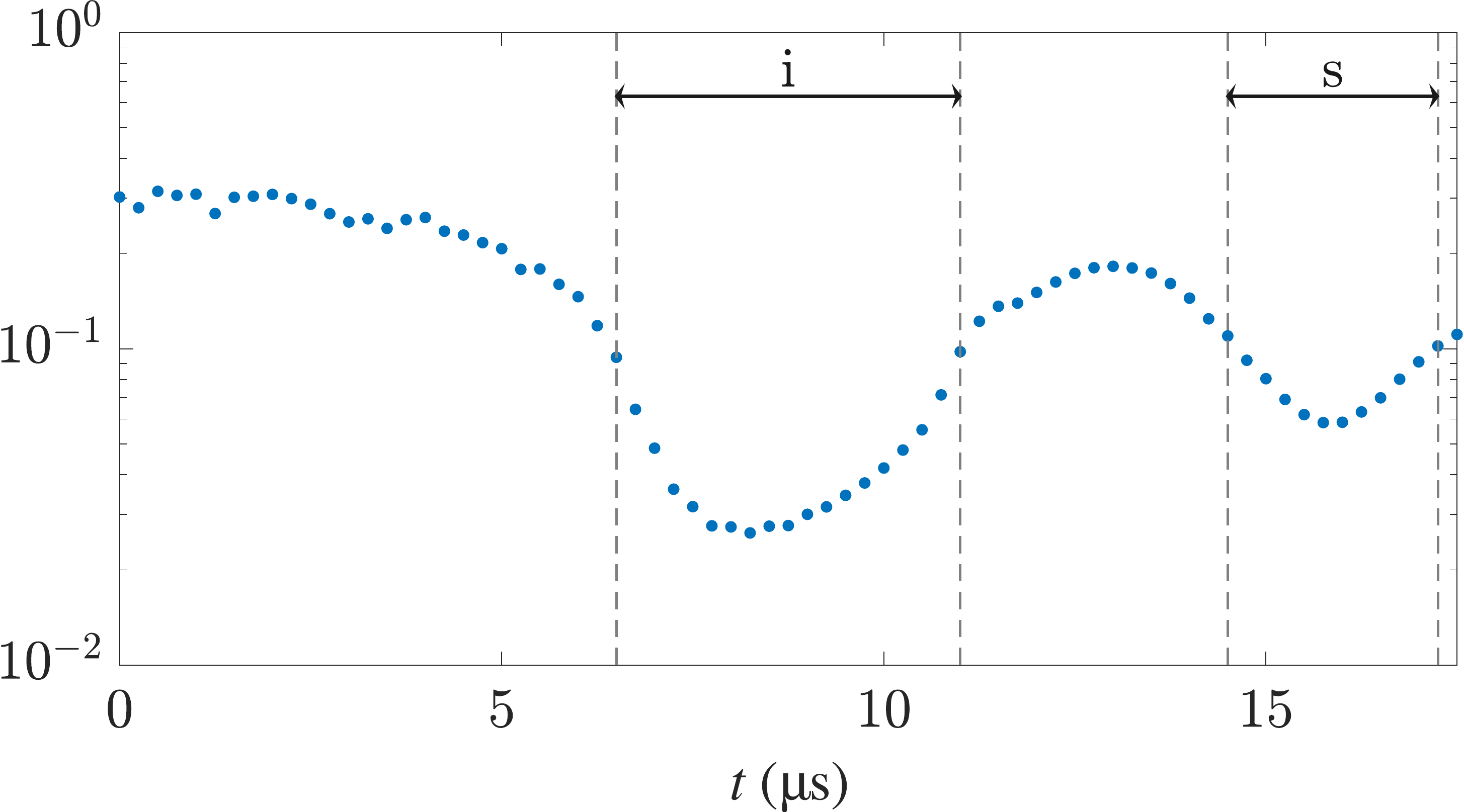}
  \\
    \raisebox{16mm}{(b)}
  & \includegraphics 
    [width=55 mm,keepaspectratio=true]  
    {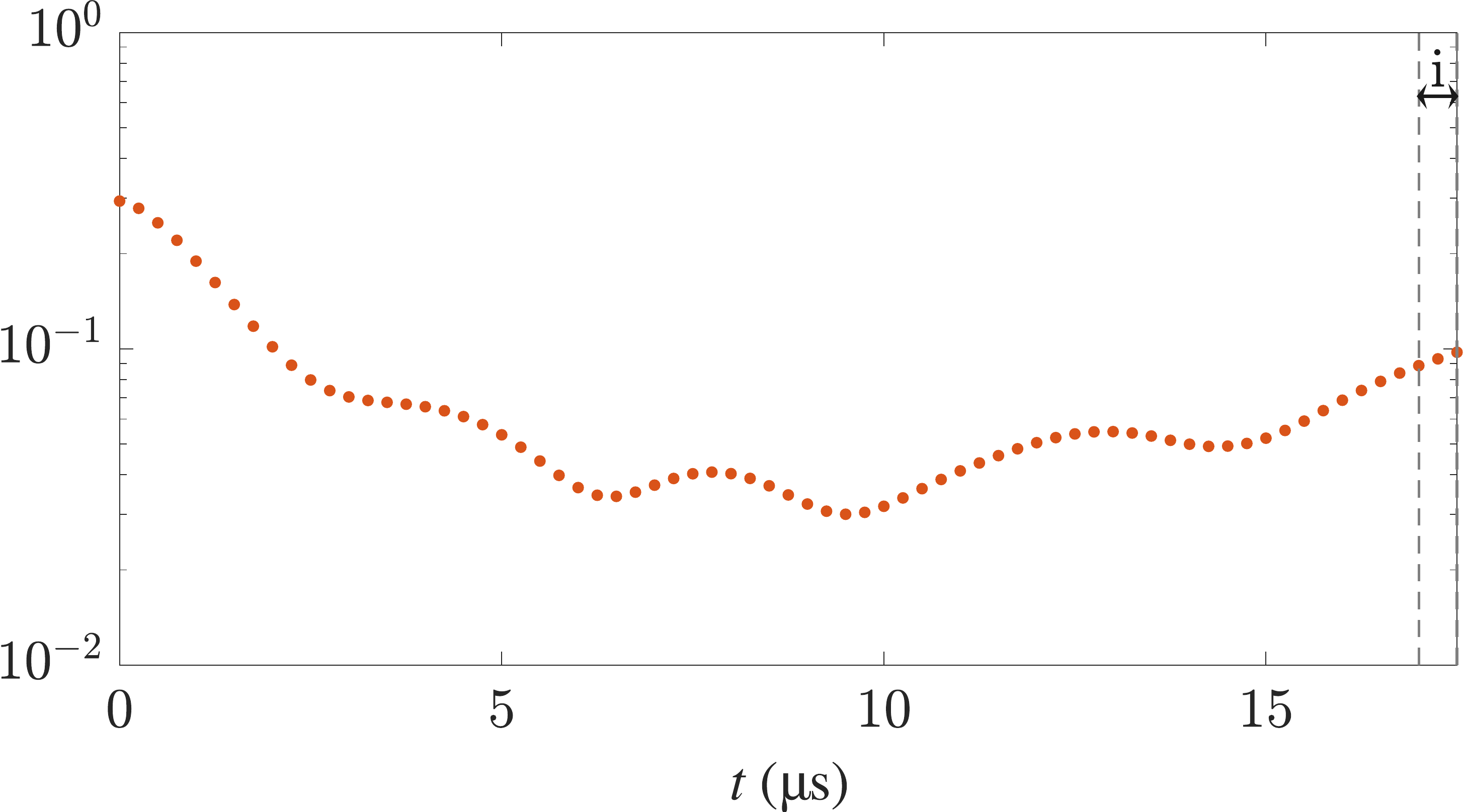}
  & \includegraphics 
   [width=55 mm,keepaspectratio=true]  
   {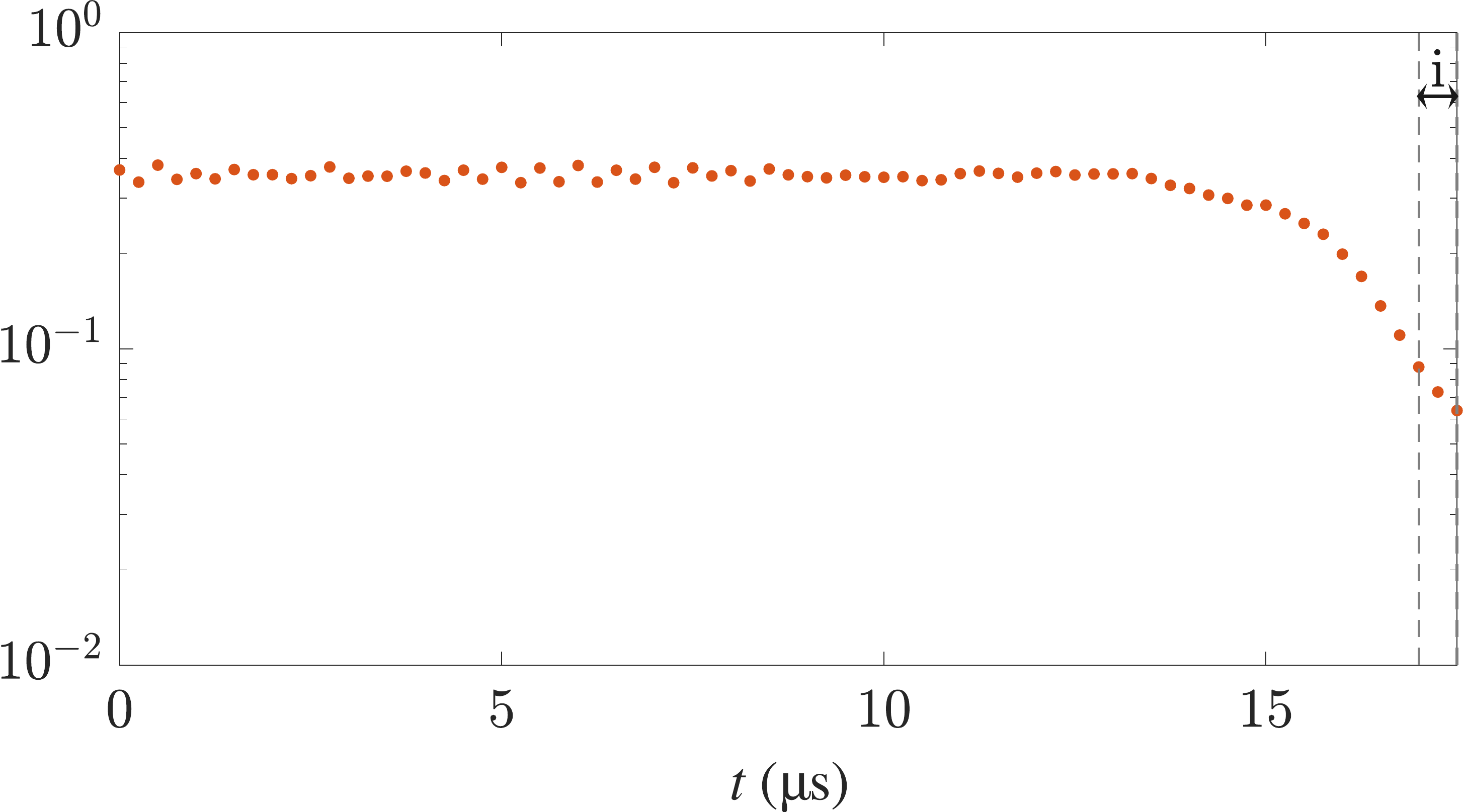}
   \\
    \raisebox{16mm}{(c)}
  & \includegraphics 
    [width=55 mm,keepaspectratio=true]  
    {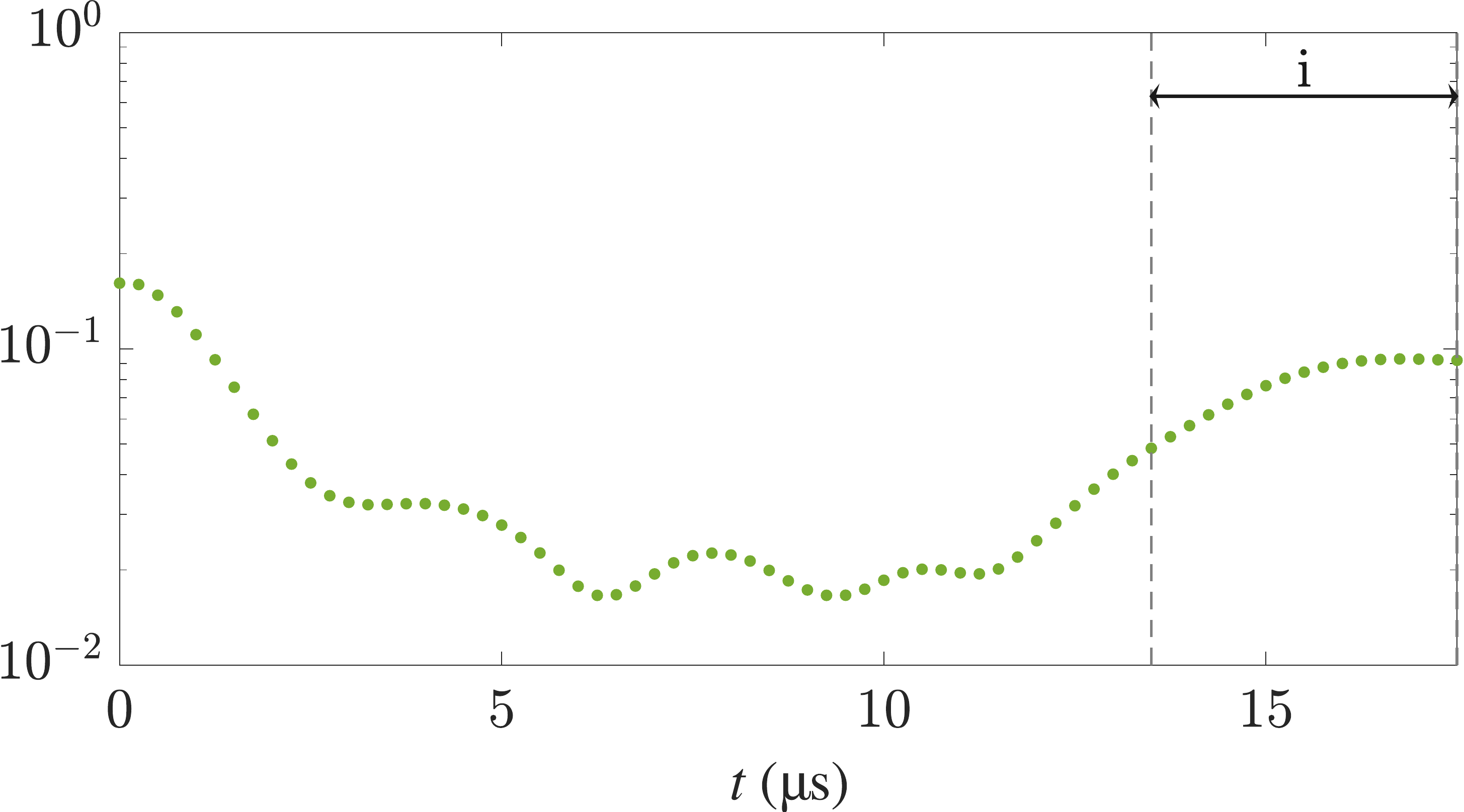}
  & \includegraphics 
    [width=55 mm,keepaspectratio=true]  
    {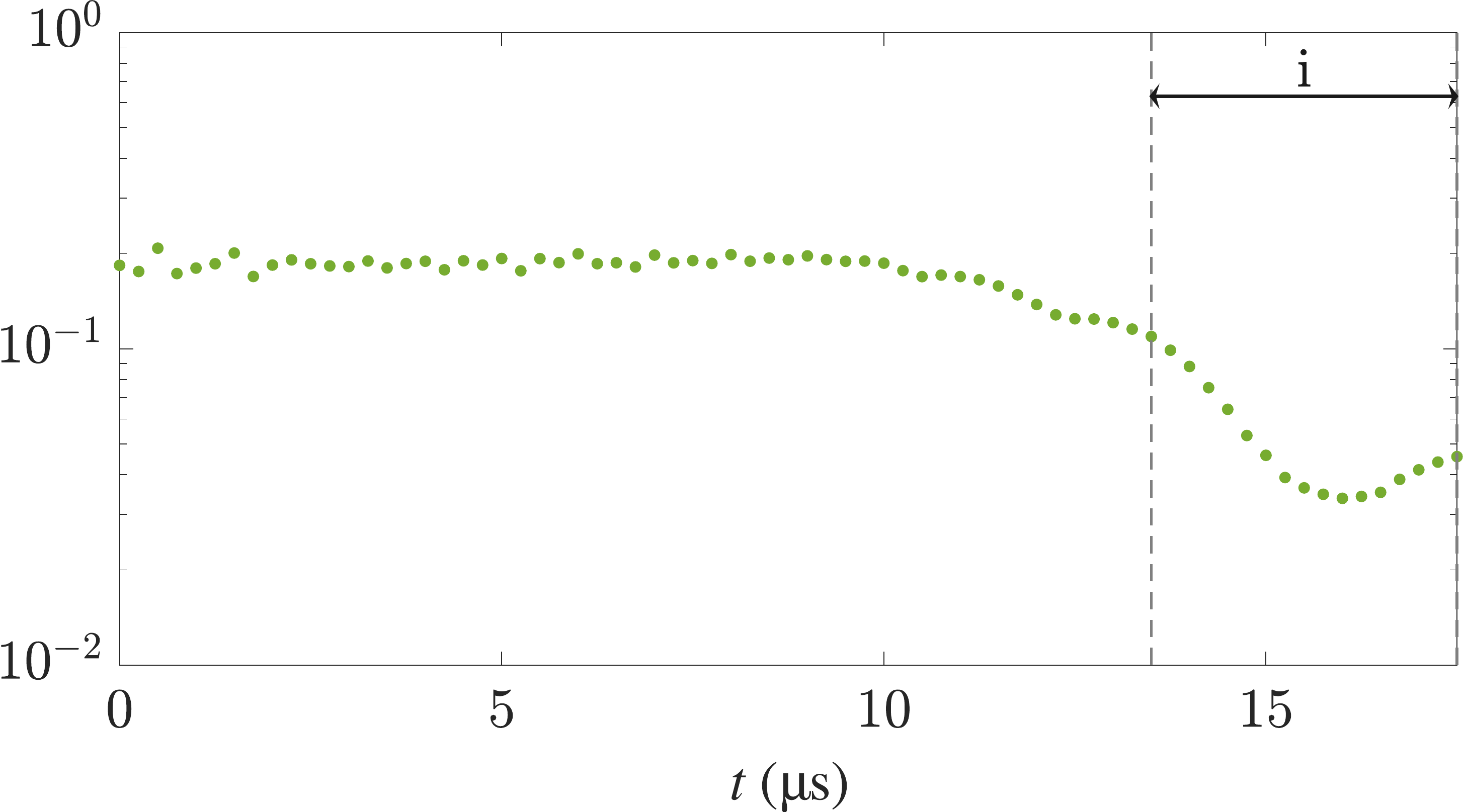}
  \\
  \raisebox{16mm}{(d)}
  & \includegraphics 
    [width=55 mm,keepaspectratio=true]  
    {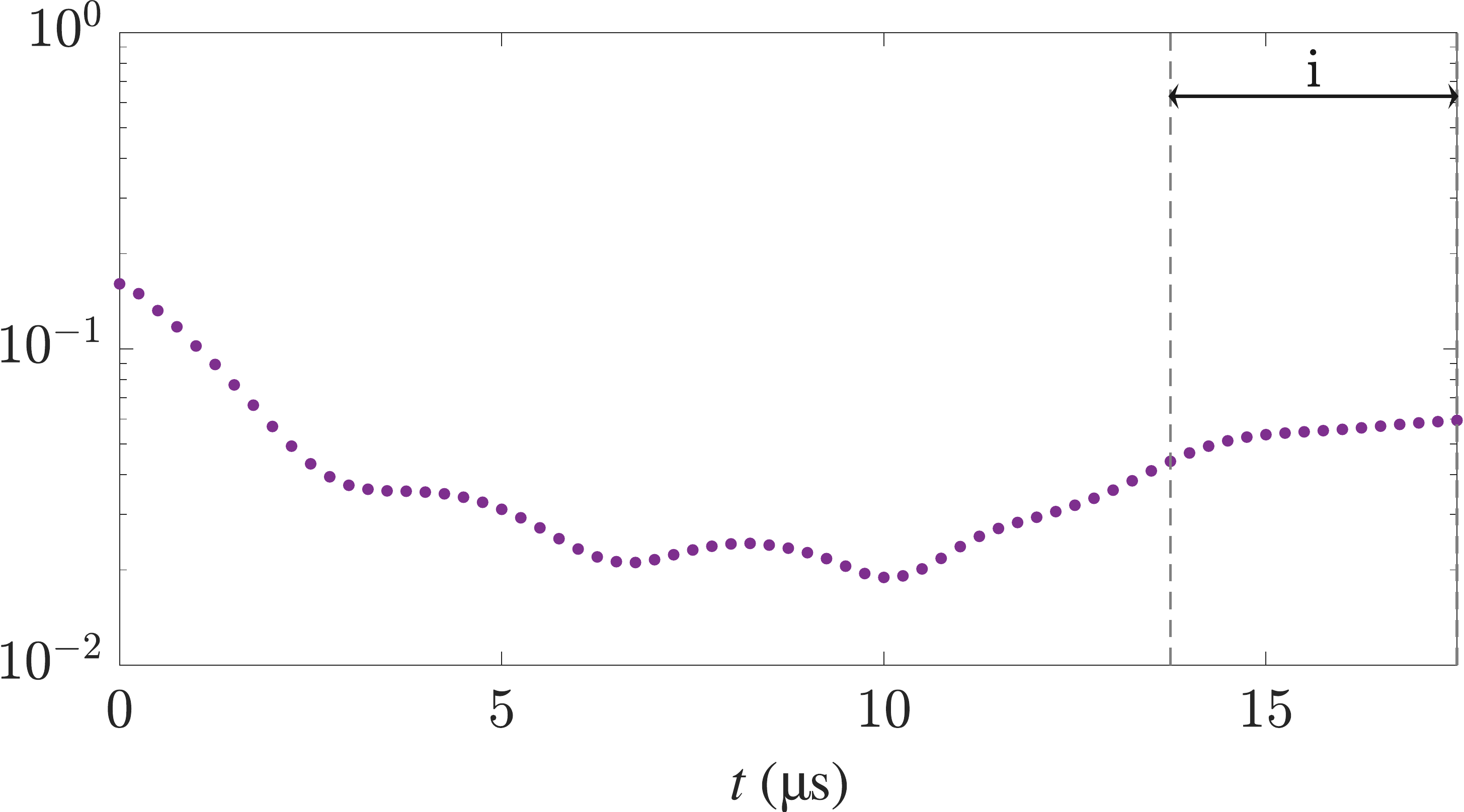}
  & \includegraphics 
    [width=55 mm,keepaspectratio=true]  
    {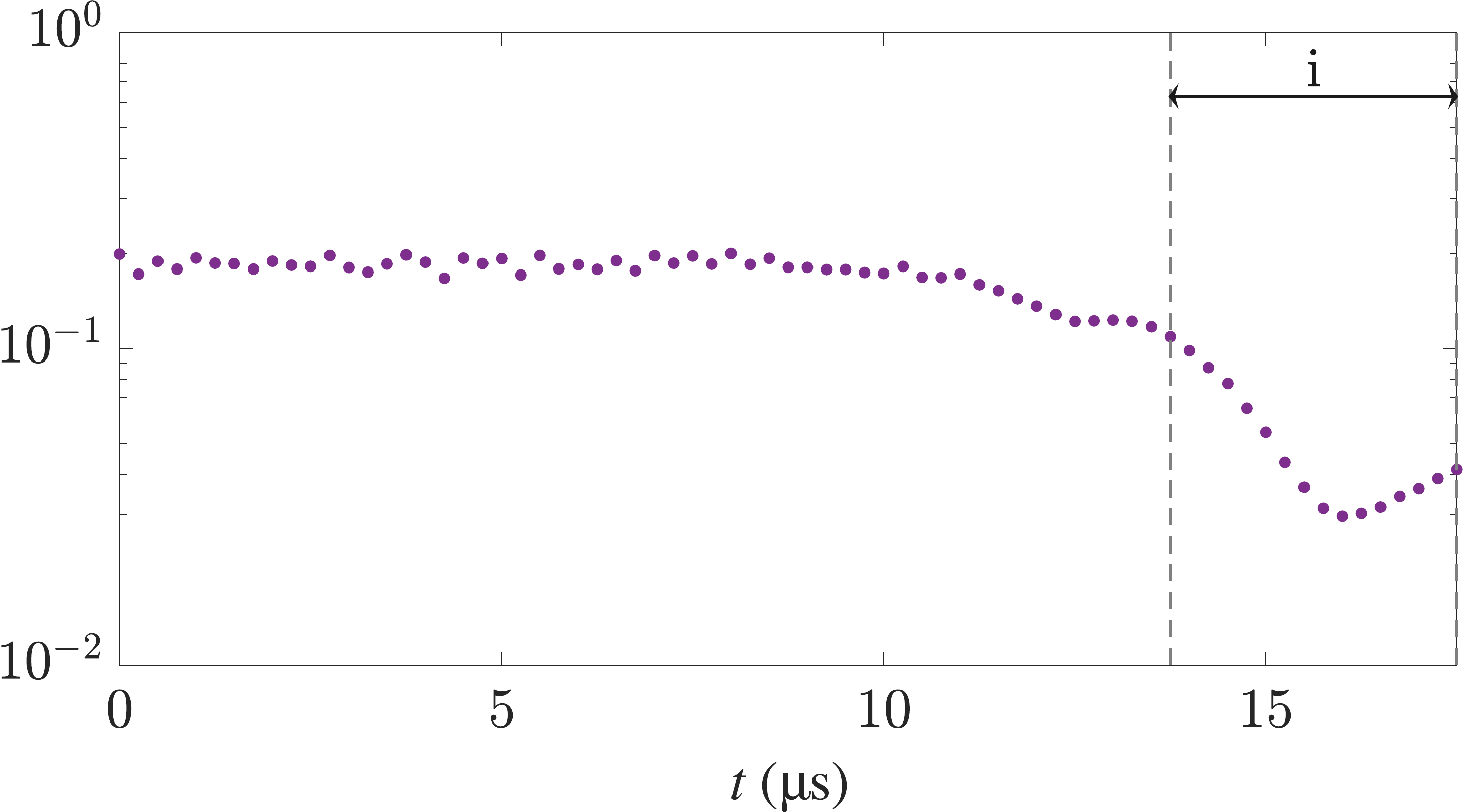}   
   \\
  &  Amplitude(C1)
  &  Phase(C2)
  \\
\end{tabular}
	\caption{Global relative total error in amplitude $\mathrm{\varepsilon}_{\mathrm{rm}}^{\mathrm{T}}\left(\mathrm{G}_{\mathrm{R}}\right)$ and phase $\mathrm{\varepsilon}_{\mathrm{rM}}^{\mathrm{T}}\left(\mathrm{G}_{\mathrm{R}}\right)$ between the problem field (numerical) and the reference field (experiment) for the complete sequence of the scattering of a transient narrow-band quasi-Rayleigh wave by cylindrical hole with $D\;=\;12\;\mathrm{mm}$ for several regions $\mathrm{G}_{\mathrm{R}}$ within the ROI corresponding to those in figure~\ref{ROI}: (a) $\mathrm{G}_{\mathrm{R}1}$, (b) $\mathrm{G}_{\mathrm{R}2}$, (c) $\mathrm{G}_{\mathrm{R}3}$ and (d) $\mathrm{G}_{\mathrm{R}4}$. The arrows identify the time intervals in which there is an incident (i) or a scattered (s) acoustic field different from zero for each region.}
\label{res_e_global_T}
\end{figure*}

This agreement between experiment and numerics found in the visual comparison has been checked over different areas in the image in a more objective quantitative way using the relative rms global errors in L2 norm in amplitude
%EQ.21************************************************************
%*****************************************************************
\begin{equation} \label{eq_e_global_T_m}
\mathrm{\varepsilon}_{\mathrm{rm}}^{\mathrm{T}}	
\left\lbrace\mathrm{R}\right\rbrace\left(t\right):=
\left[ 
      \frac{  
	       \sum_{\mathbf{x}_{h}\in \mathrm{R}}
	       \vert
           u_{3\mathrm{m}}^{\mathrm{n}}\left(\mathbf{x}_{h},t\right)	 
          -u_{3\mathrm{m}}^{\mathrm{e}}\left(\mathbf{x}_{h},t\right)                       
	      \vert^{2}
	      }
	      {
	      N_{\mathrm{R}}\;\vert u_{3\mathrm{m}}^{\mathrm{ref}}\vert^{2}    }
	      \right]^{1/2}            
\end{equation}
%**********************************************************
%*****************************************************************
and in phase
%EQ.22***********************************************************
%*****************************************************************
\begin{equation} \label{eq_e_global_T_M}
\mathrm{\varepsilon}_{\mathrm{rM}}^{\mathrm{T}}	
\left\lbrace\mathrm{R}\right\rbrace\left(t\right):=
\left[ 	
      \frac{  
	       \sum_{\mathbf{x}_{h}\in \mathrm{R}}
	       \vert
           \varphi_{3\mathrm{M}}^{\mathrm{n}}\left(\mathbf{x}_{h},t\right)            
          -\varphi_{3\mathrm{M}}^{\mathrm{e}}\left(\mathbf{x}_{h},t\right)
	      \vert^{2}
	      }
	      {
	      N_{\mathrm{R}}\;\vert \varphi_{3\mathrm{M}}^{\mathrm{ref}}\vert^{2}    }
	      \right]^{1/2}            
\end{equation}
%************************************************************
%*****************************************************************
being $N_{\mathrm{R}}$ the number of pixels in the selected region $\mathrm{R}$ and
%EQ.23************************************************************
%*****************************************************************
\begin{align} \label{eq_norm}
u_{3\mathrm{m}}^{\mathrm{ref}} & = max
\left[ 
u_{3\mathrm{m}}^{\mathrm{e}}\left(\mathbf{x}_{h},t\right)
\quad 
\forall\left(\mathbf{x}_{h},t\right)
\right] 
\nonumber \\
\varphi_{3\mathrm{M}}^{\mathrm{ref}} & = max
\left[ 
\varphi_{3\mathrm{M}}^{\mathrm{e}}\left(\mathbf{x}_{h},t\right)
\quad 
\forall\left(\mathbf{x}_{h},t\right)
\right]                                
\end{align}
%************************************************************
%*****************************************************************
the maximum values for amplitude and phase of the field over the ROI in the whole sequence. 
$\mathrm{\varepsilon}_{\mathrm{rm}}^{\mathrm{T}}$ and $\mathrm{\varepsilon}_{\mathrm{rM}}^{\mathrm{T}}$ were evaluated for all instants $t$ of the sequence in several sub-zones of interest (see figure~\ref{ROI}) and the results (figure~\ref{res_e_global_T}) show that the value of the relative rms global error is always lower than  the typical value $0.1$ of the noise-to-signal ratio, in both amplitude and phase, in the experimental PTVH system (which is mainly associated to speckle fluctuations~\cite{Tesis_PRG}). In other words, a perfect match was obtained---up to a tolerance of the order of the experimental noise. Although this quantitative comparison has been developed only in terms of the acoustic field, it is expected that the agreement between theory and experiment holds for any other relevant quantity of interest that could be derived form the field values (f.i. reflection coefficients, far-field scattering pattern, etc..). Then, the FC(Gram) elastodynamic solver can be considered as a reliable tool for the characterization of the sequences measured with our PTVH system registering the scattering of transient, high-frequency, narrow-band quasi-Rayleigh elastic waves by through-thickness holes. 

% figura 12
% figura 9 Carlos: Comparación campo incidente t1 mod y fase***********
%***************************************************************
%*****************************************************************
%*****************************************************************

% figura 13
% figura 10 Carlos: Comparación campo dispersado t2 mod y fase**********
%***************************************************************
%*****************************************************************
%*****************************************************************

% figura 14
% figura 11 Carlos: error_global_T****************************************
%***************************************************************
%*****************************************************************
%*****************************************************************

% figura 15
% figura 15 Carlos: 
% videos 3D****************************************
%***************************************************************
\begin{figure*}
\centering
   \includegraphics 
    [width=130 mm,keepaspectratio=true]  
    {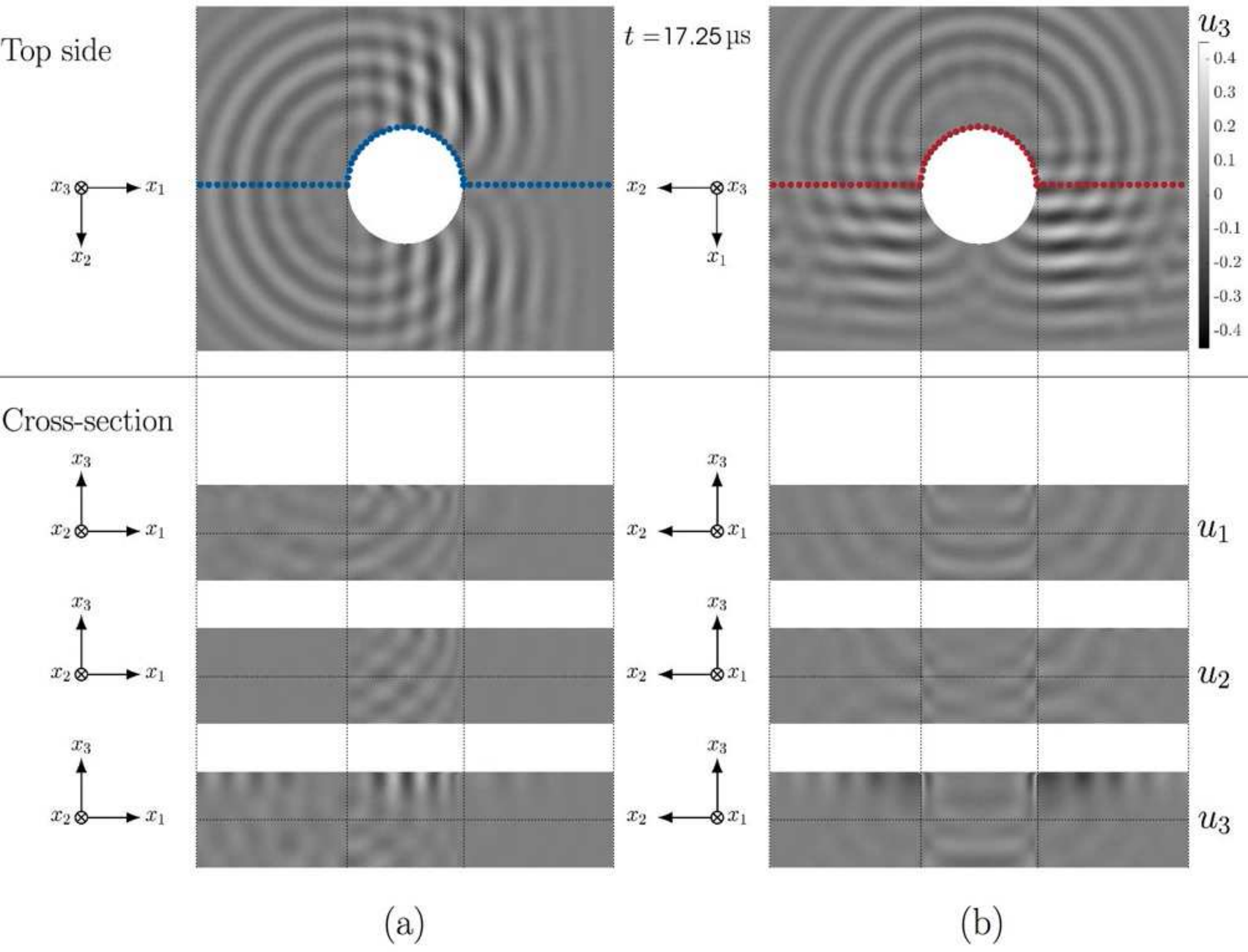}
\caption{Frame of a video sequence showing the components of the displacement vector in the scattering of a transient narrow-band quasi-Rayleigh wavetrain by a cylindrical hole with $D\;=\;12\;\mathrm{mm}$: (a) side view from the perspective of the negative $x_{2}$ axis, (b) side view from the perspective of the negative $x_{1}$ axis. The top side images show the out-of-plane component of the field in a rectangular area of $44\,\mathrm{mm} \times 36\, \mathrm{mm}$, centered at the hole within the ROI in figure~\ref{ROI}, and the height of the side view is equal to the plate thickness $2h\,=\,10$ mm. Axes
orientation corresponds to those of figures~\ref{esquema_plate} and~\ref{ROI}. Amplitude in units of $\lambda_{\mathrm{o}}/8\pi$. Phase in radians. Full sequence in Video 2, MPEG, 2.5MB [video\_2\_figure\_15.mp4].
}
\label{res_e_videos}
\end{figure*}
%*****************************************************************
%*********************************************

To illustrate the potential of the FC(Gram) solver for quantitative characterization of temporal and spatial distribution of the acoustic field ---f.i in order to optimize the experimental configuration of the PTVH system or the associated data processing and analysis strategy--- we present a full 3D sequence for the 12 mm hole that displays the temporal variation of all three components of the displacement vector in the interior volume of the plate (figure~\ref{res_e_videos}). In the backscattering area, the incident wave presents the characteristic spatial distribution of the amplitude of a quasi-Rayeligh wave, with significant values only in a layer of the order of one acoustic wavelength close to the top side of the plate---as clearly visible in the initial frames of the side view of the video corresponding to figure~\ref{res_e_videos}.a left. The scattered field, on the other hand, presents well-known features of scattering of elastic waves by defects~\cite{ews133}. Indeed, the reflected wave in the backscattering area is quasi-Rayleigh and it produces a standing wave pattern by interference with the incident field, as can be seen in the late frames of the aforementioned video. Further, the usual modal conversion to SH0 mode~\cite{Dilligent} is evidenced by the large $u_2$ displacement that appears in a vicinity of the hole in figure~\ref{res_e_videos} and in the time-interval $11.00-19.00\; \mu \mathrm{s}$ in the associated video. The video also displays a wave perturbation that travels downward along the lateral side of the hole, which is subsequently scattered at the bottom, generating an upward perturbation which interferes with the downward perturbation and thus forms a standing wave pattern~\cite{ews156} in the lateral side of the hole (see side view of figure~\ref{res_e_videos}.b, center). At the same time, a scattered wave is generated which propagates close to the bottom side of the plate---which can be viewed in the final frames of the side view portions of the video for the $u_1$ and $u_3$ components and in figure~\ref{res_e_videos}.a left.

\section{Conclusions}
\label{section5}
The scattering of transient narrow-band quasi-Rayleigh waves by through-thickness holes in aluminum plates in the high-frequency regime has been studied. Experimental sequences of scattering patterns measured with a self-developed full-field PTVH system have been compared with the corresponding simulated sequences derived by a state-of-the-art FC(Gram) elastodynamic solver. In addition to a broad agreement in the spatial distribution of the incident and scattered fields for every time instant, which is apparent by visual inspection, a detailed quantitative comparison between experiments and numerics has been presented. Consideration of the relative global errors observed show that a perfect match was obtained, both in amplitude and phase, up to a tolerance of the order of the experimental noise. It is suggested that the combined computation-and-measurement technique bears a significant potential for ultrasonic NDT applications in plate-like structures. Ongoing work focuses on generalization of these techniques to generally-shaped plate-like structures containing complex defects, such as slots, partly through-thickness holes, cracks, etc.

\section{Acknowledgments}
OB gratefully acknowledges support by NSF and AFOSR and DARPA through contracts DMS-1411876, DMS-1714169, FA9550-15-1-0043 and \linebreak HR00111720035, and the NSSEFF Vannevar Bush Fellowship under contract number N00014-16-1-2808. The authors from the University of Vigo gratefully acknowledge support by the Spanish \emph{Ministerio de Econom\'ia y Competitividad} and the European Comission (ERDF) in the context of the \emph{Plan Nacional de I+D+i} (project number DPI2011-26163), by the \emph{Subdirecci\'on Xeral de Promoci\'on Cient\'ifica e Tecnol\'oxica universitaria da Xunta de Galicia} in the context of the \emph{Plan Galego de I+D+i} (project number 10PXIB303167PR) and supplementary co-funding from the \emph{Universidade de Vigo} (contract number 09VIA07). 

%\listoffigures
%\newpage
%\appendix{\textbf{LIST OF FIGURE CAPTIONS}}
%\input{Draft_arXiv_2.lof}
% corta y pega del archivo Draft_arXiv_2.lof


%APPENDICES**********************************************************
%*******************************************************************
\appendix
\section{Real and complex description of wave propagation}
\label{appA}

For any real displacement field $\mathbf{u}\left(\mathbf{x},t\right)$  described by equations (\ref{eq_u}-\ref{eq_ui}) an associated complex field 
%Eq. A1*******************************************************************
%*******************************************************************
\begin{equation}\label{eq_hatu}
\mathbf{\hat{u}}\left(\mathbf{x},t\right) =
\sum_{i=1}^{3}\hat{u}_{i}\left(\mathbf{x},t\right)\mathbf{a}_{i}
\end{equation}
%*******************************************************************
%*******************************************************************
with complex components 
%EQ.A2*************************************************************
%*******************************************************************
\begin{multline}\label{eq_hatui}
\hat{u}_{i}\left(\mathbf{x},t\right)=
\hat{u}_{i\mathrm{m}}
           \left(\mathbf{x},t\right)
           \exp\left(-\mathrm{j}2\pi ft\right)
           \\ =
u_{i\mathrm{m}}
           \left(\mathbf{x},t\right)
           \exp\left\lbrace \mathrm{j}\left[ \varphi_{i\mathrm{m}}
           \left(\mathbf{x},t\right)-2\pi ft \right] \right\rbrace, \quad
            i=1,2,3         
\end{multline}
%*******************************************************************
%*******************************************************************
can be introduced in such a way that 
$\mathbf{u}\left(\mathbf{x},t\right)=
\mathrm{Re}\left[\mathbf{\hat{u}}\left(\mathbf{x},t\right)\right]$ 
and 
$u_{i}\left(\mathbf{x},t\right)=
\mathrm{Re}\left[\hat{u}_{i}\left(\mathbf{x},t\right)\right]$, where $\mathrm{Re}\left[\hat{a}\right]$  stands for the real part of complex number $\hat{a}$ and $\mathrm{j}$ is the imaginary unit. The complex amplitude $\hat{u}_{i\mathrm{m}}\left(\mathbf{x},t\right)$ of each complex component
$\hat{u}_{i}\left(\mathbf{x},t\right)$
 is unambiguously determined by the acoustic amplitude $u_{i\mathrm{m}}
           \left(\mathbf{x},t\right)$ and the acoustic phase $\varphi_{i\mathrm{m}}
           \left(\mathbf{x},t\right)$ of the corresponding real component 
$u_{i}\left(\mathbf{x},t\right)$           
and viceversa~\cite{Born_Wolf}. Hence, there is a one-to-one correspondence between real and complex components and, as a consequence, between the real displacement field $\mathbf{u}\left(\mathbf{x},t\right)$ and the complex displacement field $\mathbf{\hat{u}}\left(\mathbf{x},t\right)$, which can  be also written as 
$\mathbf{\hat{u}}\left(\mathbf{x},t\right)=
\mathbf{\hat{u}}_{\mathrm{m}}
           \left(\mathbf{x},t\right)
           \exp\left(-\mathrm{j}2\pi ft\right)$ where
%EQ.A3***************************************************
%*******************************************************************
\begin{equation}\label{eq_hatum}
\mathbf{\hat{u}}_{\mathrm{m}}\left(\mathbf{x},t\right) =
           \sum_{i=1}^{3}\hat{u}_{i\mathrm{m}}\left(\mathbf{x},t\right)
           \mathbf{a}_{i}           
\end{equation}
%*******************************************************************
%*******************************************************************
is its associated complex amplitude. 

Once the complex displacement is well-established, taking as a reference equations~(\ref{eq_epsilon_ij}-\ref{eq_ley_Hook_ij}) we can introduce the components of the complex strain tensor $\boldsymbol{\hat{\epsilon}}$ 
as
%EQ. A4***********************************************************
%*******************************************************************
\begin{equation} \label{eq_hatepsilon_ij}
  \hat{\epsilon}_{ij}\left(\mathbf{x},t\right)= \frac{1}{2}\left[
                                  \frac{\partial{}\hat{u}_{i}\left(\mathbf{x},t\right)}{\partial{}x_{j}}
                                 +\frac{\partial{}\hat{u}_{j}\left(\mathbf{x},t\right)}{\partial{}x_{i}}
                            \right], \quad i,j=1,2,3
\end{equation}
%*******************************************************************
%*******************************************************************
and the complex components of the complex stress tensor $\boldsymbol{\hat{\tau}}$ in the linear elastic regime as
%EQ. A5**********************************************************
%*******************************************************************
\begin{equation} \label{eq_ley_Hook_ij_c}
	\hat{\tau}_{ij}\left(\mathbf{x},t\right)=\lambda_{\mathrm{e}}\delta_{ij}
	          \sum_{k}\hat{\epsilon}_{kk}\left(\mathbf{x},t\right)+2\mu_{\mathrm{e}}\hat{\epsilon}_{ij}\left(\mathbf{x},t\right)
	          , \quad i,j=1,2,3
\end{equation} 
%*******************************************************************
%***************************************************************
From the linearity of equations~(\ref{eq_epsilon_ij}-\ref{eq_ley_Hook_ij}) 
and (\ref{eq_hatepsilon_ij}-\ref{eq_ley_Hook_ij_c}) we have that $\epsilon_{ij}\left(\mathbf{x},t\right)=
\mathrm{Re}\left[\hat{\epsilon}_{ij}\left(\mathbf{x},t\right)\right]$
and $\tau_{ij}\left(\mathbf{x},t\right)=
\mathrm{Re}\left[\hat{\tau}_{ij}\left(\mathbf{x},t\right)\right]$ and from the linearity of the Lam\'e-Navier equations~(\ref{eq_Lame-Navier_ij}) we can conclude that the dynamic of the complex displacement field is governed by the complex version of the Lam\'e-Navier equations
%EQ. A6**********************************************************
%*******************************************************************
\begin{multline} \label{eq_Lame-Navier_ij_c}
\frac{\left(\lambda_{\mathrm{e}}+\mu_{\mathrm{e}}\right)}{\rho}
    \sum_{j=1}^{3}\frac{\partial^{2}\hat{u}_{i}\left(\mathbf{x},t\right)}
    					{\partial x_{i}\partial x_{j}}
	+\frac{\mu_{\mathrm{e}}}{\rho}
	\sum_{j=1}^{3}\frac{\partial^{2} \hat{u}_{i}\left(\mathbf{x},t\right)}
						{\partial x_{j}^{2}}
\\	
	=\frac{\partial^{2}\hat{u}_{i}\left(\mathbf{x},t\right)}{\partial t^{2}}, 
	\quad i=1,2,3 \quad \mathrm{in}\quad \Omega
\end{multline} 
%*******************************************************************
%*******************************************************************
Hence, real and complex fields (with components that have the same acoustic amplitude and the same acoustic phase) provide a completely equivalent description of wave propagation with formally identical  real and complex versions of the equations and in such a way that any real field is equal to the real part of the corresponding complex field.

\section{Quasi-Rayleigh waves in plates}\label{appB}

Rayleigh waves have significant displacement amplitude only in the vicinity of the plane stress-free surface limiting a homogeneous and isotropic linear elastic semi-infinity solid. Similarly, quasi-Rayleigh waves are a singular case of guided waves in homogeneous and isotropic linear elastic plates, with significant displacement amplitudes only in the vicinity of one of the two plane stress-free surfaces limiting the plate. As quasi-Rayleigh waves results from a particular superposition of Lamb waves, using the same theoretical framework employed to describe the direct scattering problem in section \ref{section2} with the Cartesian reference frame of figure 1, we first consider the case of a plane Lamb wave traveling it the $x_{1}$ direction in harmonic regime for a single temporal frequency $f$. In this conditions it can be shown~\cite{Graff} that wave propagation can be analyzed using plane strain regime ($u_{2}=0$ and $\partial{} /  \partial{x_{2}}=0$) and  that the in-plane  $\left(\parallel \right)$ and out-of-plane $\left( \perp\right) $ components of the displacement are superposition of symmetric and anti-symmetric Lamb modes (i.e. perturbations with displacements that are, respectively, symmetric and anti-symmetric with respect to the mean plane of the plate $x_{3}=0$) with associated complex amplitudes that can be written for the symmetric case as
%EQ.B1**********************************************************
%*******************************************************************
\begin{subequations}\label{eq_u1y3m_Ls}
	\begin{align} 
	\hat{u}_{\parallel f{\mathrm{m}}}^{\mathrm{S}}
	\left(x_{3}\right)
	 = & 
	 \mathrm{j}
	 \left( k_{1}
	  B\cos{k}_{\mathrm{L3}}x_{3}
	 - {k}_{\mathrm{T3}}
	   C\cos {k}_{\mathrm{T3}}x_{3}
      \right)
\\
	\hat{u}_{\perp f{\mathrm{m}}}^{\mathrm{S}}
	\left(x_{3}\right)
	= & 
	 -{k}_{\mathrm{L3}}
	 B\sin{k}_{\mathrm{L3}}x_{3}
	 -k_{1}
	   C\sin{k}_{\mathrm{T3}}x_{3}
\end{align} 
\end{subequations}
%*******************************************************************
%*******************************************************************
and for anti-symmetric case as
%EQ. B2**********************************************************
%*******************************************************************
\begin{subequations}\label{eq_u1y3m_La}
	\begin{align} 
	\hat{u}_{\parallel f{\mathrm{m}}}^{\mathrm{A}}
	\left(x_{3}\right)
	= &
	 \mathrm{j}
	 \left(k_{1}
	   A\sin{k}_{\mathrm{L3}}x_{3}
	 +{k}_{\mathrm{T3}}
	 D\sin {k}_{\mathrm{T3}}x_{3}
	\right) 
\\
	\hat{u}_{\perp f{\mathrm{m}}}^{\mathrm{A}}
	\left(x_{3}\right)
	= &
	 {k}_{\mathrm{L3}}
	   A\cos{k}_{\mathrm{L3}}x_{3}
	 -k_{1}
	 D\cos{k}_{\mathrm{T3}}x_{3}
\end{align} 
\end{subequations}
%*******************************************************************
%*******************************************************************
being
%EQ. B3**********************************************************
%*******************************************************************
\begin{subequations}\label{eq_KL3KT3}
	\begin{align} 
k_{\mathrm{L3}}^{2}=
\frac{\omega^{2}}{c_{\mathrm{L}}^{2}}-k_{1}^{2}
\\
k_{\mathrm{T3}}^{2}=
\frac{\omega^{2}}{c_{\mathrm{T}}^{2}}-k_{1}^{2}
\end{align} 
\end{subequations}
%*******************************************************************
%***************************************
Each Lamb mode results from the superposition of longitudinal and transversal perturbations with a common  effective axial wavenumber $k_{1}$ and stationary spatial distribution of the displacement along the $x_{3}$ direction characterized by
$k_{\mathrm{L3}}$ and $k_{\mathrm{T3}}$, respectively. The complex fields for the in-plane and out-of-plane components of each mode can be obtained multiplying the corresponding complex amplitudes, given by expression (\ref{eq_u1y3m_Ls}-\ref{eq_u1y3m_La}), by the phase factor $\exp \left[ \mathrm{j}\left(k_{1}x_{1}-2 \pi ft \right)\right]$ (although sub-indexes $1$ and $3$ would be the natural notation for the quantities associated to the two non-zero components of the displacement in this case, sub-indexes $\parallel$ and $\perp$ are more adequate to consider the case of quasi-Rayleigh waves with cylindrical symmetry later on). 

Applying the stress-free boundary condition at the surface of the plate for the symmetric case (due to the symmetry it is enough to impose the boundary condition at $x_{3}=h$) an homogeneous system of equation for $B$ and $C$ is obtained that has a non trivial solution verifying
%EQ. B4**********************************************************
%*******************************************************************
\begin{equation} \label{eq_ratioBC_Ls}
	\frac{B}{C}=
	\frac
	{
	2k_{1}k_{\mathrm{T3}}
	\cos{k}_{\mathrm{T3}}h
	}
	{
	\left(k_{1}^{2}-{k}_{\mathrm{T3}}^{2}\right)
	\cos{k}_{\mathrm{L3}}h
	}=
	-\frac
	{
	\left(k_{1}^{2}-{k}_{\mathrm{T3}}^{2}\right)
	\sin{k}_{\mathrm{T3}}h
	}
	{
	2k_{1}k_{\mathrm{L3}}
	\sin{k}_{\mathrm{L3}}h
	}
\end{equation} 
%*******************************************************************
%*******************************************************************
if and only if the determinant of the coefficients is zero, which gives the dispersion relation for symmetric Lamb modes
%EQ.B5**********************************************************
%*******************************************************************
\begin{equation} \label{eq_disper_Ls}
    \left(k_{1}^{2}-{{k}_{\mathrm{T3}}}^{2}\right)^{2}
	\cos{k}_{\mathrm{L3}}h
	\sin{k}_{\mathrm{T3}}h
	+
	4k_{\mathrm{L3}}k_{\mathrm{T3}}k_{1}^{2}
	\sin{k}_{\mathrm{L3}}h
	\cos{k}_{\mathrm{T3}}h
	=0
\end{equation} 
%*******************************************************************
%*******************************************************
Following the same scheme for the anti-symmetric case we obtain that the non trivial solution verifies
%EQ. B6**********************************************************
%*******************************************************************
\begin{equation} \label{eq_ratioAD_La}
	\frac{A}{D}=
	-\frac
	{
	2k_{1}k_{\mathrm{T3}}
	\sin{k}_{\mathrm{T3}}h
	}
	{
	\left(k_{1}^{2}-{k}_{\mathrm{T3}}^{2}\right)
	\sin{k}_{\mathrm{L3}}h
	}=
	\frac
	{
	\left(k_{1}^{2}-{k}_{\mathrm{T3}}^{2}\right)
	\cos{k}_{\mathrm{T3}}h
	}
	{
	2k_{1}k_{\mathrm{L3}}
	\cos{k}_{\mathrm{L3}}h
	}
\end{equation} 
%*******************************************************************
%*******************************************************************
provided that the determinant of the coefficients is zero, which gives the dispersion relation for anti-symmetric Lamb modes
%EQ. B7**********************************************************
%*******************************************************************
\begin{equation} \label{eq_disper_La}
    \left(k_{1}^{2}-{{k}_{\mathrm{T3}}}^{2}\right)^{2}
	\sin{k}_{\mathrm{L3}}h
	\cos{k}_{\mathrm{T3}}h
	+
	4k_{\mathrm{L3}}k_{\mathrm{T3}}k_{1}^{2}
	\cos{k}_{\mathrm{L3}}h
	\sin{k}_{\mathrm{T3}}h
	=0
\end{equation} 
%*******************************************************************
%*******************************************************
For each temporal frequency $f$, an integer finite number of propagating Lamb modes and an infinite number of non-propagating Lamb modes could be excited in the plate at the same time (figure  \ref{espectro_Lamb}). Each propagating Lamb mode (symmetric or anti-symmetric) has a characteristic axial real wavenumber $k_{1}$ and travels along direction $x_{1}$ (parallel to the center plane of the plate) with a well-defined phase velocity, maintaining its transversal stationary displacement distribution within the plate section during propagation in areas without defects~\cite{Achenbach_R}. A non-propagating Lamb mode shares formally the same properties than a propagating one, but has complex axial wavenumber and phase velocity and its amplitude decays to negligible values at distances of the order of a few acoustic wavelengths.

A quasi-Rayleigh plane harmonic guided wave is just the superposition of the plane harmonic waves of a symmetric S0 Lamb mode and an anti-symmetric A0 Lamb mode with the same temporal frequency $f$ within the so-called quasi-Rayleigh range and with the same value of their amplitudes at the top surface of the plate. The S0 and A0 modes are, respectively, the symmetric and anti-symmetric propagating Lamb modes with the lowest order. They have no cut-off and in the quasi-Rayleigh region of the spectrum at high frequencies (see qR region in figure  \ref{espectro_Lamb}) become non-dispersive  and have practically the same propagation wavenumber $k_{1}$ and a common phase velocity, which is the Rayleigh phase velocity of the plate $c_{\mathrm{R}}$.
Hence, the complex amplitude of the components of the displacement of the plane harmonic quasi-Rayleigh wave can be written as
%EQ. B8*********************************************************
%*******************************************************************
\begin{subequations}\label{eq_u1y3m_qR}
	\begin{align} 
   \hat{u}_{\parallel f\mathrm{m}}
   \left(x_{3}\right)
   = 
   \left[ 
   \hat{u}_{\parallel f\mathrm{m}}^{\mathrm{S0}}
   \left(x_{3}\right)
   + 
   \hat{u}_{\parallel f\mathrm{m}}^{\mathrm{A0}}
   \left(x_{3}\right)
   \right] 
   \\
   \hat{u}_{\perp f\mathrm{m}}
   \left(x_{3}\right)
   = 
   \left[
   \hat{u}_{\perp f\mathrm{m}}^{\mathrm{S0}}
   \left(x_{3}\right)
   + 
   \hat{u}_{\perp f\mathrm{m}}^{\mathrm{A0}}
   \left(x_{3}\right)
   \right]
\end{align} 
\end{subequations}
%*******************************************************************
%*******************************************************************
verifying the condition
%EQ. B9*********************************************************
%*******************************************************************
\begin{subequations}\label{eq_u1y3m_top}
\begin{align} 
   \frac{
   \hat{u}_{\parallel f\mathrm{m}}
   \left(h\right)
   }
   {2}
   = 
   \hat{u}_{\parallel f\mathrm{m}}^{\mathrm{S0}}
   \left(h\right)
  =
   \hat{u}_{\parallel f\mathrm{m}}^{\mathrm{A0}}
   \left(h\right)
\\
   \frac{
   \hat{u}_{\perp f\mathrm{m}}
   \left(h\right)
	}
	{2}
   = 
   \hat{u}_{\perp f\mathrm{m}}^{\mathrm{S0}}
   \left(h\right)
   =
   \hat{u}_{\perp f\mathrm{m}}^{\mathrm{A0}}
   \left(h\right)
\end{align} 
\end{subequations}
%*******************************************************************
%*******************************************************************
where $\hat{u}_{\parallel f\mathrm{m}}^{\mathrm{S0}}\left(x_{3}\right)$ and $\hat{u}_{\perp f\mathrm{m}}^{\mathrm{S0}}\left(x_{3}\right)$ are given by expression (\ref{eq_u1y3m_Ls}) and
 $\hat{u}_{\parallel f\mathrm{m}}^{\mathrm{A0}}\left(x_{3}\right)$ and $\hat{u}_{\perp f\mathrm{m}}^{\mathrm{A0}}\left(x_{3}\right)$ by  expression (\ref{eq_u1y3m_La}) 
in the particular case that modes S0 and A0 have the same value of $k_{1}$ and so, share the same values of $k_{\mathrm{L3}}$ and $k_{\mathrm{T3}}$, given by equation (\ref{eq_KL3KT3}) in terms of $f$, $k_{1}$ and the phase velocities of the plate $c_{\mathrm{L}}$ and $c_{\mathrm{T}}$. 
Inserting (\ref{eq_ratioBC_Ls}) and (\ref{eq_ratioAD_La}) with these common values ($k_{\mathrm{L3}}$, $k_{\mathrm{T3}}$ and $k_{1}$) in expressions (\ref{eq_u1y3m_Ls}) and   (\ref{eq_u1y3m_La}), respectively,  and using (\ref{eq_u1y3m_top}) we can obtain  the normalized values of the complex amplitude of the in-plane and out-of-plane components of the associated displacement for the S0 and A0 modes with respect to the value of the complex amplitude of out-of-plane component of the quasi-Rayleigh harmonic plane wave at the top surface of the plate as
%EQ. B10**********************************************************
%*******************************************************************
\begin{subequations}\label{eq_u1y3m_norm_qR}
	\begin{align} 
\underline{\hat{u}}_{\parallel f\mathrm{m}}^{\mathrm{S0}}\left(x_{3}\right)	
	& = 
	\frac
	{\hat{u}_{\parallel f\mathrm{m}}^{\mathrm{S0}}\left(x_{3}\right)}
	{\hat{u}_{\perp f\mathrm{m}}\left(h\right)}
    \nonumber \\	
	& =
	\mathrm{j}
	\left[
    \left(k_{1}^{2}-{{k}_{\mathrm{T3}}}^{2}\right)
	\cos{k}_{\mathrm{L3}}x_{3}
	\sin{k}_{\mathrm{T3}}h
    \right.  
	\nonumber \\	
	& + 
	\left. 
	2k_{\mathrm{L3}}k_{\mathrm{T3}}
	\sin{k}_{\mathrm{L3}}h
	\cos{k}_{\mathrm{T3}}x_{3}
	\right]
	\nonumber \\	
	& \times 
	\left(
    2\chi_{\mathrm{RT}}^{2} k_{\mathrm{L3}}k_{1}
	\sin{k}_{\mathrm{L3}}h
	\sin{k}_{\mathrm{T3}}h
	\right)^{-1}
\\
	\underline{\hat{u}}_{\perp f\mathrm{m}}^{\mathrm{S0}}\left(x_{3}\right)	
	& =
	\frac
	{\hat{u}_{\perp f\mathrm{m}}^{\mathrm{S0}}\left(x_{3}\right)}
	{\hat{u}_{\perp f\mathrm{m}}\left(h\right)}
    \nonumber \\	
	& =
	\left[
   -\left(k_{1}^{2}-{{k}_{\mathrm{T3}}}^{2}\right)
	\sin{k}_{\mathrm{L3}}x_{3}
	\sin{k}_{\mathrm{T3}}h
	\right.
	\nonumber \\	
	& +
	\left.
	2k_{1}^{2}
	\sin{k}_{\mathrm{L3}}h
	\sin{k}_{\mathrm{T3}}x_{3}
	\right]
    \nonumber \\
    & \times 
    \left( 2\chi_{\mathrm{RT}}^{2}k_{1}^{2}
	\sin{k}_{\mathrm{L3}}h
	\sin{k}_{\mathrm{T3}}h
	\right)^{-1}
\\
\underline{\hat{u}}_{\parallel f\mathrm{m}}^{\mathrm{A0}}\left(x_{3}\right)	
     & =
	\frac
	{\hat{u}_{\parallel f\mathrm{m}}^{\mathrm{A0}}\left(x_{3}\right)}
	{\hat{u}_{\perp f\mathrm{m}}\left(h\right)}
	\nonumber \\ 
	& =
	-\mathrm{j}
	\left[
     \left(k_{1}^{2}-{{k}_{\mathrm{T3}}}^{2}\right)
	\sin{k}_{\mathrm{L3}}x_{3}
	\cos{k}_{\mathrm{T3}}h
	\right.
	\nonumber \\
	& +
	\left.
	2k_{\mathrm{L3}}k_{\mathrm{T3}}
	\cos{k}_{\mathrm{L3}}h
	\sin{k}_{\mathrm{T3}}x_{3}
	\right]
	\nonumber \\
	& \times 
	\left( 
    2\chi_{\mathrm{RT}}^{2} k_{\mathrm{L3}}k_{1}
	\cos{k}_{\mathrm{L3}}h
	\cos{k}_{\mathrm{T3}}h
	\right)^{-1}
\\
\underline{\hat{u}}_{\perp f\mathrm{m}}^{\mathrm{A0}}\left(x_{3}\right)	 
 & =
\frac
	{\hat{u}_{\perp f\mathrm{m}}^{\mathrm{A0}}\left(x_{3}\right)}
	{\hat{u}_{\perp f\mathrm{m}}\left(h\right)}
    \nonumber \\ 	
	& =
    \left[
     -\left(k_{1}^{2}-{{k}_{\mathrm{T3}}}^{2}\right)
	\cos{k}_{\mathrm{L3}}x_{3}
	\cos{k}_{\mathrm{T3}}h
	\right.
	\nonumber \\ 
	& +
	\left.
	2k_{1}^{2}
	\cos{k}_{\mathrm{L3}}h
	\cos{k}_{\mathrm{T3}}x_{3}
	\right]
	\nonumber \\ 
	& \times 
	\left(
    2\chi_{\mathrm{RT}}^{2}k_{1}^{2}
	\cos{k}_{\mathrm{L3}}h
	\cos{k}_{\mathrm{T3}}h
	\right)^{-1}
\end{align} 
\end{subequations}
%*******************************************************************
%***************************************
Once the particular characteristic values of the phase velocities of the plate $c_{\mathrm{L}}$,  $c_{\mathrm{T}}$ and $c_{\mathrm{R}}$ have been established (only two of them are independent as we have pointed out previously) these normalized distributions (figure~\ref{onda_Ray}) are completely determined by specifying, on the one hand, the thickness of the plate $2h$ and, on the other hand, the temporal frequency $f$ or the corresponding axial wavenumber $k_{1}$. Then, using (\ref{eq_u1y3m_norm_qR}) in 
(\ref{eq_u1y3m_qR}) results
%EQ.B11**************************************************************
%*******************************************************************
\begin{subequations}\label{eq_u1y3m_qR_rec}
\begin{align}
   \hat{u}_{\parallel f\mathrm{m}}
   \left(x_{3}\right)
   = 
   \left[
   \underline{\hat{u}}_{\parallel f\mathrm{m}}^{\mathrm{S0}}
   \left(x_{3}\right)
   + 
   \underline{\hat{u}}_{\parallel f\mathrm{m}}^{\mathrm{A0}}
   \left(x_{3}\right)
   \right]
   \hat{u}_{\perp f\mathrm{m}}
   \left(h\right)
\\
   \hat{u}_{\perp f\mathrm{m}}
   \left(x_{3}\right)
   = 
   \left[ 
   \underline{\hat{u}}_{\perp f\mathrm{m}}^{\mathrm{S0}}
   \left(x_{3}\right)
   + 
   \underline{\hat{u}}_{\perp f\mathrm{m}}^{\mathrm{A0}}
   \left(x_{3f\mathrm{m}}\right)
   \right]
   \hat{u}_{\perp f\mathrm{m}}
   \left(h\right)
\end{align}
\end{subequations}
%*******************************************************************
%*******************************************************************
and multiplying both sides by $\exp \left[ j\left(k_{1}x_{1}-2 \pi ft \right)\right]$ we obtain 
%EQ.B12**************************************************************
%*******************************************************************
\begin{subequations}\label{eq_u1y3_qR_rec}
\begin{align}
   \hat{u}_{\parallel f}
   \left(\mathbf{x},t\right)
   =
   \left[
   \underline{\hat{u}}_{\parallel f\mathrm{m}}^{\mathrm{S0}}
   \left(x_{3}\right)
   + 
   \underline{\hat{u}}_{\parallel f\mathrm{m}}^{\mathrm{A0}}
   \left(x_{3}\right)
   \right] 
   \hat{u}_{\perp f}
   \left(\mathbf{x}_{h},t\right)
\\
   \hat{u}_{\perp f}
   \left(\mathbf{x},t\right)
   = 
   \left[ 
   \underline{\hat{u}}_{\perp f\mathrm{m}}^{\mathrm{S0}}
   \left(x_{3}\right)
   + 
   \underline{\hat{u}}_{\perp f\mathrm{m}}^{\mathrm{A0}}
   \left(x_{3f}\right)
   \right] 
   \hat{u}_{\perp f}
   \left(\mathbf{x}_{h},t\right)
\end{align}
\end{subequations}
%*******************************************************************
%*******************************************************************
which allow us to reconstruct the in-plane and out-plane complex component of the quasi-Rayleigh plane harmonic wave  inside the plate in terms of its out-of-plane complex component at top surface of the plate and the normalized transversal distributions of S0 and A0 modes given by expression (\ref{eq_u1y3m_norm_qR}).

\setcounter{figure}{0}
% figura 16
% figura 12 Carlos: onda quasi-Rayleigh***********************************
%*****************************************************************
\begin{figure*}
\centering
\begin{tabular}{
c@{\hspace{0.2cm}}
c
}
   \multicolumn{2}{c}{\includegraphics 
     [width=53 mm,keepaspectratio=true]   
     {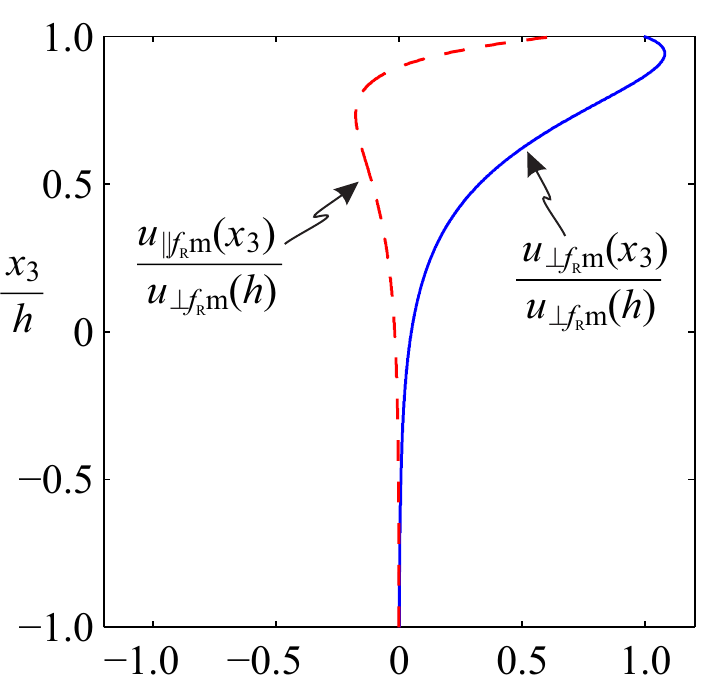}}
\\
   \multicolumn{2}{c}{(a)} 
\\
   \includegraphics 
    [width=53 mm,keepaspectratio=true]   
    {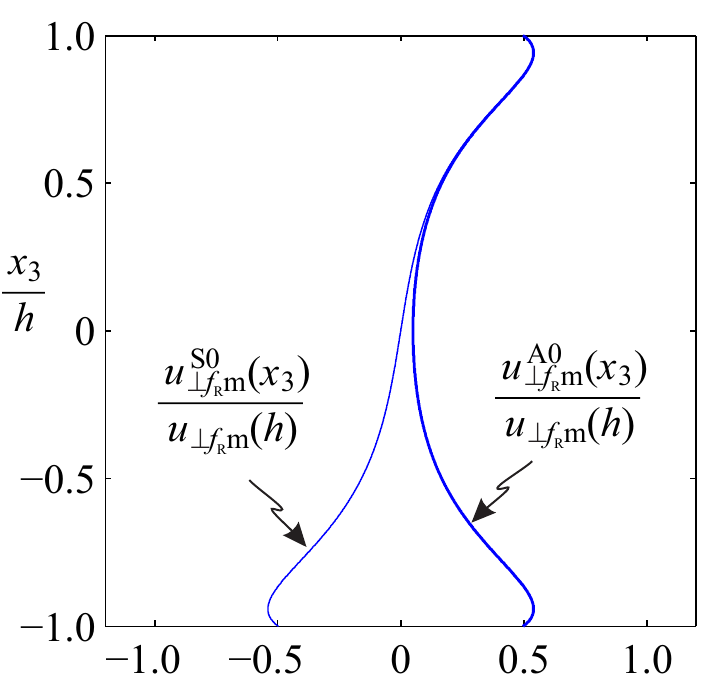}
   & \includegraphics 
    [width=53 mm,keepaspectratio=true]   
    {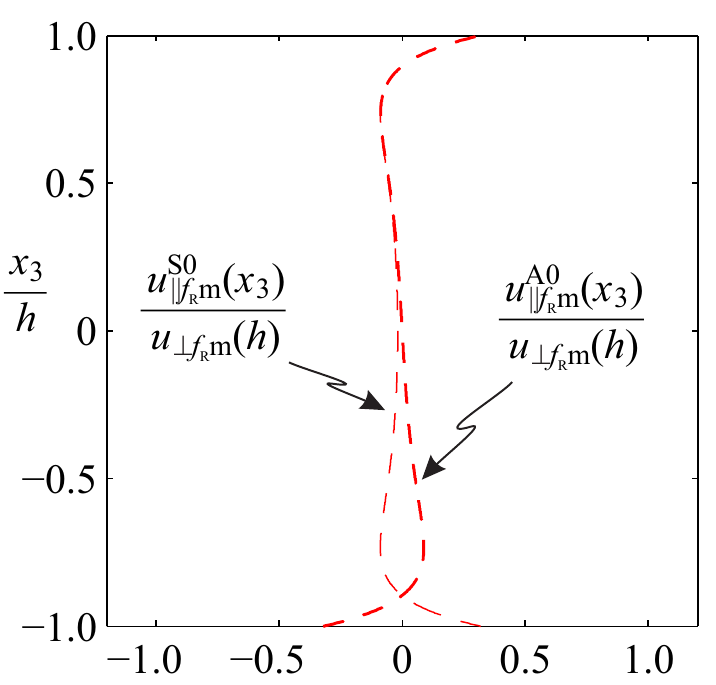}
\\
    (b)
    & (c)
\\
\end{tabular}
	\caption{Plane harmonic quasi-Rayleigh wave traveling in the $x_{1}$ direction of  a plate with thickness $2h$ for a normalized frequency $\gamma_{\mathrm{R}}=3.1$
(i.e. close to the situation in our experiments): (a) amplitudes of the in-plane $u_{\parallel f_{\mathrm{R}}\mathrm{m}}(x_{3})$ and out-of-plane $u_{\perp f_{\mathrm{R}}\mathrm{m}}(x_{3})$ displacements of the quasi-Rayleigh wave, (b) amplitude of the 
out-of-plane displacements $u_{\perp f_{\mathrm{R}}\mathrm{m}}^{\mathrm{S0}}(x_{3})$ and  $u_{\perp f_{\mathrm{R}}\mathrm{m}}^{\mathrm{A0}}(x_{3})$ of the S0 and A0 components of the quasi-Rayleigh wave, (c) amplitude of the 
in-plane displacements $u_{\parallel f_{\mathrm{R}}\mathrm{m}}^{\mathrm{S0}}(x_{3})$ and  $u_{\parallel f_{\mathrm{R}}\mathrm{m}}^{\mathrm{A0}}(x_{3})$ of the S0 and A0 components of the quasi-Rayleigh wave.
The amplitudes are normalized with respect to the amplitude of the out-of-plane displacement of the quasi-Rayleigh wave at the top surface of the pate $u_{\perp f_{\mathrm{R}}\mathrm{m}}(h)$. The amplitude is positive or negative when is in phase or out of phase by $\pi$ with respect to $u_{\perp f_{\mathrm{R}}\mathrm{m}}(h)$. The vertical coordinate is normalized with respect to $h$.}
	\label{onda_Ray}
\end{figure*}
%***************************************************************
%***************************************************************

Despite the fact we have considered in the previous analysis the case of a plane harmonic quasi-Rayleigh wave propagation along the $x_{1}$ direction, it can be shown that  expressions (\ref{eq_u1y3m_Ls})-(\ref{eq_u1y3_qR_rec}) are also valid for the case of cylindrical harmonic quasi-Rayleigh wave generated by a line source located in the $x_{3}$ axis, provided that we interchange coordinate $x_{1}$ by the radial cylindrical coordinate $r$ (i.e. the phase factor 
$\exp \left[ \mathrm{j}\left(k_{1}x_{1}-2 \pi ft \right)\right]$ by the phase factor $\exp \left[ \mathrm{j}\left(k_{1}r-2 \pi ft \right)\right]$) and taking into account that in this case the in-plane component is along the radial direction~\cite{Achenbach_R}. Hence, using the projection of the radial displacement along the $x_{1}$ and $x_{2}$ direction we obtain the Cartesian components of the complex displacement of the harmonic quasi-Rayleigh wave with cylindrical symmetry as
%EQ.B13**************************************************************
%*******************************************************************
\begin{subequations}\label{eq_ui_qQ_rec}
\begin{align}
   \hat{u}_{if}^{\mathrm{i}}
   \left(\mathbf{x},t\right)
   & =
   \left[
   \underline{\hat{u}}_{\parallel f_{\mathrm{m}}}^{\mathrm{S0}}
   \left(x_{3}\right)
   + 
   \underline{\hat{u}}_{\parallel f_{\mathrm{m}}}^{\mathrm{A0}}
   \left(x_{3}\right)
   \right]
   \nonumber \\
   & \times
   \left[
   \frac
   {
   \mathbf{x}_{h}
   }
   {
   \vert \mathbf{x}_{h}\vert
   }
   \cdot 
   \mathbf{a}_{i}
   \right]    
   \hat{u}_{3f}^{\mathrm{i}}
   \left(\mathbf{x}_{h},t\right) 
   \quad i=1,2
\\
   \hat{u}_{3f}^{\mathrm{i}}
   \left(\mathbf{x},t\right)
   & = 
   \left[ 
   \underline{\hat{u}}_{\perp f_{\mathrm{m}}}^{\mathrm{S0}}
   \left(x_{3}\right)
   + 
   \underline{\hat{u}}_{\perp f_{\mathrm{m}}}^{\mathrm{A0}}
   \left(x_{3}\right)\right] 
   \hat{u}_{3f}^{\mathrm{i}}
   \left(\mathbf{x}_{h},t\right) 
   \quad \quad \quad \quad
\end{align}
\end{subequations}
%*******************************************************************
%*******************************************************************
In the more general case that the line-source is not coincident with the $x_{3}$ axis and intersects the top surface of the plate at a virtual point source
$\mathrm{F}$ with position vector $\mathbf{x}_{h}^{\mathrm{F}}=\left({x}_{1}^{\mathrm{F}},{x}_{2}^{\mathrm{F}},h\right)$ (see figure~\ref{esquema_plate}.b) we have
%EQ.B14**************************************************************
%*******************************************************************
\begin{subequations}\label{eq_ui_qQ_gen}
\begin{align}
   \hat{u}_{if}^{\mathrm{i}}
   \left(\mathbf{x},t\right)
   & =
   \left[
   \underline{\hat{u}}_{\parallel f_{\mathrm{m}}}^{\mathrm{S0}}
   \left(x_{3}\right)
   + 
   \underline{\hat{u}}_{\parallel f_{\mathrm{m}}}^{\mathrm{A0}}
   \left(x_{3}\right)
   \right]
   \nonumber \\
   & \times
   \left[
   \frac
   {
   \mathbf{x}_{h}-\mathbf{x}^{\mathrm{^F}}
   }
   {
   \vert \mathbf{x}_{h}-\mathbf{x}^{\mathrm{^F}}\vert
   }
   \cdot 
   \mathbf{a}_{i}
   \right]    
   \hat{u}_{3f}^{\mathrm{i}}
   \left(\mathbf{x}_{h},t\right) 
   \quad i=1,2
\\
   \hat{u}_{3f}^{\mathrm{i}}
   \left(\mathbf{x},t\right)
   & = 
   \left[ 
   \underline{\hat{u}}_{\perp f_{\mathrm{m}}}^{\mathrm{S0}}
   \left(x_{3}\right)
   + 
   \underline{\hat{u}}_{\perp f_{\mathrm{m}}}^{\mathrm{A0}}
   \left(x_{3}\right)\right] 
   \hat{u}_{3f}^{\mathrm{i}}
   \left(\mathbf{x}_{h},t\right) 
   \quad \quad \quad \quad
\end{align}
\end{subequations}
%*******************************************************************
%*******************************************************************

In the case of a narrow band quasi-Rayleigh wavetrain, as the one described in subsection \ref{subsection2_B}, we can assume that the normalized transversal distribution of anyone of its harmonic components is nearly equal to the normalized transversal distribution of the harmonic component with frequency and wavenumber equal to the central frequency $f_{\mathrm{R}}$ and central wavenumber $k_{\mathrm{R}}$ of the train. In this conditions, the complex field of the each harmonic component can be approximated as
%EQ.B15**************************************************************
%*******************************************************************
\begin{subequations}\label{eq_ui_qQ_com}
\begin{align}
   \hat{u}_{if}^{\mathrm{i}}
   \left(\mathbf{x},t\right)
   & =
   \left[
   \underline{\hat{u}}_{\parallel f_{\mathrm{R}}\mathrm{m}}^{\mathrm{S0}}
   \left(x_{3}\right)
   + 
   \underline{\hat{u}}_{\parallel f_{\mathrm{R}}\mathrm{m}}^{\mathrm{A0}}
   \left(x_{3}\right)
   \right]
   \nonumber \\
   & \times
   \left[
   \frac
   {
   \mathbf{x}_{h}-\mathbf{x}^{\mathrm{^F}}_{h}
   }
   {
   \vert \mathbf{x}_{h}-\mathbf{x}^{\mathrm{^F}}_{h}\vert
   }
   \cdot 
   \mathbf{a}_{i}
   \right]    
   \hat{u}_{3f}^{\mathrm{i}}
   \left(\mathbf{x}_{h},t\right) 
   \quad i=1,2
\\
   \hat{u}_{3f}^{\mathrm{i}}
   \left(\mathbf{x},t\right)
   & = 
   \left[ 
   \underline{\hat{u}}_{\perp f_{\mathrm{R}}\mathrm{m}}^{\mathrm{S0}}
   \left(x_{3}\right)
   + 
   \underline{\hat{u}}_{\perp f_{\mathrm{R}}\mathrm{m}}^{\mathrm{A0}}
   \left(x_{3}\right)\right] 
   \hat{u}_{3f}^{\mathrm{i}}
   \left(\mathbf{x}_{h},t\right) 
   \quad \quad \quad \quad
\end{align}
\end{subequations}
%*******************************************************************
%**************************************
Using (\ref{eq_ui_qQ_com}) with (\ref{eq_uiQR}-\ref{eq_uifQR}) we get expression (\ref{eq_uqR_uqRh}).

%Bibliografia****************************************************
%%\bibliographystyle{elsarticle-num}  
%% \bibliography{MSSP_1}

%\begin{thebibliography}{10}
% Inserta la biblio sin bibtex una vez terminada la redaccion
% corta y pega del fichero draft_arXiv_3.bbl
%\end{thebibliography}



%% lista de figuras y tablas
%%\newpage
%%\listoffigures
%%\listoftables

\end{document}